\documentclass[aps,eqsecnum,preprint,floats,epsf,epsfig,nofootinbib,letter]{revtex4}
\usepackage{amsfonts}
\usepackage{amsmath}
\usepackage{amssymb}
\usepackage{mathptmx}
\usepackage[dvips]{color}
\usepackage{epsfig}
\usepackage{graphicx}
\begin{document}
\def\be{\begin{eqnarray}}
\def\en{\end{eqnarray}}
\def\la{\langle}
\def\ra{\rangle}
\def\non{\nonumber}
\def\A{{\cal A}}
\def\B{{\cal B}}
\def\ov{\overline}
\def\up{\uparrow}
\def\dw{\downarrow}
\def\vma{{_{V-A}}}
\def\vpa{{_{V+A}}}
\def\smp{{_{S-P}}}
\def\spp{{_{S+P}}}
\def\vp{\varepsilon}
\def\CP{{\it CP}~}
\def\pr{{Phys. Rev.}~}
\def\prl{{ Phys. Rev. Lett.}~}
\def\pl{{ Phys. Lett.}~}
\def\np{{ Nucl. Phys.}~}
\def\zp{{ Z. Phys.}~}
\def\lsim{ {\ \lower-1.2pt\vbox{\hbox{\rlap{$<$}\lower5pt\vbox{\hbox{$\sim$}
}}}\ } }
\def\gsim{ {\ \lower-1.2pt\vbox{\hbox{\rlap{$>$}\lower5pt\vbox{\hbox{$\sim$}
}}}\ } }

\newcommand{\acp}{\ensuremath{A_{ CP}}}

\font\el=cmbx10 scaled \magstep2{\obeylines\hfill September, 2009}

\vskip 1.5 cm

\centerline{\large\bf Revisiting  Charmless Hadronic $B_{u,d}$ Decays in QCD Factorization}
\bigskip
\centerline{\bf Hai-Yang Cheng,$^{1,2}$ Chun-Khiang Chua$^3$}
\medskip
\centerline{$^1$ Institute of Physics, Academia Sinica}
\centerline{Taipei, Taiwan 115, Republic of China}
\medskip
\medskip
\centerline{$^2$ Physics Department, Brookhaven National
Laboratory} \centerline{Upton, New York 11973}
\medskip
\medskip
\centerline{$^3$ Department of Physics, Chung Yuan Christian University}
\centerline{Chung-Li, Taiwan 320, Republic of China}
\medskip

\centerline{\bf Abstract}
\bigskip
\small

Within the framework of QCD factorization (QCDF), we consider two different types of power correction effects in order to resolve the \CP puzzles and rate deficit problems with penguin-dominated two-body decays of $B$ mesons  and color-suppressed tree-dominated $\pi^0\pi^0$ and $\rho^0\pi^0$ modes: penguin annihilation and soft corrections to
the color-suppressed tree amplitude. We emphasize that the electroweak penguin solution to the $B\to K\pi$ \CP puzzle via New Physics is irrelevant for solving the \CP and rate puzzles related to tree-dominated decays.  While some channels e.g. $K^-\pi^+,K^-\rho^0,\pi^+\pi^-,\rho^\pm\pi^\mp$ need penguin annihilation to induce the correct magnitudes and signs for their \CP violation, some other decays such as $B^-\to K^-\pi^0,\pi^-\eta, K^-\eta$ and $\bar B^0\to \bar K^{*0}\eta,\pi^0\pi^0$ require the presence of both power corrections to account for the measured \CP asymmetries. In general, QCDF predictions for the branching fractions and direct \CP asymmetries of $\bar B\to PP,VP,VV$ decays are in good agreement with experiment. The predictions of pQCD and soft-collinear effective theory are included for comparison.

\pagebreak

\section {Introduction}

Although the underlying dynamics for the hadronic $B$ decays is extremely complicated, it is greatly simplified in the heavy quark limit.
In the $m_b\to\infty$ limit, hadronic matrix elements can be expressed in terms of certain nonperturbative input quantities such as light cone distribution amplitudes and transition form factors. Consequently,
the decay amplitudes of charmless two-body decays of $B$ mesons can be described  in terms of decay constants and form factors. However, the leading-order $1/m_b$ predictions encounter three major difficulties: (i) the predicted branching fractions for penguin-dominated $\bar B\to PP,VP,VV$ decays are systematically below the measurements \cite{BN} and the rates for color-suppressed tree-dominated decays $\bar B^0\to\pi^0\pi^0,\rho^0\pi^0$ are too small,   (ii) direct {\it CP}-violating asymmetries for $\bar B\to K^-\pi^+$, $\bar B\to K^{*-}\pi^+$, $B^-\to K^-\rho^0$, $\bar B\to \pi^+\pi^-$ and $\bar B_s\to K^+\pi^-$ disagree with experiment in signs, and (iii) the transverse polarization fraction in penguin-dominated charmless $B\to VV$ decays is predicted to be very small, while experimentally it is comparable to the longitudinal polarization one. All these indicate the necessity of going beyond zeroth $1/m_b$ power expansion.

In the QCD factorization (QCDF) approach \cite{BBNS}, power corrections often involve endpoint
divergences. For example,  the hard spectator
scattering diagram at twist-3 order is power suppressed and posses
soft and collinear divergences arising from the soft spectator
quark and the $1/m_b$ annihilation amplitude has
endpoint divergences even at twist-2 level. Since the treatment of endpoint divergences is model
dependent, subleading power corrections generally can be studied
only in a phenomenological way. Therefore, $1/m_b$ power suppressed effects are generally nonperturbative in nature and hence not calculable by the perturbative method.

As a first step, let us consider power corrections to the QCD penguin amplitude of the $\bar B\to PP$ decay which has the generic expression
\be \label{eq:P}
P&=& P_{\rm SD}+P_{\rm LD}, \non \\
&=& A_{PP}[\lambda_u(a_4^u+r_\chi^P a_6^u)+\lambda_c(a_4^c+r_\chi^P a_6^c)]+1/m_b~{\rm corrections},
\en
where $\lambda_p^{(q)}=V_{pb}V_{pq}^*$ with $q=s,d$, $a_{4,6}$ are the effective parameters to be defined below and  $r^P_\chi$ is a chiral factor of order unity.
Strictly speaking, the penguin contributions associated with the chiral factor $r_\chi^P$  are formerly $1/m_b$ suppressed but chirally enhanced. Since they are of order $1/m_b^0$ numerically, their effects are included in the zeroth order calculation. Possible power corrections to penguin amplitudes include long-distance charming penguins, final-state interactions and  penguin annihilation characterized by the parameters $\beta_3^{u,c}$. Because of possible ``double counting" problems, one should not take into account all power correction effects simultaneously.  As we shall see below in Sec. IV.B, \CP violation of $K^-\pi^+$ and $\pi^+\pi^-$ arise from the interference between the tree amplitude $\lambda_u^{(q)}a_1$ and the penguin amplitude $\lambda_c^{(q)}(a_4^c+r_\chi^P a_6^c)$ with $q=s$ for the former and $q=d$ for the latter. The short-distance contribution to $a_4^c+r_\chi^P a_6$ will yield a positive $\acp(K^-\pi^+)$ and a negative $\acp(\pi^+\pi^-)$.  Both are wrong in signs when confronted with experiment.
In the so-called ``S4" scenario of QCDF \cite{BN}, power corrections to the penguin annihilation topology characterized by $\lambda_u\beta_3^u+\lambda_c\beta_3^c$ are added to Eq. (\ref{eq:P}). By adjusting the magnitude and phase of $\beta_3$ in this scenario, all the above-mentioned discrepancies except for the rate deficit problem with the decays $\bar B^0\to\pi^0\pi^0,\rho^0\pi^0$ can be resolved.

However, a scrutiny of the QCDF predictions reveals more puzzles in the regard of direct \CP violation. While the signs of \CP asymmetries in $K^-\pi^+,K^-\rho^0$ modes are flipped to the right ones in the presence of power corrections from penguin annihilation,
the signs of $\acp$ in $B^-\to K^-\pi^0,~K^-\eta,~\pi^-\eta$ and $\bar B^0\to\pi^0\pi^0,~\bar K^{*0}\eta$ will also get reversed in such a way that they disagree with experiment. In other words, in the heavy quark limit the \CP asymmetries of these five  modes  have the right signs when compared with experiment.

The so-called $B\to K\pi$ {\it CP}-puzzle is related to the difference of \CP asymmetries of $B^-\to K^-\pi^0$ and $\bar B^0\to K^-\pi^+$.
This can be illustrated by considering the decay amplitudes of $\bar B\to \bar K\pi$ in terms of topological diagrams
 \begin{eqnarray} \label{eq:ampBKpi}
 A(\bar B^0\to K^-\pi^+) &=& P'+T'+{2\over 3}P'^c_{\rm EW}+P'_A, \nonumber \\
 A(\bar B^0\to \bar K^0\pi^0) &=& {-1\over \sqrt{2}}(P'-C'-P'_{\rm EW}-{1\over 3}P'^c_{\rm EW}+P'_A), \\
 A(B^-\to \bar K^0\pi^-) &=& P'-{1\over 3}P'^c_{\rm EW}+A'+P'_A,  \nonumber \\
 A(B^-\to K^-\pi^0) &=& {1\over\sqrt{2}}(P'+T'+C'+P'_{\rm
 EW}+{2\over 3}P'^c_{\rm EW}+A'+P'_A), \non
 \end{eqnarray}
where $T$, $C$, $E$, $A$, $P_{\rm EW}$ and $P^c_{\rm EW}$ are color-allowed tree, color-suppressed tree, $W$-exchange, $W$-annihilation, color-allowed and color-suppressed electroweak
penguin amplitudes, respectively, and
$P_A$ is the penguin-induced weak annihilation amplitude. We use unprimed and
primed symbols to denote $\Delta S=0$ and $|\Delta S|=1$ transitions, respectively. We notice that if $C'$, $P'_{\rm EW}$
and $A'$ are negligible compared with $T'$, it is
clear from Eq. (\ref{eq:ampBKpi}) that the decay amplitudes of $K^-\pi^0$ and
$K^-\pi^+$ will be the same apart from a trivial factor of $1/\sqrt{2}$. Hence,
one will expect that $\acp(K^-\pi^0)\approx \acp(K^-\pi^+)$, while they
differ by 5.3$\sigma$ experimentally,
$\Delta A_{K\pi}\equiv\acp(K^-\pi^0)-\acp(K^-\pi^+)=0.148\pm0.028$ \cite{HFAG}.

The aforementioned direct \CP puzzles indicate that it is necessary to consider subleading power corrections other than penguin annihilation. For example, the large power corrections due to $P'$ cannot explain the $\Delta A_{K\pi}$ puzzle as they contribute equally to both $B^-\to K^-\pi^0$ and $\bar B^0\to K^-\pi^+$. The additional power correction should have little effect on the decay rates of penguin-dominated decays but will manifest in the measurement of direct \CP asymmetries.
Note that all the "problematic" modes receive a contribution from $c^{(')}=C^{(')}+P_{\rm EW}^{(')}$. Since
$A(B^-\to K^-\pi^0)\propto t'+c'+p'$ and $A(\bar B^0\to K^-\pi^+)\propto t'+p'$ with $t'=T'+P'^c_{\rm EW}$ and
$p'=P'-{1\over 3}P'^c_{\rm EW}+P'_A$, we can consider this puzzle resolved, provided that $c'/t'$ is of order
$1.3\sim 1.4$ with a large negative phase ($|c'/t'|\sim 0.9$ in the standard short-distance effective Hamiltonian approach). There are several  possibilities for a
large complex $c'$: either a large complex $C'$ or a large complex electroweak penguin
$P'_{\rm EW}$ or a combination of them. Various scenarios
for accommodating large $C'$ \cite{Li05,Kim,Rosner,Chua,Ciuchini,Kagan,Li09,Baek09} or $P'_{\rm EW}$ \cite{LargeEWP,Hou} have
been proposed. To get a large complex $C'$, one can resort to spectator scattering or final-state interactions (see discussions in Sec.3.E). However, the general consensus for a large complex $P'_{\rm EW}$ is that one needs New Physics beyond the Standard Model because it is well known that $P'_{\rm EW}$ is essentially real in the SM as it does not carry a nontrivial strong phase \cite{NR}.
In principle, one cannot discriminate between these two  possibilities in penguin-dominated decays as it is always the combination $c'=C'+P'_{\rm EW}$ that enters into the decay  amplitude except for the decays involving
$\eta$ and/or $\eta'$ in the final state where
%
%
both $c'$ and $P'_{\rm EW}$ present in the
amplitudes~\cite{Chiang}.  Nevertheless, these two scenarios will lead to very distinct predictions for tree-dominated decays where $P_{\rm EW}\ll C$. (In penguin-dominated decays, $P'_{\rm EW}$ is comparable to $C'$ due to the fact that $\lambda_c^{(s)}\gg\lambda_u^{(s)}$.) The decay rates of $\bar B^0\to \pi^0\pi^0,\rho^0\pi^0$ will be substantially enhanced for a large $C$ but remain intact for a large $P_{\rm EW}$. Since $P_{\rm EW}\ll C$ in tree-dominated
channels, \CP puzzles with $\pi^-\eta$ and $\pi^0\pi^0$ cannot be
resolved with a large $P_{\rm EW}$. Therefore, it is most likely
that the  color-suppressed tree amplitude is large and complex.
In other words, the $B\to K\pi$ \CP puzzle can be resolved without invoking New Physics.

In this work we shall consider
the possibility of having a large color-suppressed tree amplitude with a sizable strong phase relative to the color-allowed tree amplitude \cite{CCcp}
\be
C=[\lambda_u a_2^u]_{\rm SD}+[\lambda_u a_2^u]_{\rm LD}+{\rm FSIs}+\cdots.
\en
As will be discussed below, the long-distance contribution to $a_2$ can come from the twist-3 effects in spectator rescattering, while an example of final-state rescattering contribution to $C$ will be illustrated below.

Note that our phenomenological study of power corrections to penguin annihilation and to color-suppressed tree topology is in the same spirit of S4 and S2 scenarios, respectively, considered by Beneke and Neubert \cite{BN}. In the ``large $\alpha_2$" S2 scenario, the ratio $a_2/a_1$ is enhanced basically by having a smaller $\lambda_B$ and a smaller strange quark mass. It turns out that the \CP asymmetries of $K^-\pi^+,K^{*-}\pi^+,K^-\eta,K^-\rho^0,\pi^+\pi^-$ have correct signs in S4 but not so in S2, whereas the signs of $\acp(K^-\pi^0), \acp(K^-\eta),\acp(\pi^0\pi^0)$ in S2 (or in the heavy quark limit) agree with experiment but not in S4. In a sense, our study is a combination of S4 and S2. However, there is a crucial difference between our work and \cite{BN}, namely, our $a_2$ is not only large in the magnitude but also has a large strong phase. As we shall see, a large and {\it complex} $a_2$ is needed to account for all the remaining \CP puzzles.

It should be remarked that the aforementioned $B$-{\it CP} puzzles with the $K^-\pi^0,~K^-\eta,~\pi^-\eta,~\bar K^{*0}\eta,~\pi^0\pi^0$ modes also occur in the approach of soft-collinear effective theory (SCET) \cite{SCET} where the penguin annihilation effect in QCDF is replaced by the long-distance charming penguins. Owing to a different treatment of endpoint divergence in penguin annihilation diagrams, some of the \CP puzzles do not occur in the approach of pQCD \cite{pQCD}. For example, pQCD predicts the right sign for \CP asymmetries of $\bar B^0\to \pi^0\pi^0$ and $B^-\to \pi^-\eta$ as we shall see below.
In this work, we shall show that soft power correction to the color-suppressed tree amplitude will bring the signs of $\acp$ back to the right track. As a bonus, the rates of $\bar B^0\to\pi^0\pi^0,~\rho^0\pi^0$ can be accommodated.

In the past decade, nearly 100 charmless decays of $B_{u,d}$ mesons  have been observed at $B$ factories with
a statistical significance of at least four standard deviations (for a review, see \cite{ChengSmith}). Before moving to the era of LHCb and Super $B$ factories in the next few years, it is timing to have an overview on charmless hadronic $B$ decays to see what we have learned from the fruitful experimental results obained by BaBar and Belle. In this work, we will update QCDF calculations and compare with experiment and  other theoretical predictions.

This work is organized as follows. We outline the QCDF framework in Sec. 2 and specify various input parameters, such as form factors, LCDAs and the parameters for power corrections in Sec. 3. Then $B_{u,d}\to PP,VP,VV$ decays are analyzed in details in Secs. 4, 5 and 6, respectively.  Conclusions are given in Sec. 7.

\section{$B$ decays in QCD factorization}
Within the framework of QCD factorization \cite{BBNS}, the effective
Hamiltonian matrix elements are written in the form
\begin{equation}\label{fac}
   \langle M_1M_2 |{\cal H}_{\rm eff}|\overline B\rangle
  \! =\! \frac{G_F}{\sqrt2}\sum_{p=u,c} \! \lambda_p^{(q)}\,
\!   \langle M_1M_2 |{\cal T_A}^{h,p}\!+\!{\cal
T_B}^{h,p}|\overline B\rangle \,,
\end{equation}
where  the
superscript $h $ denotes the helicity of the final-state meson.
For $PP$ and $VP$ final states, $h=0$. ${\cal
T_A}^{h,p}$ describes contributions from naive factorization, vertex
corrections, penguin contractions and spectator scattering expressed
in terms of the flavor operators $a_i^{p,h}$, while ${\cal T_B}$
contains annihilation topology amplitudes characterized by  the
annihilation operators $b_i^{p,h}$. Specifically \cite{BBNS},
\be
{\cal T_A}^h &=& a_1^p(M_1M_2)\delta_{pu}(\bar u b)_\vma\otimes (\bar qu)_\vma + a_2^p(M_1M_2)\delta_{pu}(\bar q b)_\vma\otimes (\bar uu)_\vma   \non \\
&+& a_3^p(M_1M_2)\sum(\bar q b)_\vma\otimes (\bar q'q')_\vma + a_4^p(M_1M_2)\sum(\bar q' b)_\vma\otimes (\bar qq')_\vma   \non \\
&+& a_5^p(M_1M_2)\sum(\bar q b)_\vma\otimes (\bar q'q')_\vpa + a_6^p(M_1M_2)\sum(-2)(\bar q' b)_\smp\otimes (\bar qq')_\spp   \\
&+& a_7^p(M_1M_2)\sum(\bar q b)_\vma\otimes {3\over 2}e_q(\bar q'q')_\vpa + a_8^p(M_1M_2)\sum(-2)(\bar q' b)_\smp\otimes {3\over 2}(\bar qq')_\spp   \non \\
&+& a_9^p(M_1M_2)\sum(\bar q b)_\vma\otimes {3\over 2}e_q(\bar q'q')_\vma + a_{10}^p(M_1M_2)\sum(\bar q' b)_\vma\otimes {3\over 2}e_q(\bar qq')_\vma,  \non
\en
where $(\bar q_1q_2)_{_{V\pm A}}\equiv \bar q_1\gamma_\mu(1\pm\gamma_5)q_2$ and $(\bar q_1q_2)_{_{S\pm P}}\equiv\bar q_1(1\pm\gamma_5)q_2$ and the summation is over $q'=u,d,s$. The symbol $\otimes$ indicates that the matrix elements of the operators in ${\cal T_A}$ are to be evaluated in the factorized form. For the decays $\bar B\to PP,VP,VV$, the relevant factorizable matrix elements are
\be
\label{eq:X}
X^{(\bar B P_1,P_2)} &\equiv& \la P_2|J^{\mu}|0\ra\la P_1|J'_{\mu}|\ov B\ra=if_{P_2}(m_{B}^2-m^2_{P_1}) F_0^{ B P_1}(m_{P_2}^2),  \non \\
X^{(\bar BP,V)} &\equiv & \la V| J^{\mu}|0\ra\la
P|J'_{\mu}|\ov B \ra=2f_V\,m_Bp_c F_1^{ B
P}(m_{V}^2),   \non \\
X^{( \bar BV,P)} &\equiv &
\la P | J^{\mu}|0\ra\la V|J'_{\mu}|\ov B
\ra=2f_P\,m_Bp_cA_0^{B V}(m_{P}^2),  \non \\
X_h^{( \bar BV_1,V_2)} &\equiv & \la V_2 |J^{\mu}|0\ra\la
V_1|J'_{\mu}|\ov B \ra =- if_{V_2}m_2\Bigg[
(\vp^*_1\cdot\vp^*_2) (m_{B}+m_{V_1})A_1^{ BV_1}(m_{V_2}^2)  \non \\
&-& (\vp^*_1\cdot p_{_{B}})(\vp^*_2 \cdot p_{_{B}}){2A_2^{
BV_1}(m_{V_2}^2)\over (m_{B}+m_{V_1}) } +
i\epsilon_{\mu\nu\alpha\beta}\vp^{*\mu}_2\vp^{*\nu}_1p^\alpha_{_{B}}
p^\beta_1\,{2V^{ BV_1}(m_{V_2}^2)\over (m_{B}+m_{V_1}) }\Bigg],
\en
where we have followed the conventional definition for form factors \cite{BSW}. For $B\to VP,PV$ amplitudes, we have applied the
replacement $m_V(\vp^*\cdot p_B)\to
m_Bp_c$ with $p_c$ being the c.m. momentum. The longitudinal ($h=0$) and transverse ($h=\pm$) components of $X^{( \bar BV_1,V_2)}_h$ are given by
 \be \label{eq:Xh}
 X_0^{(\ov BV_1,V_2)} &=& {if_{V_2}\over 2m_{V_1}}\left[
 (m_B^2-m_{V_1}^2-m_{V_2}^2)(m_B+m_{V_1})A_1^{BV_1}(q^2)-{4m_B^2p_c^2\over
 m_B+m_{V_1}}A_2^{BV_1}(q^2)\right], \non \\
 X_\pm^{(\ov BV_1,V_2)} &=& -if_{V_2}m_Bm_{V_2}\left[
 \left(1+{m_{V_1}\over m_B}\right)A_1^{BV_1}(q^2)\mp{2p_c\over
 m_B+m_{V_1}}V^{BV_1}(q^2)\right].
 \en

The flavor operators $a_i^{p,h}$ are basically the Wilson coefficients
in conjunction with short-distance nonfactorizable corrections such
as vertex corrections and hard spectator interactions. In general,
they have the expressions \cite{BBNS,BN}
 \be \label{eq:ai}
  a_i^{p,h}(M_1M_2) =
 \left(c_i+{c_{i\pm1}\over N_c}\right)N_i^h(M_2)
  + {c_{i\pm1}\over N_c}\,{C_F\alpha_s\over
 4\pi}\Big[V_i^h(M_2)+{4\pi^2\over N_c}H_i^h(M_1M_2)\Big]+P_i^{h,p}(M_2),
 \en
where $i=1,\cdots,10$,  the upper (lower) signs apply when $i$ is
odd (even), $c_i$ are the Wilson coefficients,
$C_F=(N_c^2-1)/(2N_c)$ with $N_c=3$, $M_2$ is the emitted meson
and $M_1$ shares the same spectator quark with the $B$ meson. The
quantities $V_i^h(M_2)$ account for vertex corrections,
$H_i^h(M_1M_2)$ for hard spectator interactions with a hard gluon
exchange between the emitted meson and the spectator quark of the
$B$ meson and $P_i(M_2)$ for penguin contractions.   The expression
of the quantities $N_i^h(M_2)$ reads
 \be
 N_i^h(M_2)=\begin{cases} 0, & $i=6,8$, \cr
                 1, & {\rm else}. \cr \end{cases}
 \en

 The weak annihilation contributions to the decay  $\overline B\to
M_{1}M_2$ can be described in terms of the building blocks $b_i^{p,h}$ and $b_{i,{\rm EW}}^{p,h}$
\begin{eqnarray}\label{eq:h1ksann}
\frac{G_F}{\sqrt2} \sum_{p=u,c} \! \lambda_p\, \!\langle M_{1}M_2
|{\cal T_B}^{h,p} |\overline B^0\rangle &=&
i\frac{G_F}{\sqrt{2}}\sum_{p=u,c} \lambda_p
 f_B f_{M_1} f_{M_{2}}\sum_i (d_ib_i^{p,h}+d'_ib_{i,{\rm EW}}^{p,h}).
\end{eqnarray}
The building blocks have the expressions \cite{BN}
 \be \label{eq:bi}
 b_1 &=& {C_F\over N_c^2}c_1A_1^i, \qquad\quad b_3={C_F\over
 N_c^2}\left[c_3A_1^i+c_5(A_3^i+A_3^f)+N_cc_6A_3^f\right], \non \\
 b_2 &=& {C_F\over N_c^2}c_2A_1^i, \qquad\quad b_4={C_F\over
 N_c^2}\left[c_4A_1^i+c_6A_2^f\right], \non \\
 b_{\rm 3,EW} &=& {C_F\over
 N_c^2}\left[c_9A_1^{i}+c_7(A_3^{i}+A_3^{f})+N_cc_8A_3^{i}\right],
 \non \\
 b_{\rm 4,EW} &=& {C_F\over
 N_c^2}\left[c_{10}A_1^{i}+c_8A_2^{i}\right].
 \en
Here for simplicity we have omitted the superscripts $p$ and $h$ in above expressions.  The subscripts 1,2,3 of $A_n^{i,f}$ denote the annihilation
amplitudes induced from $(V-A)(V-A)$, $(V-A)(V+A)$ and $(S-P)(S+P)$ operators,
respectively, and the superscripts $i$ and $f$ refer to gluon emission from the
initial and final-state quarks, respectively. Following \cite{BN} we choose the
convention that $M_1$  contains an antiquark from the weak vertex and $M_2$
contains a quark from the weak vertex.

For the explicit expressions of vertex, hard spectator corrections and annihilation contributions, the reader is referred to \cite{BBNS,BN,BRY} for details. The decay amplitudes of $B\to PP,VP$ are given in Appendix A of \cite{BN} and can be easily generalized to $B\to VV$ (see
\cite{BuchallaVV} for explicit expressions of $B\to VV$ amplitudes). In practice, it is more convenient to express the decay amplitudes in terms of the flavor operators $\alpha_i^{h,p}$ and the annihilation operators  $\beta_i^p$ which are related to the coefficients $a_i^{h,p}$ and $b_i^p$ by
\begin{eqnarray}\label{eq:alphai}
   \alpha_1^{h}(M_1 M_2) &=& a_1^{h}(M_1 M_2) \,, \nonumber\\
   \alpha_2^{h}(M_1 M_2) &=& a_2^{h}(M_1 M_2) \,, \nonumber\\
   \alpha_3^{h,p}(M_1 M_2) &=& \left\{
    \begin{array}{cl}
     a_3^{h,p}(M_1 M_2) - a_5^{h,p}(M_1 M_2)
      & \quad \mbox{for~} M_1 M_2=PP, \, VP , \\
     a_3^{h,p}(M_1 M_2) + a_5^{h,p}(M_1 M_2)
      & \quad \mbox{for~} M_1 M_2=VV,\, PV ,
    \end{array}\right. \nonumber\\
   \alpha_4^{h,p}(M_1 M_2) &=& \left\{
    \begin{array}{cl}
     a_4^{h,p}(M_1 M_2) + r_{\chi}^{M_2}\,a_6^{h,p}(M_1 M_2)
      & \quad \mbox{for~} M_1 M_2=PP, \, PV , \\
     a_4^{h,p}(M_1 M_2) - r_{\chi}^{M_2}\,a_6^{h,p}(M_1 M_2)
      & \quad \mbox{for~} M_1 M_2=VP\,, VV,
    \end{array}\right.\\
   \alpha_{3,\rm EW}^{h,p}(M_1 M_2) &=& \left\{
    \begin{array}{cl}
     a_9^{h,p}(M_1 M_2) - a_7^{h,p}(M_1 M_2)
      & \quad \mbox{for~} M_1 M_2=PP, \, VP , \\
     a_9^{h,p}(M_1 M_2) + a_7^{h,p}(M_1 M_2)
      & \quad \mbox{for~} M_1 M_2=VV,\, PV ,
    \end{array}\right. \nonumber\\
   \alpha_{4,\rm EW}^{h,p}(M_1 M_2) &=& \left\{
    \begin{array}{cl}
     a_{10}^{h,p}(M_1 M_2) + r_{\chi}^{M_2}\,a_8^{h,p}(M_1 M_2)
      & \quad \mbox{for~} M_1 M_2=PP, \, PV , \\
     a_{10}^{h,p}(M_1 M_2) - r_{\chi}^{M_2}\,a_8^{h,p}(M_1 M_2)
      & \quad \mbox{for~} M_1 M_2=VP\,, VV,
     \end{array}\right. \nonumber
\end{eqnarray}
and
\be \label{eq:beta}
 \beta_i^p (M_1 M_2) =\frac{i f_B f_{M_1}
f_{M_2}}{X^{(\overline B M_1,M_2)}}b_i^p.
\en
The order of the arguments of $\alpha_i^p (M_1 M_2)$ and $\beta_i^p(M_1
M_2)$ is consistent with the order of the arguments of
$X^{(\overline B M_1, M_2)}\equiv A_{M_1M_2}$.
The chiral factor $r_\chi$ is given by
\be \label{eq:rchi}
 r_\chi^P(\mu)={2m_P^2\over m_b(\mu)(m_2+m_1)(\mu)},  \qquad r_\chi^V(\mu) = \frac{2m_V}{m_b(\mu)}\,\frac{f_V^\perp(\mu)}{f_V} \,.
\en

The Wilson coefficients $c_i(\mu)$ at various scales, $\mu=4.4$ GeV, 2.1 GeV,
1.45 GeV and 1 GeV are taken from \cite{Groot}. For the renormalization scale
of the decay amplitude, we choose $\mu=m_b(m_b)$. \footnote{In principle, physics should
be independent of the choice of $\mu$, but in practice there exists some
residual $\mu$ dependence in the truncated calculations. We have checked
explicitly that the decay rates without annihilation are indeed essentially
stable against $\mu$. However, when penguin annihilation is turned on, it is
sensitive to the choice of the renormalization scale because the penguin
annihilation contribution characterized by the parameter $b_3$ is dominantly proportional to $\alpha_s(\mu_h)c_6(\mu_h)$  at the hard-collinear scale
$\mu_h=\sqrt{\mu\Lambda_h}$. In our study of $B\to VV$ decays \cite{ChengVV},  we found that if the renormalization scale is chosen to be $\mu=m_b(m_b)/2=2.1$ GeV, we cannot fit the branching ratios and polarization
fractions simultaneously for both $B\to K^*\phi$ and $B\to K^*\rho$ decays. In order to ensure the validity of the
penguin-annihilation mechanism for describing $B\to VV$ decays, we will confine
ourselves to the renormalization scale $\mu=m_b(m_b)$ in the ensuing study.}
However, as stressed in \cite{BBNS}, the hard spectator and annihilation contributions should be evaluated at
the hard-collinear scale $\mu_h=\sqrt{\mu\Lambda_h}$ with $\Lambda_h\approx 500$ MeV.

\section{Input parameters}

\subsection{Form factors}

There exist many model calculations of form factors for $B\to P,V$ transitions.  For $B\to P$ transitions, recent light-cone sum rule results for form factors at $q^2=0$ are collected in Table \ref{tab:FF:BP}. A small $F_0^{B\pi}$ of order 0.25 is also preferred by the measurement of $B^-\to \pi^-\pi^0$.
It is more convenient to express the form factors for $B\to \eta^{(')}$ transitions  in terms of the flavor states   $q\bar q\equiv (u\bar u+d\bar
d)/\sqrt{2}$, $s\bar s$ and $c\bar c$ labeled by the $\eta_q$, $\eta_s$ and $\eta_{c}^0$, respectively. Neglecting the small mixing with $\eta_c^0$, we have
\be \label{eq:FFeta}
F^{B\eta}=F^{B\eta_q}\cos\theta, \quad
F^{B\eta'}=F^{B\eta_q}\sin\theta,
\en
where $\theta$ is the $\eta_q-\eta_s$ mixing angle defined by
\be
|\eta\ra &=& \cos\theta|\eta_q\ra-\sin\theta|\eta_s\ra, \non \\
|\eta'\ra &=& \sin\theta|\eta_q\ra+\cos\theta|\eta_s\ra,
\en
with $\theta=(39.3\pm1.0)^\circ$ in the Feldmann-Kroll-Stech mixing scheme \cite{FKS}. From the sum rule results shown in Table \ref{tab:FF:BP} we obtain $F_0^{B\eta_q}(0)=0.296$.  The flavor-singlet contribution to the $B\to\eta^{(')}$ form factors is characterized by the parameter $B_2^g$, a gluonic Gegenbauer moment. It appears that the singlet contribution to the form factor is small unless $B_2^g$ assumes extreme values $\sim 40$ \cite{Ball:Beta}.

The $B\to \pi,K,\eta_q$ transition form factors to be used in this work are dispalyed in Table \ref{tab:input}.
We shall use the form factors determined from QCD sum rules for $B\to V$ transitions \cite{Ball:BV}.

\begin{table}[h!]
 \caption{Form factors for $B\to P$ transitions obtained in the QCD sum rules with $B_2^g$ being the gluonic Gegenbauer moment.
 } \label{tab:FF:BP}
\begin{center}
 \begin{tabular}{| c | l c |} \hline
$F_0^{B\pi}(0)$ & ~~$0.258\pm0.031$ \cite{Ball:PP} & ~~$0.26^{+0.04}_{-0.03}$  \cite{Melic:Bpi}~~~ \\ \hline
$F_0^{BK}(0)$ & ~~$0.331\pm0.041$ \cite{Ball:PP} & $0.36^{+0.05}_{-0.04}$ \cite{Melic:BK} \\ \hline
$F_0^{B\eta}(0)$ & ~~$0.229\pm0.024\pm0.011$ \cite{Ball:Beta}~~~ & \\ \hline $F_0^{B\eta'}(0)$ & ~~$0.188\pm 0.002B_2^g\pm0.019\pm0.009$  \cite{Ball:Beta} & \\   \hline
 \end{tabular}
 \end{center}
 \end{table}

\subsection{Decay constants}

 Decay constants of various vector mesons defined by
 \be \label{eq:decayc}
 \la V(p,\epsilon)|\bar q_2\gamma_\mu q_1|0\ra &=& -if_V m_V\epsilon^*_\mu, \non \\
   \langle V(p,\epsilon) |\bar q_2 \sigma_{\mu\nu}q_1 |0\rangle
 &=& -f_V^\perp (\epsilon_{\mu}^* p^\nu -
\epsilon_{\nu}^* p^\mu)\,,
 \en
are listed in Table \ref{tab:input}. They are taken from \cite{BallfV}. For pseudoscalar mesons, we use $f_\pi=132$ MeV and $f_K=160$ MeV. Decay constants $f^{q}_{\eta^{(')}}$, $f_{\eta^{(')}}^{s}$ and $f_{\eta^{(')}}^c$ defined by
\be
\la 0|\bar q\gamma_\mu\gamma_5q|\eta^{(')}\ra=i{1\over\sqrt{2}}f_{\eta^{(')}}^q  q_\mu, \quad \la 0|\bar s\gamma_\mu\gamma_5s|\eta^{(')}\ra=if_{\eta^{(')}}^s q_\mu, \quad \la 0|\bar c\gamma_\mu\gamma_5c|\eta^{(')}\ra=if_{\eta^{(')}}^c q_\mu
\en
are also needed in calculations.
For the decay constants $f_{\eta^{(')}}^q$ and $f_{\eta^{(')}}^s$, we shall use the values
\be   \label{eq:fetaqs}
f_{\eta}^q=107\,{\rm
MeV}, \quad f_\eta^s=-112\,{\rm MeV}, \quad f_{\eta'}^q= 89\,{\rm
MeV}, \quad f_{\eta'}^s=137\,{\rm MeV}
\en
obtained in \cite{FKS}. As for $f_{\eta^{(')}}^c$,
a straightforward perturbative calculation gives \cite{fetac}
\be \label{eq:fetac}
f_{\eta^{(')}}^c=-{m_{\eta^{(')}}^2\over 12 m_c^2}\,{f_{\eta^{(')}}^q\over\sqrt{2}}.
\en

\begin{table}[t]
\caption{
Input parameters.
The values of the scale dependent quantities $f^\perp_V(\mu)$ and $a^{\bot,V}_{1,2}(\mu)$ are
given for $\mu=1\,\rm{GeV}$. The values of Gegenbauer moments are taken from \cite{Ball2007} and Wolfenstein parameters  from \cite{CKMfitter}.
} \label{tab:input}
\begin{center}
\begin{tabular}{|c||c|c|c|c|c|c|c|c|}
\hline\hline
\multicolumn{7}{|c|}{Light vector mesons} \\
\hline
$V$  & $f_V({\rm MeV})$ &
 $f^\perp_V({\rm MeV})$ & $a^V_1$ & $a^V_2$ & $a^{\bot,V}_1$ & $a^{\bot,V}_2$\\
\hline
$\rho$ & $216\pm3$ & $165\pm 9$ & 0 & $0.15\pm0.07$ &  0 & $0.14\pm0.06$ \\
$\omega$ & $187\pm5$ & $151\pm 9$ & 0 & $0.15\pm 0.07$ & 0 & $0.14\pm0.06$ \\
$\phi$ & $215\pm5$ & $186\pm 9$ & 0 & $0.18\pm 0.08$ & 0 & $0.14\pm0.07$\\
$K^*$ & $220\pm5$ & $185\pm 10$ & $0.03\pm 0.02$ & $0.11\pm 0.09$ & $0.04\pm0.03$ & $0.10\pm0.08$ \\
\hline\hline
\multicolumn{7}{|c|}{Light pseudoscalar mesons} \\
\hline
 \multicolumn{2}{|c|}{$a_1^\pi$} & \multicolumn{2}{|c|}{$a_2^\pi$} &
\multicolumn{2}{|c|}{$a_1^K$} & $a_2^K$ \\
\hline
 \multicolumn{2}{|c|}{0} & \multicolumn{2}{|c|}{$0.25\pm0.15$} & \multicolumn{2}{|c|}{$0.06\pm0.03$} & $0.25\pm0.15$ \\
\hline\hline
\multicolumn{7}{|c|}{$B$ mesons} \\
\hline
$B$ & \multicolumn{2}{|c|}{$m_B({\rm GeV}$)} & $\tau_B({\rm ps})$ &
\multicolumn{2}{|c|}{$f_B({\rm MeV})$} & $\lambda_B({\rm MeV})$ \\
\hline
$B_u$ & \multicolumn{2}{|c|}{$5.279$} & $1.638$ &
\multicolumn{2}{|c|}{$210\pm 20$} & $300\pm 100$ \\
\hline
$B_d$ & \multicolumn{2}{|c|}{$5.279$} & $1.525$ &
\multicolumn{2}{|c|}{$210\pm 20$} & $300\pm 100$ \\
\hline
$B_s$ & \multicolumn{2}{|c|}{$5.366$} & $1.472$ &
\multicolumn{2}{|c|}{$230\pm 20$} & $300\pm 100$ \\
\hline\hline
\multicolumn{7}{|c|}{Form factors at $q^2=0$} \\
\hline
\multicolumn{2}{|c|}{$F^{BK}_0(0)$} &
$A^{B K^*}_0(0)$ &
$A^{B K^*}_1(0)$ &
$A^{B K^*}_2(0)$ &
\multicolumn{2}{|c|}{$V^{B K^*}_0(0)$}  \\
\hline
\multicolumn{2}{|c|}{$0.35\pm0.04$} &
$0.374\pm0.033$ &
$0.292\pm0.028$ &
$0.259\pm0.027$ &
\multicolumn{2}{|c|}{$0.411\pm0.033$}  \\
\hline
\multicolumn{2}{|c|}{$F^{B\pi}_0(0)$} &
$A^{B \rho}_0(0)$ &
$A^{B \rho}_1(0)$ &
$A^{B \rho}_2(0)$ &
\multicolumn{2}{|c|}{$V^{B \rho}_0(0)$}  \\
\hline
\multicolumn{2}{|c|}{$0.25\pm0.03$} &
$0.303\pm0.029$ &
$0.242\pm0.023$ &
$0.221\pm0.023$ &
\multicolumn{2}{|c|}{$0.323\pm0.030$}  \\
\hline
\multicolumn{2}{|c|}{$F^{B\eta_q}_0(0)$} &
$A^{B \omega}_0(0)$ &
$A^{B \omega}_1(0)$ &
$A^{B \omega}_2(0)$ &
\multicolumn{2}{|c|}{$V^{B \omega}_0(0)$}  \\
\hline
\multicolumn{2}{|c|}{$0.296\pm0.028$} &
$0.281\pm0.030$ &
$0.219\pm0.024$ &
$0.198\pm0.023$ &
\multicolumn{2}{|c|}{$0.293\pm0.029$}  \\
\hline
\hline
\multicolumn{7}{|c|}{Quark masses} \\
\hline
\multicolumn{2}{|c|}{$m_b(m_b)/{\rm GeV}$} &
$m_c(m_b)/{\rm GeV}$ & \multicolumn{2}{|c|}{$m_c^{\rm pole}/m_b^{\rm pole}$} &
\multicolumn{2}{|c|}{$m_s(2.1~{\rm GeV})/{\rm GeV}$}   \\
\hline
\multicolumn{2}{|c|}{$4.2$} &
$0.91$ & \multicolumn{2}{|c|}{$0.3$} &
\multicolumn{2}{|c|}{$0.095\pm0.020$}  \\
\hline\hline
\multicolumn{7}{|c|}{Wolfenstein parameters} \\
\hline
\multicolumn{2}{|c|}{$A$} & $\lambda$ &
$\bar\rho$ & $\bar\eta$ &
\multicolumn{2}{|c|}{$\gamma$}  \\
\hline
\multicolumn{2}{|c|}{$0.8116$} & $0.2252$ &
$0.139$ & $0.341$ &
\multicolumn{2}{|c|}{$(67.8^{+4.2}_{-3.9})^\circ$}  \\
\hline
\hline
\end{tabular}
\end{center}
\end{table}

\subsection{LCDAs}
We next specify the light-cone distribution amplitudes (LCDAs) for pseudoscalar and vector mesons. The
general expressions of twist-2 LCDAs are
 \be
 \Phi_{P}(x,\mu) &=& 6x(1-x)\left[1+\sum_{n=1}^\infty
 a_n^{P}(\mu)C_n^{3/2}(2x-1)\right], \non \\
 \Phi^{V}_{\parallel}(x,\mu) &=& 6x(1-x)\left[1+\sum_{n=1}^\infty
 a_n^{V}(\mu)C_n^{3/2}(2x-1)\right],  \non \\
 \Phi^{V}_{\perp}(x,\mu) &=& 6x(1-x)\left[1+\sum_{n=1}^\infty
 a_n^{\perp,V}(\mu)C_n^{3/2}(2x-1)\right],
 \en
and twist-3 ones
 \be
&& \Phi_p(x)=1, \qquad \Phi_\sigma(x)=6x(1-x), \non \\
&&  \Phi_v(x,\mu)=3\left[2x-1+\sum_{n=1}^\infty
 a_{n}^{\bot,V}(\mu)P_{n+1}(2x-1)\right],
 \en
where $C_n(x)$ and $P_n(x)$ are the Gegenbauer and Legendre polynomials, respectively. When three-particle amplitudes are neglected, the twist-3 $\Phi_v(x)$ can be expressed in terms of $\Phi_\perp$
\be
\Phi_v(x)=\int_0^x
\frac{\Phi_\perp(u)}{\bar u}du -\int_x^1
\frac{\Phi_\perp(u)}{u}du.
\en
The normalization of LCDAs is
 \be
 \int^1_0dx \Phi_V(x)=1, \qquad \int^1_0dx \Phi_v(x)=0.
 \en
Note that the Gegenbauer moments $a_i^{(\perp),K^*}$ displayed in Table \ref{tab:input} taken from \cite{Ball2007}
are for the mesons containing a strange quark.

The integral of the $B$
meson wave function is parameterized as \cite{BBNS}
 \begin{eqnarray}
 \int_0^1 \frac{d\rho}{1-\rho}\Phi_1^B(\rho) \equiv
 \frac{m_B}{\lambda_B}\,,
 \end{eqnarray}
where $1-\rho$ is the momentum fraction carried by the light
spectator quark in the $B$ meson. The study of hadronic $B$ decays favors a smaller first inverse moment $\lambda_B$: a value of $350\pm150$ MeV was employed in  \cite{BBNS} and $200_{-0}^{+250}$ MeV in \cite{BRY}, though QCD sum rule and other studies prefer a larger $\lambda_B\sim 460$ MeV \cite{lambdaB}.
We shall use $\lambda_B=300\pm100$ MeV.

For the running quark masses we  shall use \cite{PDG,Xing}
 \be \label{eq:quarkmass}
 && m_b(m_b)=4.2\,{\rm GeV}, \qquad~~~~ m_b(2.1\,{\rm GeV})=4.94\,{\rm
 GeV}, \qquad m_b(1\,{\rm GeV})=6.34\,{\rm
 GeV}, \non \\
 && m_c(m_b)=0.91\,{\rm GeV}, \qquad~~~ m_c(2.1\,{\rm GeV})=1.06\,{\rm  GeV},
 \qquad m_c(1\,{\rm GeV})=1.32\,{\rm
 GeV}, \non \\
 && m_s(2.1\,{\rm GeV})=95\,{\rm MeV}, \quad~ m_s(1\,{\rm GeV})=118\,{\rm
 MeV}, \non\\
 && m_d(2.1\,{\rm GeV})=5.0\,{\rm  MeV}, \quad~ m_u(2.1\,{\rm GeV})=2.2\,{\rm
 MeV}.
 \en
Note that the charm quark masses here are smaller than the one $m_c(m_b)=1.3\pm 0.2$ GeV adopted in \cite{BN,BuchallaVV} and
consistent with the high precision mass determination from lattice QCD \cite{LQCDmc}:
$m_c(3\,{\rm GeV})=0.986\pm0.010$ GeV and $m_c(m_c)=1.267\pm0.009$ GeV (see also \cite{Kuhn}).
Among the quarks, the strange quark gives the major theoretical uncertainty to
the decay amplitude. Hence, we will only consider the uncertainty in the
strange quark mass given by $m_s(2.1\,{\rm GeV})=95\pm20$ MeV.
Notice that for the one-loop penguin contribution, the relevant quark mass is the pole mass rather than the current one \cite{Li:2006jb}. Since the penguin loop correction is governed by the ratio of the pole masses squared $s_i\equiv(m_i^{\rm pole}/m_b^{\rm pole})^2$ and since the pole mass is meaningful only for heavy quarks, we only need to consider the ratio of $c$ and $b$ quark pole masses given by $s_c\approx (0.3)^2$.

\subsection{Penguin annihilation}
In the QCDF approach, the hadronic $B$ decay amplitude receives contributions from tree, penguin, electroweak penguin and weak annihilation topologies. In the absence of $1/m_b$ power corrections except for the chiral enhanced penguin contributions, the leading QCDF predictions encounter three major difficulties as discussed in the Introduction. This implies the necessity of introducing $1/m_b$ power corrections. Soft corrections due to penguin annihilation have been proposed to resolve the rate deficit problem for penguin-dominated decays and the \CP puzzle for $\bar B^0\to K^-\pi^+$. \footnote{Besides the mechanisms of penguin annihilation, charming penguins and final-state rescattering, another possibility of solving the rate and \CP puzzle for $\bar B^0\to K^-\pi^+$ was advocated recently in \cite{PhamKpi} by adding to the $B\to K\pi$ QCDF amplitude a real and an absorptive part with a strength 10\% and 30\% of the penguin amplitude, respectively.}
However, the penguin annihilation amplitude involve troublesome endpoint divergences. Hence, subleading power corrections generally can be studied only in a
phenomenological way. We shall follow \cite{BBNS} to model the endpoint divergence $X\equiv\int^1_0 dx/\bar x$ in the annihilation and hard spectator
scattering diagrams as
 \be \label{eq:XA}
 X_A=\ln\left({m_B\over \Lambda_h}\right)(1+\rho_A e^{i\phi_A}),
 \en
with $\Lambda_h$ being a typical scale of order
500 MeV, and  $\rho_{A}$, $\phi_{A}$ being the unknown real parameters.

A fit to the data of $B_{u,d}\to PP,VP,PV$ and $VV$ decays yields the values of $\rho_A$ and $\phi_A$ shown in Table \ref{tab:rhoA}. Basically, it is very similar to the so-called ``S4 scenario" presented in \cite{BN}. Within the framework of QCDF,
one cannot account for all charmless two-body $B$ decay data by a universal set of $\rho_A$ and $\phi_A$ parameters.   Since the penguin annihilation effects are different for $B\to VP$ and $B\to PV$ decays,
\be
&& A_1^i\approx -A_2^i\approx 6\pi\alpha_s\left[3\left(X_A^{VP}-4+{\pi^2\over 3}\right)+r_\chi^V r_\chi^P\Big((X_A^{VP})^2-2X_A^{VP}\Big)\right], \non \\
&& A_3^i\approx  6\pi\alpha_s\left[-3r_\chi^V\left((X_A^{VP})^2-2X_A^{VP}+4-{\pi^2\over 3}\right)+r_\chi^P \left((X_A^{VP})^2-2X_A^{VP}+{\pi^2\over 3}\right)\right], \non \\
&& A_3^f\approx  6\pi\alpha_s\left[3r_\chi^V(2X_A^{VP}-1)(2-X_A^{VP})-r_\chi^P \Big(2(X_A^{VP})^2-X_A^{VP}\Big)\right],
\en
for $M_1M_2=VP$  and
\be
&& A_1^i\approx -A_2^i\approx 6\pi\alpha_s\left[3\left(X_A^{PV}-4+{\pi^2\over 3}\right)+r_\chi^V r_\chi^P\Big((X_A^{PV})^2-2X_A^{PV}\Big)\right], \non \\
&& A_3^i\approx  6\pi\alpha_s\left[-3r_\chi^P\left((X_A^{PV})^2-2X_A^{PV}+4-{\pi^2\over 3}\right)+r_\chi^V \left((X_A^{PV})^2-2X_A^{PV}+{\pi^2\over 3}\right)\right], \non \\
&& A_3^f\approx  6\pi\alpha_s\left[-3r_\chi^P(2X_A^{PV}-1)(2-X_A^{PV})+r_\chi^V \Big(2(X_A^{PV})^2-X_A^{PV}\Big)\right],
\en
for $M_1M_2=PV$, the parameters $X_A^{VP}$ and $X_A^{PV}$ are not necessarily the same. Indeed, a fit to the $B\to VP,PV$ decays yields $\rho_A^{VP}\approx 1.07$, $\phi_A^{VP}\approx -70^\circ$ and $\rho_A^{PV}\approx 0.87$, $\phi_A^{PV}\approx -30^\circ$ (see Table \ref{tab:rhoA}).
For the estimate of theoretical uncertainties, we shall assign an error of $\pm0.1$ to $\rho_A$ and $\pm 20^\circ$ to $\phi_A$. Note that penguin annihilation contributions to $K\phi$ ($K^*\phi$) are smaller than other $PV$ ($VV$) modes. In general, penguin annihilation is dominated by $b_3$ or $\beta_3$ through $(S-P)(S+P)$ interactions.

\begin{table}[h!]
 \caption{The parameters $\rho_A$ and $\phi_A$ for penguin annihilation. The fitted $\rho_A$ and $\phi_A$ for $B\to VV$ decays are taken from \cite{ChengVV}.
 } \label{tab:rhoA}
\begin{center}
 \begin{tabular}{| l c c |l c c |} \hline
 {Mode}~~~~~~~~ & $\rho_A$  \qquad &  $\phi_A$ \quad & ~~Mode ~~~~~~~~ & $\rho_A$\qquad  & $\phi_A$ \\ \hline
 $B\to PP$ & 1.10 & $-50^\circ$~~ & ~~$B\to VP$ & 1.07 & $-70^\circ$ \\
 $B\to PV$ & 0.87 & $-30^\circ$~~ & ~~$B\to K\phi$ & 0.70 & $-40^\circ$ \\
 $B\to K^*\rho $ & 0.78 & $-43^\circ$~~ &
 ~~$B\to K^*\phi $ & 0.65 & $-53^\circ$ \\ \hline
 \end{tabular}
 \end{center}
 \end{table}

\subsection{Power corrections to $a_2$}

As pointed out in \cite{CCcp}, while the discrepancies between theory and experiment for the  rates of penguin-dominated two-body decays of $B$ mesons and direct \CP asymmetries of $\bar B_d\to K^-\pi^+$, $B^-\to K^-\rho^0$ and $\bar B_d\to \pi^+\pi^-$ are resolved by the power corrections due to penguin annihilation, the signs of  direct {\it CP}-violating effects in  $B^-\to K^-\pi^0, B^-\to K^-\eta$ and $\bar B^0\to\pi^0\pi^0$ are flipped to the wrong ones when confronted with experiment. These new $B$-{\it CP} puzzles in QCDF can be explained by the subleading power corrections to the color-suppressed tree amplitudes due to hard spectator interactions and/or final-state interactions that yield not only correct signs for  aforementioned \CP asymmetries but also accommodate the observed $\bar B_d\to \pi^0\pi^0$ and $\rho^0\pi^0$ rates simultaneously.

Following \cite{CCcp}, power corrections to the color-suppressed topology are parametrized as \be \label{eq:a2}
a_2 \to a_2(1+\rho_C e^{i\phi_C}),
\en
with the unknown parameters $\rho_C$ and $\phi_C$ to be inferred from experiment. We shall use \cite{CCcp}
\be
 \rho_C\approx 1.3\,,~0.8\,,~0, \qquad
 \phi_C\approx -70^\circ\,,-80^\circ\,,~0,
\en
for $\bar B\to PP,VP,VV$ decays, respectively.
This pattern that soft power corrections to $a_2$ are large for $PP$ modes, moderate for $VP$ ones and very small for $VV$ cases is consistent with the observation made in \cite{Kagan} that soft power correction dominance is much larger for $PP$ than $VP$ and $VV$ final states.
It has been argued that this has to do with the special nature of the pion which is a $q\bar q$ bound state on the one hand and a nearly massless Nambu-Goldstone boson on the other hand \cite{Kagan}.

What is the origin of power corrections to $a_2$ ? There are two possible sources: hard spectator interactions and final-state interactions. From Eq. (\ref{eq:ai}) we have the expression
\be \label{eq:ai}
  a_2(M_1M_2) =
 c_2+{c_1\over N_c}
  + {c_1\over N_c}\,{C_F\alpha_s\over
 4\pi}\Big[V_2(M_2)+{4\pi^2\over N_c}H_2(M_1M_2)\Big]+a_2(M_1M_2)_{\rm LD},
 \en
for $a_2$.
The hard spectator term $H_2(M_1 M_2)$ reads
\begin{eqnarray}\label{eq:hardspec}
  H_2(M_1 M_2)= {if_B f_{M_1} f_{M_2} \over X^{(\overline{B} M_1,
  M_2)}}\,{m_B\over\lambda_B} \int^1_0 d x d y \,
 \Bigg( \frac{\Phi_{M_1}(x) \Phi_{M_2}(y)}{\bar x\bar y} + r_\chi^{M_1}
  \frac{\Phi_{m_1} (x) \Phi_{M_2}(y)}{\bar x y}\Bigg),
 \hspace{0.5cm}
 \end{eqnarray}
where $X^{(\overline{B} M_1,
  M_2)}$ is the factorizable amplitude for $\ov B\to M_1M_2$, $\bar x=1-x$.
Power corrections from the twist-3 amplitude $\Phi_m$ are divergent and can be parameterized as
 \be \label{eq:XH}
 X_H\equiv \int^1_0{dy\over y}={\rm ln}{m_B\over
 \Lambda_h}(1+\rho_H e^{i\phi_H}).
 \end{eqnarray}
Since $c_1\sim {\cal O}(1)$ and $c_9\sim {\cal O}(-1.3)$ in units of $\alpha_{em}$, it is clear that hard spectator contributions to $a_i$ are usually very small except for $a_2$ and $a_{10}$. Indeed, there is a huge cancelation between the vertex and naive factorizable terms so that the real part of $a_2$ is governed
by spectator interactions, while its imaginary part comes mainly from the vertex corrections \cite{BBNS2}. The value of $a_2(K\pi)\approx 0.51\,e^{-i58^\circ}$ needed to solve the $B\to K\pi$ \CP puzzle [see Eq. (\ref{eq:a2PP})] corresponds to $\rho_H\approx 4.9$ and $\phi_H\approx -77^\circ$. Therefore, there is no reason to restrict $\rho_H$ to the range $0\leq \rho_H\leq 1$. A sizable color-suppressed tree amplitude also can be induced via color-allowed decay $B^-\to K^-\eta'$ followed by the rescattering of $K^-\eta'$ into $K^-\pi^0$ as depicted in Fig. 1.
Recall that among the 2-body $B$ decays, $B\to K\eta'$ has the largest branching fraction, of order $70\times 10^{-6}$.
This final-state rescattering has the same topology as the color-suppressed tree diagram \cite{CCSfsi}. One of us (CKC) has studied the FSI effects through residual rescattering among $PP$ states and resolved the $B$-$CP$ puzzles \cite{Chua}. As stressed by Neubert sometime ago, in the presence of soft final-state interactions, there is no color suppression of $C$ with respect to $T$ \cite{Neubert}.

Since the chiral factor $r_\chi^V$ for the vector meson is substantially smaller than $r_\chi^P$ for the pseudoscalar meson (typically, $r_\chi^P={\cal O}(0.8)$ and $r_\chi^V={\cal O}(0.2)$ at the hard-collinear scale $\mu_h=\sqrt{\Lambda_h m_b}$), one may argue that
Eq. (\ref{eq:hardspec}) provides a natural explanation as to  why the power corrections to $a_2$ is smaller when $M_1$ is a vector meson, provided that soft corrections arise from spectator rescattering. Unfortunately, this is not the case. Numerically, we found that, for example, $H(K^*\pi)$ is comparable to $H(K\pi)$. This is due to the fact that $\int_0^1 dx \,r_\chi^M\Phi_m(x)/(1-x)$ is equal to $X_Hr_\chi^P$ for $M=P$ and approximated to $3(X_H-2)r_\chi^V$ for $M=V$.

We use NLO results for $a_2$ in Eq. (\ref{eq:a2}) as a benchmark to define the parameters $\rho_C$ and $\phi_C$. The NNLO calculations of spectator-scattering tree amplitudes  and vertex corrections at order $\alpha_s^2$ have been carried out in \cite{Beneke} and \cite{Bell}, respectively. As pointed out in \cite{BenekeFPCP,Bell09}, a smaller value of $\lambda_B$ can enhance the hard spectator interaction and hence $a_2$ substantially. For example, $a_2(\pi\pi)\sim 0.375-0.076i$ for $\lambda_B=200$ MeV was found in \cite{Bell09}.
However, the recent BaBar data on $B\to\gamma \ell\bar\nu$
\cite{BaBar:gammalnu} seems to imply a larger $\lambda_B$ ($>
300$ MeV at the 90\% CL). While NNLO corrections can in principle push the magnitude of $a_2(\pi\pi)$ up to the order of 0.40 by lowering the value of the $B$ meson parameter $\lambda_B$, the strong phase of $a_2$ relative to $a_1$ cannot be larger than $15^\circ$ \cite{BenekeFPCP}. In this work we reply on $\rho_C$ and $\phi_C$ to get a large magnitude and strong phase for $a_2$.

\begin{figure}[t]
\begin{center}
\includegraphics[width=0.6\textwidth]{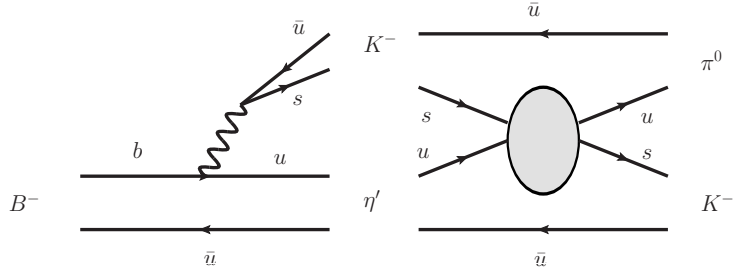}
\vspace{0.0cm}
\caption{Contribution to the color-suppressed tree amplitude of $B^-\to K^-\pi^0$ from the weak decay $B^-\to K^-\eta'$ followed by the final-state rescattering of $K^-\eta'$ into $\bar K^0\pi^0$. This has the same topology as the color-suppressed tree diagram.} \label{fig:trigluon} \end{center}
\end{figure}

\section{$B\to PP$ Decays}
Effects of power corrections on penguin annihilation and the color-suppressed tree amplitude for some selective $B\to PP$ decays are shown in Table \ref{tab:PP}. The implications will be discussed below.
Branching fractions and \CP asymmetries for  all $B\to PP$ decays are shown in Tables \ref{tab:PPBr} and \ref{tab:PPCP}, resepctively. The theoretical errors
correspond to the uncertainties due to the variation of (i) the Gegenbauer moments,
the  decay constants, (ii) the heavy-to-light form factors and the strange
quark mass, and (iii) the wave function of the $B$ meson characterized by the
parameter $\lambda_B$, the power corrections due to weak annihilation and hard
spectator interactions described by the parameters $\rho_{A,H}$, $\phi_{A,H}$,
respectively. To obtain the errors shown in these tables, we first scan randomly the points in the
allowed ranges of the above nine parameters (specifically, the ranges $\rho_A^0-0.1\leq \rho_A\leq \rho_A^0+0.1$, $\phi_A^0-20^\circ\leq \phi_A\leq \phi_A^0+20^\circ$, $0\leq \rho_H\leq 1$ and $0\leq \phi_{H}\leq 2\pi$ are used in this work, where the values of $\rho_A^0$ and $\phi_A^0$ are displayed in Table \ref{tab:rhoA}) and then add errors in quadrature. More specifically, the second error in the table is referred  to the
uncertainties caused by the variation of  $\rho_{A,H}$ and $\phi_{A,H}$, where
all other uncertainties are lumped into the first error.  Power corrections beyond the heavy
quark limit generally give the major theoretical uncertainties.

 \begin{table}[t]
 \caption{$CP$-averaged branching fractions (in units of $10^{-6}$) and direct \CP asymmetries (in \%) of some selective $B\to PP$ decays obtained in QCD factorization for three distinct cases: (i) without any power corrections, (ii) with power corrections to penguin annihilation, and (iii) with power corrections to both penguin annihilation and color-suppressed tree amplitudes. The parameters $\rho_A$ and $\phi_A$ are taken from Table \ref{tab:rhoA},
 $\rho_C=1.3$ and $\phi_C=-70^\circ$. The theoretical errors correspond to the uncertainties due to the variation of (i) Gegenbauer moments, decay constants, quark masses, form factors, the $\lambda_B$ parameter for the $B$ meson wave function, and (ii) $\rho_{A,H}$, $\phi_{A,H}$, respectively.
 } \label{tab:PP}
\begin{ruledtabular}
 \begin{tabular}{l  c c c c c}
 {Mode}
   &   {\rm W/o}~$\rho_{A,C},\phi_{A,C}$ &  {\rm With}~$\rho_A,\phi_A$ &  {\rm With}~$\rho_{A,C},\phi_{A,C}$ & Expt. \cite{HFAG} \\  \hline
   $\B({\overline{B}}^{0}\to K^{-}\pi^+)$
        & $13.1^{+5.8+0.7}_{-3.5-0.7}$
                                                                &
                                                                $19.3^{+7.9+8.2}_{-4.8-6.2}$
                                                                & $19.3^{+7.9+8.2}_{-4.8-6.2}$  & $19.4\pm0.6$
                                                                \\
   $\B({\overline{B}}^{0} \to \bar K^{0}{\pi}^{0})$
     & $5.5^{+2.8+0.3}_{-1.7-0.3}$
                                                                 &
                                                                 $8.4^{+3.8+3.8}_{-2.3-2.9}$
                                                                 & $8.6^{+3.8+3.8}_{-2.2-2.9}$  & $9.8\pm0.6$
                                                                 \\
  $\B(B^-\to \bar K^0\pi^-)$ & $14.9^{+6.9+0.9}_{-4.5-1.0}$ & $21.7^{+9.2+9.0}_{-6.0-6.9}$ & $21.7^{+9.2+9.0}_{-6.0-6.9}$ & $23.1\pm1.0$ \\
  $\B(B^-\to K^-\pi^0)$ & $9.1^{+3.6+0.5}_{-2.3+0.5}$ & $12.6^{+4.7+4.8}_{-3.0-3.7}$ &  $12.5^{+4.7+4.9}_{-3.0-3.8}$ & $12.9\pm0.6$ \\
  $\B(B^-\to K^-\eta)$ & $1.6^{+1.1+0.3}_{-0.7-0.4}$ & $2.4^{+1.8+1.3}_{-1.1-1.0}$ & $2.4^{+1.8+1.3}_{-1.1-1.0}$ & $2.36\pm0.27$  \\
  $\B({\overline{B}}^{0}\to \pi^{+}\pi^-)$ & $6.2^{+0.4+0.2}_{-0.6-0.4}$ & $7.0^{+0.4+0.7}_{-0.7-0.7}$ & $7.0^{+0.4+0.7}_{-0.7-0.7}$ & $5.16\pm0.22$ \\
  $\B({\overline{B}}^{0}\to \pi^{0}\pi^0)$ & $0.42^{+0.29+0.18}_{-0.11-0.08}$ & $0.52^{+0.26+0.21}_{-0.10-0.10}$ & $1.1^{+1.0+0.7}_{-0.4-0.3}$ & $1.55\pm0.19$ \footnotemark[1] \\
  $\B({{B}}^{-}\to \pi^{-}\pi^0)$ & $4.9^{+0.9+0.6}_{-0.5-0.3}$ & $4.9^{+0.9+0.6}_{-0.5-0.3}$ & $5.9^{+2.2+1.4}_{-1.1-1.1}$ & $5.59^{+0.41}_{-0.40}$ \\
  $\B(B^-\to \pi^-\eta)$ & $4.4^{+0.6+0.4}_{-0.3-0.2}$ & $4.5^{+0.6+0.5}_{-0.3-0.3}$ &  $5.0^{+1.2+0.9}_{-0.6-0.7}$ & $4.1\pm0.3$ \\
  \hline
   $\acp({\overline{B}}^{0}\to K^{-}\pi^+)$
        & $4.0^{+0.6+1.1}_{-0.7-1.1}$
                                                                &
                                                                $-7.4^{+1.7+4.3}_{-1.5-4.8}$
                                                                & $-7.4^{+1.7+4.3}_{-1.5-4.8}$  & $-9.8^{+1.2}_{-1.1}$
                                                                \\
   $\acp({\overline{B}}^{0} \to \bar K^{0}{\pi}^{0})$
     & $-4.0^{+1.2+3.5}_{-1.8-3.0}$
                                                                 &
                                                                 $0.75^{+1.88+2.56}_{-0.94-3.32}$
                                                                 & $-10.6^{+2.7+5.6}_{-3.8-4.3}$  & $-1\pm10$
                                                                 \\
  $\acp(B^-\to \bar K^0\pi^-)$ & $0.72^{+0.06+0.05}_{-0.05-0.05}$ & $0.28^{+0.03+0.09}_{-0.03-0.10}$ & $0.28^{+0.03+0.09}_{-0.03-0.10}$ & $0.9\pm2.5$ \\
  $\acp(B^-\to K^-\pi^0)$ & $7.3^{+1.6+2.3}_{-1.2-2.7}$ & $-5.5^{+1.3+4.9}_{-1.8-4.6}$ & $4.9^{+3.9+4.4}_{-2.1-5.4}$ & $5.0\pm2.5$ \\
  $\acp(B^-\to K^-\eta)$ & $-22.1^{+~7.7+14.0}_{-16.7-~7.3}$ & $12.7^{+7.7+13.4}_{-5.0-15.0}$ & $-11.0^{+~8.4+14.9}_{-21.6-10.1}$ & $-37\pm9$ \\
  $\acp({\overline{B}}^{0}\to \pi^{+}\pi^-)$ &  $-6.2^{+0.4+2.0}_{-0.5-1.8}$ & $17.0^{+1.3+4.3}_{-1.2-8.7}$ & $17.0^{+1.3+4.3}_{-1.2-8.7}$ & $38\pm6$ \\
  $\acp({\overline{B}}^{0}\to \pi^{0}\pi^0)$ & $33.4^{+~6.8+34.8}_{-10.6-37.7}$ & $-26.9^{+8.4+48.5}_{-6.0-37.5}$ & $57.2^{+14.8+30.3}_{-20.8-34.6}$ & $43^{+25}_{-24}$ \\
  $\acp({{B}}^{-}\to \pi^{-}\pi^0)$ & $-0.06^{+0.01+0.01}_{-0.01-0.02}$ & $-0.06^{+0.01+0.01}_{-0.01-0.02}$ &
  $-0.11^{+0.01+0.06}_{-0.01-0.03}$ & $6\pm5$ \\
  $\acp(B^-\to \pi^-\eta)$ & $-11.4^{+1.1+2.3}_{-1.0-2.7}$ & $11.4^{+0.9+4.5}_{-0.9-9.1}$ & $-5.0^{+2.4+~8.4}_{-3.4-10.3}$ & $-13\pm7$  \\
 \end{tabular}
\footnotetext[1]{If an $S$ factor is included, the average will become $1.55\pm0.35$\,.}
 \end{ruledtabular}
 \end{table}

\subsection{Branching fractions}
\vskip 0.05in \noindent{\it \underline{$B\to K\pi$}} \vskip 0.1in
The $B\to K\pi$ decays are dominated by penguin contributions because of
$|V_{us}V_{ub}^*|\ll |V_{cs}V_{cb}^*|\approx |V_{ts}V_{tb}^*|$ and the large
top quark mass. For the ratios defined by
\begin{eqnarray}
R_c\equiv {2\Gamma(B^-\to K^-\pi^0)\over \Gamma(B^-\to \bar K^0\pi^-)},
\qquad R_n\equiv {\Gamma(\bar B^0\to K^-\pi^+)\over 2\Gamma(\bar B^0\to \bar K^0\pi^0)},
\end{eqnarray}
we have $R_c=R_n\approx 1$ if the other quark-diagram amplitudes are negligible compared with $P'$. The current experimental measurements give $R_c=1.12\pm0.07$ and $R_n=0.99\pm0.07$. In  QCDF we have $R_c=1.15\pm 0.03$ and $R_n=1.12\pm 0.03$, which are consistent with experiment.

From Table \ref{tab:PP}, we see that
the predicted rates for penguin-dominated $B\to PP$ decays to the zeroth order of $1/m_b$ expansion are usually $(30\sim 45)\%$ below measurements (see the second column of Table \ref{tab:PP}). Also the direct \CP asymmetry $\acp(K^-\pi^+)$ is wrong in sign. We use penguin annihilation dictated by $\rho_A=1.10$ and $\phi_A=-50^\circ$ to fix both problems.

\vskip 0.2in \noindent{\it \underline{$B\to K\eta^{(')}$}} \vskip 0.1in
Among the 2-body $B$ decays, $B\to K\eta'$ has the largest branching fraction,
of order $70\times 10^{-6}$, while ${\cal B}(B\to\eta K)$ is only
$(1-3)\times 10^{-6}$. This can be qualitatively  understood as follows. Since the $\eta-\eta'$ mixing angle in the quark-flavor basis $\eta_q=(u\bar u+d\bar d)/\sqrt{2}$ and $\eta_s=s\bar s$
\begin{eqnarray}
\eta=\cos\phi\eta_q-\sin\phi\eta_s, \qquad \eta'=\sin\phi\eta_q+\cos\phi\eta_s,
\end{eqnarray}
is extracted from the data to be $\phi=39.3^\circ$ \cite{FKS}, it is clear that the interference between the $B\to K\eta_q$ amplitude induced by the $b\to sq\bar q$ penguin and the $B\to K\eta_s$ amplitude induced by $b\to ss\bar s$  is constructive for $B\to K\eta'$ and destructive for $B\to\eta K$. This explains the large rate of the former and the suppression of the latter. However, most of the model calculations still fall short of the data for $\B(B\to K\eta')$.

Many possible solutions to the puzzle for the abnormally large $K\eta'$ rate have been proposed in the past: (i) a significant flavor-singlet contribution
\cite{Chiang,BenekeETA}, (ii) a large $B\to \eta'$ form factor \cite{Pham}, (iii) a contribution from the charm content of the $\eta'$, (iv) an enhanced hadronic matrix element
$\langle 0|\bar s\gamma_5 s|\eta'\rangle$ due to the axial U(1) anomaly \cite{Gerard}, (v) a large chiral scale $m_0^q$ associated with the $\eta_q$ \cite{ChenKeta,LiKeta}, (vi) a long-distance charming penguin in SCET \cite{Zupan}, and (vii) a large contribution from the two-gluon fusion mechanism \cite{Ahmady}.

Numerically, Beneke and Neubert already obtained $\B(B^-\to K^-\eta')\sim {\cal O}(50\times 10^{-6})$ in QCDF using the default values $\rho_A=\rho_H=0$ \cite{BN}. Here we found similar results $57\times 10^{-6}$ ($53\times 10^{-6}$) with (without) the contributions from the ``charm content" of the $\eta'$. In the presence of penguin annihilation, we obtain $\B(B^-\to K^-\eta')\sim 78\times 10^{-6}$ ($71\times 10^{-6}$)
with (without) the ``charm content" contributions.
Therefore, the observed large $B\to K\eta'$ rates are naturally explained in QCDF without invoking, for example, flavor-singlet contributions. Data on $B\to K\eta$ modes are also well accounted for by QCDF.

 \begin{table}[htbp!]
 \caption{$CP$-averaged branching fractions (in units of $10^{-6}$) of  $B\to PP$ decays obtained in various approaches. The pQCD results are taken from \cite{LiPP,LuKK,XiaoKeta,Xiaoetaeta,Xiaopieta}. Note that there exist several pQCD calculations for $B\to K\eta^{(')}$ \cite{Kou,XiaoKeta,ChenKeta,LiKeta} and here we cite the pQCD results with partial NLO corrections \cite{XiaoKeta}. There are two solution sets with SCET predictions for decays involving $\eta$ and/or $\eta'$ \cite{Zupan}.
 } \label{tab:PPBr}
\begin{ruledtabular}
 \begin{tabular}{l  c c c c c}
 {Mode}
   &  QCDF (this work) &  pQCD &  SCET  & Expt. \cite{HFAG} \\  \hline
  $B^-\to \bar K^0\pi^-$                                         &
                                                                 $21.7^{+9.2+9.0}_{-6.0-6.9}$ & $23.6^{+14.5}_{-~8.4}$ & $20.8\pm7.9\pm0.6\pm0.7$ & $23.1\pm1.0$ \\
  $B^-\to K^-\pi^0$                                              &
                                                                 $12.5^{+4.7+4.9}_{-3.0-3.8}$ & $13.6^{+10.3}_{-~5.7}$ &  $11.3\pm4.1\pm1.0\pm0.3$ & $12.9\pm0.6$ \\
   ${\overline{B}}^{0}\to K^{-}\pi^+$
        & $19.3^{+7.9+8.2}_{-4.8-6.2}$
                                                                &
                                                                $20.4^{+16.1}_{-~8.4}$
                                                                & $20.1\pm7.4\pm1.3\pm0.6$  & $19.4\pm0.6$
                                                                \\
   ${\overline{B}}^{0} \to \bar K^{0}{\pi}^{0}$
     &  $8.6^{+3.8+3.8}_{-2.2-2.9}$
                                                                 &
                                                                 $8.7^{+6.0}_{-3.4}$
                                                                 & $9.4\pm3.6\pm0.2\pm0.3$  & $9.8\pm0.6$
                                                                 \\
  ${{B}}^{-}\to \pi^{-}\pi^0$                                    &
                                                                 $5.9^{+2.2+1.4}_{-1.1-1.1}$      & $4.0^{+3.4}_{-1.9}$ & $5.2\pm1.6\pm2.1\pm0.6$ & $5.59^{+0.41}_{-0.40}$ \\
  ${\overline{B}}^{0}\to \pi^{+}\pi^-$                           &
                                                                 $7.0^{+0.4+0.7}_{-0.7-0.7}$ & $6.5^{+6.7}_{-3.8}$ & $5.4\pm1.3\pm1.4\pm0.4$ & $5.16\pm0.22$ \\
  ${\overline{B}}^{0}\to \pi^{0}\pi^0$                           &
                                                                 $1.1^{+1.0+0.7}_{-0.4-0.3}$ & $0.29^{+0.50}_{-0.20}$ & $0.84\pm0.29\pm0.30\pm0.19$ & $1.55\pm0.19$  \\
  $B^{-}\to K^-K^0$                                              &
                                                                 $1.8^{+0.9+0.7}_{-0.5-0.5}$ & 1.66 &$1.1\pm0.4\pm1.4\pm0.03$ & $1.36^{+0.29}_{-0.27}$ \\
  ${\overline{B}}^{0}\to K^+K^-$                                 &
                                                                 $0.10^{+0.03+0.03}_{-0.02-0.03}$ & 0.046 & $--$ & $0.15^{+0.11}_{-0.10}$ \\
  ${\overline{B}}^{0}\to K^0\bar K^0$                            &
                                                                 $2.1^{+1.0+0.8}_{-0.6-0.6}$ & 1.75 & $1.0\pm0.4\pm1.4\pm0.03$ & $0.96^{+0.21}_{-0.19}$ \\ \hline
  $B^-\to K^-\eta$                                               &
                                                                 $2.3^{+1.8+1.3}_{-1.1-1.0}$ & $3.2^{+1.2+2.7+1.1}_{-0.9-1.2-1.0}$ & $2.7\pm4.8\pm0.4\pm0.3$ & $2.36\pm0.27$  \\
                                                                 &&&$2.3\pm4.5\pm0.4\pm0.3$ &\\
  $B^-\to K^-\eta'$
                                                                 &   $78.4^{+61.2+26.4}_{-26.8-19.5}$
                                                                 & $51.0^{+13.5+11.2+4.2}_{-~8.2-~\,6.2-3.5}$ & $69.5\pm27.0\pm4.4\pm7.7$ & $71.1\pm2.6$  \\
                                                                 &&&$69.3\pm26.0\pm7.1\pm6.3$
                                                                 &\\
  $\bar B^0\to \bar K^0\eta$
                                                                 & $1.6^{+1.5+1.1}_{-0.9-0.8}$ & $2.1^{+0.8+2.3+1.0}_{-0.6-1.0-0.9}$ & $2.4\pm4.4\pm0.2\pm0.3$& $1.12^{+0.30}_{-0.28}$  \\
                                                                 &&&$2.3\pm4.4\pm0.2\pm0.5$
                                                                 &\\
  $\bar B^0\to \bar K^0\eta'$                                    &
                                                                 $74.2^{+56.5+24.7}_{-24.9-18.4}$ & $50.3^{+11.8+11.1+4.5}_{-~8.2-~6.2-2.7}$ &$63.2\pm24.7\pm4.2\pm8.1$ & $66.1\pm3.1$ & \\
                                                                 &&&$62.2\pm23.7\pm5.5\pm7.2$
                                                                 &\\
  $B^-\to \pi^-\eta$                                             &
                                                                 $5.0^{+1.2+0.9}_{-0.6-0.7}$ & $4.1^{+1.3+0.4+0.6}_{-0.9-0.3-0.5}$ &  $4.9\pm1.7\pm1.0\pm0.5$ & $4.07\pm0.32$ \\
                                                                 &&&$5.0\pm1.7\pm1.2\pm0.4$
                                                                 &\\
  $B^-\to \pi^-\eta'$                                            &
                                                                 $3.8^{+1.3+0.9}_{-0.6-0.6}$ & $2.4^{+0.8}_{-0.5}\pm0.2\pm0.3$ &  $2.4\pm1.2\pm0.2\pm0.4$ & $2.7^{+0.5}_{-0.4}$
                                                                 \\
                                                                 &&&$2.8\pm1.2\pm0.3\pm0.3$
                                                                 &\\
  $\bar B^0\to \pi^0\eta$                                        &
                                                                 $0.36^{+0.03+0.13}_{-0.02-0.10}$ & $0.23^{+0.04+0.04}_{-0.03-0.03}\pm0.05$ &  $0.88\pm0.54\pm0.06\pm0.42$ & $<1.5$ \\
                                                                 &&&$0.68\pm0.46\pm0.03\pm0.41$
                                                                 &\\
  $\bar B^0\to \pi^0\eta'$                                       &
                                                                 $0.42^{+0.21+0.18}_{-0.09-0.12}$ & $0.19\pm0.02\pm0.03^{+0.04}_{-0.05}$ &  $2.3\pm0.8\pm0.3\pm2.7$ & $1.2\pm0.4$ \\
                                                                 &&&$1.3\pm0.5\pm0.1\pm0.3$
                                                                 &\\
  $\bar B^0\to \eta\eta$                                         &
                                                                 $0.32^{+0.13+0.07}_{-0.05-0.06}$ & $0.67^{+0.32}_{-0.25}$ &  $0.69\pm0.38\pm0.13\pm0.58$ & $<1.0$ \\
                                                                 &&&$1.0\pm0.4\pm0.3\pm1.4$
                                                                 &\\
  $\bar B^0\to \eta\eta'$                                        &
                                                                 $0.36^{+0.24+0.12}_{-0.10-0.08}$ & $0.18\pm0.11$ &  $1.0\pm0.5\pm0.1\pm1.5$ & $<1.2$ \\
                                                                 &&&$2.2\pm0.7\pm0.6\pm5.4$
                                                                 &\\
  $\bar B^0\to \eta'\eta'$                                       &
                                                                 $0.22^{+0.14+0.08}_{-0.06-0.06}$ & $0.11^{+0.12}_{-0.09}$ &  $0.57\pm0.23\pm0.03\pm0.69$ & $<1.7$ \\
                                                                 &&&$1.2\pm0.4\pm0.3\pm3.7$
                                                                 &\\
 \end{tabular}
\end{ruledtabular}
 \end{table}

\vskip 0.2in \noindent{\it \underline{$B\to \pi\pi$}} \vskip 0.1in

From
Table \ref{tab:PP} we see that power corrections to the color-suppressed tree amplitude have almost no effect on the decay rates of penguin-dominated decays, but will enhance the color-suppressed tree dominated decay
$B\to\pi^0\pi^0$ substantially owing to the enhancement of $|a_2|\sim {\cal O}(0.6)$ [see Eq. (\ref{eq:a2PP}) below]. Since $|P_{\rm EW}/C|$ is of order 0.06 before any power corrections, it is very unlikely that an enhancement of $P_{\rm EW}$ through New Physics effects can render $c=C+P_{\rm EW}$ large and complex.
Notice that the central values of the branching fractions of $B^0\to\pi^0\pi^0$ measured by BaBar \cite{BaBarpi0pi0} and Belle \cite{Bellepi0pi0}, $(1.83\pm0.21\pm0.13)\times 10^{-6}$ and $(1.1\pm0.3\pm0.1)\times 10^{-6}$ respectively, are somewhat different in their central values. The charged mode $B^-\to \pi^-\pi^0$ also gets an enhancement as its amplitude is proportional to $a_1+a_2$. The prediction of QCDF or pQCD (see Table \ref{tab:PPBr}) for $\B(B^0\to \pi^+\pi^-)$ is slightly too large compared to the data. This is a long standing issue. One possibility for the remedy is that there exists $\pi\pi\to\pi\pi$ meson annihilation contributions in which two initial quark pairs in the zero isospin configuration are destroyed and then created. Indeed, in the topological quark diagram approach, this corresponds to the vertical $W$-loop diagram \cite{CC87}. As shown in \cite{CCSfsi,Chua}, this additional long-distance contribution may lower the $\pi^+\pi^-$ rate. In the final-state rescattering model considered by Hou and Yang \cite{HouYang} and elaborated more by one of us (CKC) \cite{Chua}, $\bar B^0\to \pi^+\pi^-$ and $\pi^0\pi^0$ rates are reduced and enhanced roughly by a factor of 2, respectively, through FSIs.
It should be remarked that in the pQCD approach, it has been shown recently that the color-suppressed tree amplitude will be enhanced by a soft factor arising from the uncanceled soft divergences in the $k_T$ factorization for nonfactorizable hadronic $B$ decays \cite{Li09}. As a consequence, the $B^0\to\pi^0\pi^0$ rate can be enhanced to the right magnitude.

\vskip 0.2in \noindent{\it \underline{$B\to K\bar K$}} \vskip 0.1in
The decays $B^-\to K^-K^0$ and $\bar B^0\to \bar K^0 K^0$ receive $b\to d$ penguin contributions and $\bar B^0\to K^+K^-$ proceeds only through weak annihilation. Hence, the first two modes have branching fractions of order $10^{-6}$, while the last one is suppressed to the order of $10^{-8}$.

 \begin{table}[htbp!]
 \caption{$CP$-averaged branching fractions (in units of $10^{-6}$) of $B\to \pi\eta^{(')}$ decays.
 } \label{tab:Br:pieta}
\begin{ruledtabular}
 \begin{tabular}{l  c c c c }

   &  $\pi^-\eta'$ & $\pi^0\eta'$ &  $\pi^-\eta$  & $\pi^0\eta$  \\  \hline
BaBar & $3.5\pm0.6\pm0.2$ \cite{BaBar:Beta} & $0.9\pm0.4\pm0.1$ \cite{BaBar:pi0eta} & $4.00\pm0.40\pm0.24$ \cite{BaBar:Beta} & $<1.5$ \cite{BaBar:pi0eta}\\
Belle & $1.8^{+0.7}_{-0.6}\pm0.1$ \cite{Belle:pieta'} & $2.8\pm1.0\pm0.3$ \cite{Belle:pieta'} & $4.2\pm0.4\pm0.2$ \cite{Belle:Beta2} & $<2.5$ \cite{Belle:Beta1} \\
Average & $2.7^{+0.5}_{-0.4}$ & $1.2\pm0.4$ & $4.1\pm0.3$ & $<1.5$ \\
 \end{tabular}
\end{ruledtabular}
 \end{table}

\vskip 0.2in \noindent{\it \underline{$B\to \pi\eta^{(')}$}} \vskip 0.1in

The decay amplitudes of $B\to \pi\eta$ are
\begin{eqnarray} \label{eq:Btopieta}
\sqrt{2}A(B^-\to\pi^-\eta) &\approx& A_{\pi\eta_q}\left[\delta_{pu}(\alpha_2+\beta_2)+2 \alpha_3^p+\hat\alpha_4^p\right]+A_{\eta_q\pi}\left[\delta_{pu}(\alpha_1+\beta_2)+\hat\alpha_4^p\right], \non \\
-2A(\bar B^0\to\pi^0\eta) &\approx& A_{\pi\eta_q}\left[\delta_{pu}(\alpha_2-\beta_1)+2 \alpha_3^p+\hat\alpha_4^p\right]+A_{\eta_q\pi}\left[\delta_{pu}(-\alpha_2-\beta_1)+\hat\alpha_4^p\right],
\end{eqnarray}
with $\hat \alpha_4=\alpha_4+\beta_3$ and similar expressions for $B\to\pi\eta'$.
It is clear that the decays $\bar B^0\to\eta^{(')}\pi^0$ have very small rates because of near cancelation of the the color-suppressed tree amplitudes, while the charged modes $\eta^{(')}\pi^-$ receive color-allowed tree contributions.
From the experimental data shown in Table \ref{tab:Br:pieta}, it is clear that the BaBar's measurement of $\B(B^-\to\pi^-\eta')\gg \B(\bar B^0\to\pi^0\eta')$ is in accordance with the theoretical expectation, whereas the Belle's results indicate the other way around. Nevertheless, BaBar and Bell agree with each other on $\B(B\to\pi\eta)$.
QCDF predictions for $B\to\pi\eta^{(')}$ agree well with the BaBar data. As for the pQCD approach, it appears that its prediction for $\B(\bar B^0\to\pi^0\eta')$ is too small. At any rate, it is important to have more accurate measurements of $B\to \pi\eta^{(')}$.

 \begin{table}[tbp!]
 \caption{Same as Table \ref{tab:PPBr} except for direct \CP asymmetries (in \%) of $B\to PP$ decays obtained in various approaches.
 } \label{tab:PPCP}
\begin{ruledtabular}
 \begin{tabular}{l  c c c c c}
 {Mode}
   &  QCDF (this work) &  pQCD &  SCET  & Expt. \cite{HFAG} \\  \hline
  $B^-\to \bar K^0\pi^-$                                         &
                                                                 $0.28^{+0.03+0.09}_{-0.03-0.10}$ & $0$ & $<5$ & $0.9\pm2.5$ \\
  $B^-\to K^-\pi^0$                                              &
                                                                 $4.9^{+3.9+4.4}_{-2.1-5.4}$ & $-1^{+3}_{-6}$ & $-11\pm9\pm11\pm2$ & $5.0\pm2.5$ \\
   ${\overline{B}}^{0}\to K^{-}\pi^+$ &
       $-7.4^{+1.7+4.3}_{-1.5-4.8}$
                                                                &
                                                                $-10^{+7}_{-8}$
                                                                & $-6\pm5\pm6\pm2$  & $-9.8^{+1.2}_{-1.1}$
                                                                \\
   ${\overline{B}}^{0} \to \bar K^{0}{\pi}^{0}$
     &  $-10.6^{+2.7+5.6}_{-3.8-4.3}$
                                                                 &
                                                                 $-7^{+3}_{-4}$
                                                                 & $5\pm4\pm4\pm1$  & $-1\pm10$
                                                                 \\
  ${{B}}^{-}\to \pi^{-}\pi^0$                                    &
                                                                 $-0.11^{+0.01+0.06}_{-0.01-0.03}$ & $0$ & $<4$ & $6\pm5$ \\
  ${\overline{B}}^{0}\to \pi^{+}\pi^-$
                                                                 & $17.0^{+1.3+4.3}_{-1.2-8.7}$
                                                                 & $18^{+20}_{-12}$ & $20\pm17\pm19\pm5$ & $38\pm6$ \\
  ${\overline{B}}^{0}\to \pi^{0}\pi^0$
                                                                 & $57.2^{+14.8+30.3}_{-20.8-34.6}$ & $63^{+35}_{-34}$ & $-58\pm39\pm39\pm13$ & $43^{+25}_{-24}$ \\
  $B^{-}\to K^-K^0$                                              &
                                                                 $-6.4^{+0.8+1.8}_{-0.6-1.8}$ & 11 &$1.1\pm0.4\pm1.4\pm0.03$ & $12^{+17}_{-18}$ \\
 ${\overline{B}}^{0}\to K^+K^-$                                 &
                                                                 $0$ & 29 & & $$ \\
  ${\overline{B}}^{0}\to K^0\bar K^0$                            &
                                                                 $-10.0^{+0.7+1.0}_{-0.7-1.9}$ & 0 & $1.0\pm0.4\pm1.4\pm0.03$ & $$ \\
                                                                 \hline
  $B^-\to K^-\eta$                                               &
                                                                 $-11.2^{+~8.5+15.2}_{-22.0-10.3}$ & $-11.7^{+6.8+3.9+2.9}_{-9.6-4.2-5.6}$ & $33\pm30\pm7\pm3$ & $-37\pm9$  \\
                                                                 &&&$-33\pm39\pm10\pm4$
                                                                 &\\
  $B^-\to K^-\eta'$                                              &
                                                                 $0.52^{+0.66+1.14}_{-0.53-0.90}$ & $-6.2^{+1.2+1.3+1.3}_{-1.1-1.0-1.0}$ &$-10\pm6\pm7\pm5$ & $1.3^{+1.6}_{-1.7}$ \\
                                                                 &&&$0.7\pm0.5\pm0.2\pm0.9$
                                                                 &\\
  $\bar B^0\to \bar K^0\eta$                                     &
                                                                 $-21.4^{+~8.6+11.8}_{-22.9-11.3}$ & $-12.7^{+4.1+3.2+3.2}_{-4.1-1.5-6.7}$ & $21\pm20\pm4\pm3$ & \\
                                                                 &&&$-18\pm22\pm6\pm4$
                                                                 &\\
  $\bar B^0\to \bar K^0\eta'$                                    &
                                                                 $3.0^{+0.6+0.7}_{-0.5-0.8}$ & $2.3^{+0.5+0.3+0.2}_{-0.4-0.6-0.1}$ &$11\pm6\pm12\pm2$ & $5\pm5$ \\ &&&$-27\pm7\pm8\pm5$ &\\
  $B^-\to \pi^-\eta$                                             &
                                                                 $-5.0^{+2.4+~8.4}_{-3.4-10.3}$  & $-37^{+8+4+0}_{-6-4-1}$ & $5\pm19\pm21\pm5$ & $-13\pm7$ \\
                                                                 &&&$37\pm19\pm21\pm5$
                                                                 &\\
  $B^-\to \pi^-\eta'$                                            &
                                                                 $1.6^{+5.0+~9.4}_{-8.2-11.1}$ & $-33^{+6+4+0}_{-4-6-2}$ &  $21\pm12\pm10\pm14$ & $6\pm15$\\
                                                                 &&&$2\pm10\pm4\pm15$
                                                                 &\\
  $\bar B^0\to \pi^0\eta$                                        &
                                                                 $-5.2^{+2.8+24.6}_{-5.0-15.6}$ & $-42^{+~9+3+1}_{-12-2-3}$ &  $3\pm10\pm12\pm5$ & \\
                                                                 &&&$-7\pm16\pm4\pm90$
                                                                 &\\
  $\bar B^0\to \pi^0\eta'$                                       &
                                                                 $-7.3^{+1.0+17.6}_{-1.8-14.0}$ & $-36^{+10+2+2}_{-~9-1-3}$ &  $-24\pm10\pm19\pm24$ & \\ &&&$--$& \\
  $\bar B^0\to \eta\eta$                                         &
                                                                 $-63.5^{+10.4+~9.8}_{-~6.4-12.4}$ & $-33^{+2.6+4.1+3.5}_{-2.8-3.8-0.0}$ &  $-9\pm24\pm21\pm4$ & $$ \\
                                                                 &&&$48\pm22\pm20\pm13$
                                                                 &\\
  $\bar B^0\to \eta\eta'$ &
                                                                 $-59.2^{+7.2+3.8}_{-6.8-4.8}$ & $77.4^{+0.0+~6.9+8.0}_{-5.6-11.2-9.0}$ &  $--$ & $$ \\
                                                                 &&&$70\pm13\pm20\pm4$
                                                                 &\\
  $\bar B^0\to \eta'\eta'$                                       &
                                                                 $-44.9^{+3.1+8.5}_{-3.1-9.2}$ & $23.7^{+10.0+18.5+6.0}_{-~6.9-16.9-8.5}$ &  $--$ & $$ \\
                                                                 &&&$60\pm11\pm22\pm29$
                                                                 &\\
 \end{tabular}
 \end{ruledtabular}
 \end{table}

\subsection{Direct \CP asymmetries}
For $\rho_C\approx 1.3 $ and $\phi_C\approx -70^\circ$, we find that all the \CP puzzles in $B\to PP$ decays are resolved as shown in fourth column of Table \ref{tab:PP}. The corresponding $a_2$'s are
\be \label{eq:a2PP}
a_2(\pi \pi)\approx 0.60\,e^{-i55^\circ}, &\quad& a_2(K\pi)\approx 0.51\,e^{-i58^\circ}.
\en
They are consistent with the phenomenological determination of $C^{(')}/T^{(')}\sim a_2/a_1$ from a global fit to the available data \cite{Chiang}.
Due to the interference between the penguin and the large complex color-suppressed tree amplitudes, it is clear from Table \ref{tab:PP} that theoretical predictions for direct \CP asymmetries now agree with experiment in signs even for those modes with the significance of $\acp$ less than 3$\sigma$. We shall discuss each case one by one.

\vskip 0.2in \noindent{\it \underline{$\acp(K^-\pi^+)$}} \vskip 0.1in
Neglecting electroweak penguin contributions, the decay amplitude of $\bar B^0\to K^-\pi^+$ reads
\be \label{eq:BtoKpi}
A(\bar B^0\to K^-\pi^+) &=& A_{\pi\bar K}(\delta_u \alpha_1+\alpha_4^p+\beta_3^p).
\en
Following \cite{BN}, the \CP asymmetry of $\bar B^0\to K^-\pi^+$ can be expressed as
\be \label{eq:acpKpi}
\acp(\bar B^0\to K^-\pi^+)\,R_{\rm FM}=-2\sin\gamma\,{\rm Im}\,r_{\rm FM},
\en
with
\be \label{eq:rFM}
R_{\rm FM} &\equiv& {\Gamma(\bar B^0\to K^-\pi^+)\over \Gamma(B^-\to \bar K^0\pi^-)}=1-2\cos\gamma\,{\rm Re}\, r_{\rm FM}+|r_{\rm FM}|^2, \non \\
r_{\rm FM} &=& \left|{\lambda_u^{(s)}\over \lambda_c^{(s)}}\right|\,{\alpha_1(\pi \bar K)\over -\alpha_4^c(\pi\bar K)-\beta_3^c(\pi\bar K)},
\en
where the small contribution from $\hat \alpha_4^u$ has been neglected and the decay amplitude of $B^-\to \bar K^0\pi^-$ is given in Eq. (\ref{eq:AmpK0pi0}).   Theoretically, we obtain $r_{\rm FM}=0.14$ for $\gamma=67.8^\circ$ with a small imaginary part and $R_{\rm FM}=0.91$, to be compared with the experimental value $R_{\rm FM} =0.84\pm0.04$. In the absence of penguin annihilation, direct \CP violation of $\bar B^0\to K^-\pi^+$ is positive as Im\,$\alpha_4^c\approx 0.013$. When the power correction to penguin annihilation is turned on, we have Im$(\alpha_4^c+\beta^c_3)\approx -0.039$ and hence a negative $\acp(K^-\pi^+)$. This also explains why \CP asymmetries of penguin-dominated decays in the QCDF framework will often reverse their signs in the presence of penguin annihilation.

\vskip 0.2in \noindent{\it \underline{$\acp(K^-\pi^0)$}} \vskip 0.1in
The decay amplitude is
\be
\sqrt{2}A(B^-\to K^-\pi^0) &=& A_{\pi\bar K}(\delta_u \alpha_1+\alpha_4^p+\beta_3^p)+A_{\bar K\pi}(\delta_{pu}\alpha_2+{3\over 2}\alpha^p_{\rm 3,EW}).
\en
If the color-suppressed tree and electroweak penguin amplitudes are negligible, it is obvious that the amplitude of $K^-\pi^0$ will be the same as that of $K^-\pi^+$ except for a trivial factor of $1/\sqrt{2}$. The \CP asymmetry difference $\Delta A_{K\pi}\equiv\acp(K^-\pi^0)-\acp(K^-\pi^+)$ arising from the interference between $P'$ and $C'$ and between $P_{\rm EW}'$ and $T'$  is expected to be small, while it is $0.148\pm0.028$ experimentally \cite{HFAG}. To identify the effect due to the color-suppressed tree amplitude, we write
\be \label{eq:DeltaA}
\Delta A_{K\pi}=0.015^{+0.006+0.008}_{-0.006-0.013}-2\sin\gamma\, {\rm Im}\,r_C+\cdots,
\en
where the first term on the r.h.s. is due to the interference of the electroweak penguin  with color-allowed tree and QCD penguin amplitudes and
\be
r_C=\left|{\lambda_u^{(s)}\over \lambda_c^{(s)}}\right|\,{f_\pi F_0^{BK}(0)\over f_K F_0^{B\pi}(0)}{\alpha_2(\pi \bar K)\over -\alpha_4^c(\pi\bar K)-\beta_3^c(\pi\bar K)}.
\en
The imaginary part of $r_C$ is rather small because of the cancelation of the phases between $\alpha_2$ and $\alpha_4^c+\beta_3^c$. When soft corrections to $a_2$ are included, we have $r_C\approx 0.078-0.063i$\,. It follows from Eq. (\ref{eq:DeltaA}) that $\Delta A_{K\pi}$ will become of order 0.13\,.

As first emphasized by Lunghi and Soni \cite{Lunghi}, in the QCDF analysis of the quantity $\Delta A_{K\pi}$, although the theoretical uncertainties due to power corrections from penguin annihilation are large for individual asymmetries $\acp(K^-\pi^0)$ and $\acp(K^-\pi^+)$, they essentially cancel out in their difference, rendering the theoretical prediction more reliable.
We find $\Delta A_{K\pi}=(12.3^{+2.2+2.1}_{-0.9-4.7})\%$, while it is only $(1.9^{+0.5+1.6}_{-0.4-1.0})\%$ in the absence of power corrections to $a_2$ or to the topological amplitude $C'$.

\vskip 0.2in \noindent{\it \underline{$\acp(\bar K^0\pi^0)$ and $\acp(\bar K^0\pi^-)$}} \vskip 0.1in
The decay amplitudes are
\be \label{eq:AmpK0pi0}
\sqrt{2}A(\bar B^0\to \bar K^0\pi^0) &=& A_{\pi\bar K}(-\alpha_4^p-\beta_3^p)+A_{\bar K\pi}(\delta_{pu}\alpha_2+{3\over 2}\alpha^p_{\rm 3,EW})=-p'+c', \non \\
A(B^-\to \bar K^0\pi^-) &=& A_{\pi\bar K}(\alpha_4^p+\beta_3^p)=p',
\en
where the amplitudes $p'=P'-{1\over 3}P'^c_{\rm EW}+P'_A$, and $c'=C'+P'_{\rm EW}$ have been introduced in Sec. 1.
\CP violation of $B^-\to \bar K^0\pi^-$ is expected to be very small as it is a pure penguin process. Indeed, QCDF predicts $\acp(\bar K^0\pi^-)\approx 0.003$. If $c'$ is negligible compared to $p'$, $\acp(\bar K^0\pi^0)$ will be very small.
Just as the previous case, the \CP asymmetry difference of the $\bar K^0\pi^0$ and $\bar K^0\pi^-$ modes reads
\be
\Delta A'_{K\pi}\equiv \acp(\bar K^0\pi^0)-\acp(\bar K^0\pi^-)=(0.57^{+0.04+0.14}_{-0.04-0.06})\%+
2\sin\gamma\, {\rm Im}\,r_C+\cdots,
\en
where the first term on the r.h.s. is  due to the interference between the electroweak and QCD penguin amplitudes. To a good approximation, we have $\Delta A'_{K\pi}\sim -\Delta A_{K\pi}$. This together with the measured value of $\Delta A_{K\pi}$ and the smallness of $\acp(\bar K^0\pi^-)$ indicates that $\acp(\bar K^0\pi^0)$ should be roughly of order $-0.15$.
Using Im\,$r_C\approx -0.063$ as discussed before, it follows from the above equation that
$\acp(\bar K^0\pi^0)$ is of order $-11\%$.
More precisely, we predict $\acp(\bar K^0\pi^0)=(-10.6^{+2.7+5.6}_{-3.8-4.3})\%$ and $\Delta A'_{K\pi}=(-11.0^{+2.7+5.8}_{-3.8-4.3})\%$, while they are of order 0.0075 and 0.0057, respectively, in the absence of $\rho_C$ and $\phi_C$.
Therefore, an observation of $\acp(\bar K^0\pi^0)$ at the level of $-(10\sim 15)\%$ will be a strong support for the presence of power corrections to $c'$.  This is essentially a model independent statement.

Experimentally, the current world average $-0.01\pm0.10$ is consistent with no \CP violation because the BaBar and Belle measurements $-0.13\pm0.13\pm0.03$ \cite{BaBarK0pi0} and $0.14\pm0.13\pm0.06$ \cite{BelleK0pi0}, respectively, are of opposite sign.  Nevertheless, there exist several model-independent determinations of this asymmetry: one is the SU(3) relation $\Delta\Gamma(\pi^0\pi^0)=-\Delta \Gamma(\bar K^0\pi^0)$ \cite{Deshpande}
    and the other is the approximate sum rule for \CP rate asymmetries \cite{AS98}
\begin{eqnarray} \label{eq:SR}
\Delta\Gamma(K^-\pi^+)+\Delta \Gamma(\bar K^0\pi^-)\approx 2[\Delta \Gamma(K^-\pi^0)+\Delta \Gamma(\bar K^0\pi^0)],
\end{eqnarray}
based on isospin symmetry, where
$\Delta \Gamma(K\pi)\equiv \Gamma(\bar B\to\bar K\bar\pi)-\Gamma(B\to K\pi)$. This sum rule allows us to extract $\acp(\bar K^0\pi^0)$ in terms of the other three asymmetries of $K^-\pi^+,K^-\pi^0,\bar K^0\pi^-$ modes that have been measured.
From the current data of branching fractions and \CP asymmetries, the above SU(3) relation and {\it CP}-asymmetry sum rule lead to $\acp(\bar K^0\pi^0)=-0.073^{+0.042}_{-0.041}$ and
$\acp(\bar K^0\pi^0)=-0.15\pm 0.04$, respectively. An analysis based on the topological quark diagrams also yields a similar result $-0.08\sim -0.12$ \cite{Chiang09}. All these indicate that the direct \CP violation  $\acp(\bar K^0\pi^0)$ should be negative and has a magnitude  of order 0.10\,.

\vskip 0.2in \noindent{\it \underline{$\acp(K\eta^{(')})$}} \vskip 0.1in
The world average of $\acp(B^-\to K^-\eta)=-0.37\pm0.09$ due to the measurements $-0.36\pm0.11\pm0.03$ from BaBar \cite{BaBar:Beta} and $-0.39\pm0.16\pm0.03$ from Belle \cite{Belle:Beta2} differs from zero by 4.1$\sigma$ deviations.
The decay amplitude of $B^-\to K^-\eta$ is given by \cite{BN}
\be \label{eq:BtoKeta}
\sqrt{2}A(B^-\to K^-\eta) &=& A_{\bar K\eta_q}\left[\delta_{pu}\alpha_2+2\alpha_3^p\right]
+\sqrt{2}A_{\bar K\eta_s}\left[\delta_{pu}\beta_2+\alpha_3^p+\alpha_4^p+\beta_3^p\right]  \\
&+&\sqrt{2}A_{\bar K\eta_c}\left[\delta_{pc}\alpha_2+\alpha_3^p\right]
 +A_{\eta_q \bar K}\left[\delta_{pu}(\alpha_1+\beta_2)+\alpha_4^p+\beta_3^p\right], \non
\en
where the flavor states of the $\eta$ meson,  $q\bar q\equiv (u\bar u+d\bar
d)/\sqrt{2}$, $s\bar s$ and $c\bar c$ are labeled by the $\eta_q$, $\eta_s$ and $\eta_{c}^0$, respectively.  Since the two penguin processes $b\to ss\bar s$ and $b\to sq\bar q$ contribute
destructively to $B\to K\eta$ (i.e. $A_{\bar K\eta_s}=X^{(\bar B\bar K,\eta_s)}$ has an opposite sign to $A_{\bar K\eta_q}$ and $A_{\eta_q\bar K}$), the penguin amplitude is comparable in magnitude to the tree amplitude induced from $b\to us\bar u$, contrary to the decay $B\to K\eta'$ which is dominated by large penguin amplitudes. Consequently, a sizable direct \CP asymmetry is expected in $B^-\to K^-\eta$
but not in $K^-\eta'$ \cite{BSS}.

The decay constants $f_\eta^{q}$, $f_\eta^{s}$ and $f_\eta^c$
are given before in Eqs. (\ref{eq:fetaqs}) and (\ref{eq:fetac}).
Although $f_\eta^c\approx -2$ MeV is much smaller than $f_\eta^{q,s}$, its effect is CKM enhanced by $V_{cb}V_{cs}^*/(V_{ub}V_{us}^*)$. In the presence of penguin annihilation, $\acp(K^-\eta)$ is found to be of order $0.127$ (see Table \ref{tab:PP}). When $\rho_C$ and $\phi_C$ are turned on, $\acp(K^-\eta)$ will be reduced to 0.004
if there is no  intrinsic charm content of the $\eta$. When the effect of $f_\eta^c$ is taken into account, $\acp(K^-\eta)$ finally reaches the level of $-11\%$ and has a sign in agreement with experiment. Hence, \CP violation in $B^-\to K^-\eta$ is the place where the charm content of the $\eta$ plays a role.

Two remarks are in order.
First, the pQCD prediction for $\acp(K^-\eta)$ is very sensitive to $m_{qq}$, the mass of the $\eta_q$, which is generally taken to be of order $m_\pi$. It was found in \cite{ChenKeta} that for $m_{qq}=0.14$, 0.18 and 0.22 GeV, $\acp(K^-\eta)$ becomes 0.0562, 0.0588 and $-0.3064$, respectively. There are two issues here: (i) Is it reasonable to have a large value of $m_{qq}$ ? and (ii) The fact that $\acp(K^-\eta)$ is so sensitive to $m_{qq}$ implies that the pQCD prediction is not stable. Within the framework of pQCD, the authors of \cite{Xiao} rely on the NLO corrections to get a negative \CP asymmetry and avoid the aforementioned issues. At the lowest order, pQCD predicts $\acp(K^-\eta)\approx 9.3\%$. Then NLO corrections will flip the sign and give rise to $\acp(K^-\eta)=(-11.7^{+~8.4}_{-11.4})\%$. In view of the sign change of $\acp$ by NLO effects here, this indicates that pQCD calculations should be carried out systematically to NLO in order to have a reliable estimate of \CP asymmetries.
Second, while both QCDF and pQCD can manage to lead to a correct sign for $\acp(K^-\eta)$, the predicted magnitude still falls short of the measurement $-0.37\pm0.09$. At first sight, it appears that the QCDF prediction $\acp(K^-\eta)=-0.221^{+0.160}_{-0.182}$ (see Table \ref{tab:PP}) obtained in the leading $1/m_b$ expansion already agrees well with the data. However, the agreement is just an accident. Recall that in the absence of power corrections, the calculated \CP asymmetries for $K^-\pi^+$ and $\pi^+\pi^-$ modes are wrong in signs. That is why it is important to consider the major power corrections step by step.
The QCDF results in the heavy quark limit should  not be considered as the final QCDF predictions to be compared with experiment.

\vskip 0.2in \noindent{\it \underline{$\acp(\pi^-\eta)$}} \vskip 0.1in
As for the decay $B^-\to\pi^-\eta$, it is interesting to see that penguin annihilation will flip the sign of $\acp(\pi^-\eta)$ into a wrong one without affecting its magnitude (see Table \ref{tab:PP}). Again, soft corrections to $a_2$ will bring the \CP asymmetry back to the right track.
Contrary to the previous case of $B^-\to K^-\eta$, the charm content of the $\eta$ here does not play a role as it does not get a CKM enhancement.

\vskip 0.2in \noindent{\it \underline{$\acp(\pi^+\pi^-)$}} \vskip 0.1in
It is well known that based on SU(3) flavor symmetry, direct \CP\ asymmetries in $K\pi$ and
$\pi\pi$ systems are related as \cite{Deshpande}:
\begin{eqnarray}
\Delta\Gamma (K^-\pi^+)=-\Delta\Gamma (\pi^+\pi^-), \qquad \Delta\Gamma (\bar K^0\pi^0)=-\Delta\Gamma (\pi^0\pi^0).
\end{eqnarray}
The first relation leads to
$\acp(\pi^+\pi^-)=[{\cal B}(K^-\pi^+)/{\cal
B}(\pi^+\pi^-)]\acp(K^-\pi^+)\approx 0.37$\,, which is in good
agreement with the current world average of $0.38\pm0.06$ \cite{HFAG}.

The decay amplitude is
\be
A(\bar B^0\to \pi^+\pi^-)=A_{\pi\pi} \left[ \delta_{pu}(\alpha_1+\beta_1)+\alpha_4^p+\beta_3^p+\cdots\right],
\en
which is very similar to the amplitude of the $K^-\pi^+$ mode (see Eq. (\ref{eq:BtoKpi})) except for the CKM matrix elements. Since the penguin contribution is small compared to the tree one, its \CP asymmetry is approximately given by
\be
\acp(\pi^+\pi^-)\approx 2\sin\gamma\,{\rm Im}\,r_{\pi\pi},
\en
with
\be
 r_{\pi\pi}=\left|{\lambda_c^{(d)}\over \lambda_u^{(d)}}\right|\,{\alpha_4^c(\pi \pi)+\beta_3^c(\pi\pi)\over \alpha_1(\pi\pi)}.
\en
Numerically, we obtain Im$r_{\pi\pi}$= 0.107 ($-0.033$) with (without) the annihilation term $\beta_3^c$. Hence, one needs penguin annihilation in order to have a correct sign for $\acp(\pi^+\pi^-)$. However, the dynamical calculation of both QCDF and pQCD yields $\acp(\pi^+\pi^-)\approx 0.17\sim 0.20$. It is hard to push the \CP asymmetry to the level of 0.38\,. Note that the central values of current $B$ factory measurements of \CP asymmetry: $-0.25\pm0.08\pm0.02$ by BaBar \cite{BaBar:pipiCP} and $-0.55\pm0.08\pm0.05$ by Belle \cite{Belle:pipiCP}, differ by a factor of 2.

\vskip 0.2in \noindent{\it \underline{$\acp(\pi^0\pi^0)$}} \vskip 0.1in
Just like the $\pi^0\eta$ mode, penguin annihilation will flip the sign of $\acp(\pi^0\pi^0)$ into a wrong one (see Table \ref{tab:PP}). If the amplitude $c=C+P_{\rm EW}$ is large and complex, its interference with the QCD penguin will bring the sign of \CP asymmetry into the right one. As mentioned before,
$|P_{\rm EW}/C|$ is of order 0.06 before any power corrections. It is thus very unlikely that an enhancement of $P_{\rm EW}$ through New Physics  can render $c$ large and complex. For the $a_2(\pi\pi)$ given by Eq. (\ref{eq:a2PP}), we find that $\acp(\pi^0\pi^0)$ is of order 0.55, to be compared with the current average, $0.43^{+0.25}_{-0.24}$ \cite{HFAG}.

\vskip 0.2in \noindent{\it \underline{$\acp(\pi^-\pi^0)$}} \vskip 0.1in

It is generally believed that direct \CP violation of $B^-\to\pi^-\pi^0$ is very small. This is because the isospin of
the $\pi^-\pi^0$ state is
$I=2$ and hence it does not receive QCD penguin contributions and
receives only the loop contributions from electroweak penguins.
Since this decay is tree dominated, SM predicts an almost null
\CP asymmetry, of order $10^{-3}\sim 10^{-4}$. What will happen if $a_2$ has a large magnitude and strong phase ? We find that power corrections to the color-suppressed tree amplitude
will enhance $\acp(\pi^-\pi^0)$ substantially to the level of 2\%. Similar conclusions were also obtained by the analysis based on the diagrammatic approach \cite{Chiang}. However, one must be very cautious about this. The point is that power corrections will affect not only $a_2$, but also other parameters $a_i$ with $i\neq 2$. Since the isospin of $\pi^-\pi^0$ is $I=2$, the soft corrections to $a_2$ and $a_i$  must be conspired in such a way that $\pi^-\pi^0$  is always an $I=2$ state.
As explained below, there are two possible sources of power corrections to $a_2$: spectator scattering and final-state interactions.  For final-state rescattering, it is found in \cite{CCSfsi} that effects of FSIs on $\acp(\pi^-\pi^0)$ are small, consistent with the requirement followed from the CPT theorem.
In the specific residual scattering model considered by one of us (CKC) \cite{Chua}, $\pi^-\pi^0$ can only rescatter into itself, and as a consequence, direct \CP violation will not receive any contribution from residual final-state interactions.  Likewise, if
large $\rho_H$ and $\phi_H$ are turned on to mimic Eq. (\ref{eq:a2PP}), we find $\acp(\pi^-\pi^0)$ is at most of order $10^{-3}$. (The result of $\acp(\pi^-\pi^0)$ in QCDF listed in Tables \ref{tab:PP} and \ref{tab:PPCP} is obtained in this manner.) This is because spectator scattering will contribute to not only $a_{2}$ but also $a_1$ and the electroweak penguin parameters $a_{7-10}$. Therefore,  a
measurement of direct \CP violation in $B^-\to \pi^-\pi^0$
provides a nice test of the Standard Model and New Physics.

\vskip 0.2in \noindent{\it \underline{\CP asymmetries in pQCD and SCET}} \vskip 0.1in
For most of the $B\to PP$ decays, pQCD predictions of \CP asymmetries are similar to the QCDF ones at least in signs except for $K\bar K,~K^-\pi^0,~K^-\eta',~\pi\eta,~\eta\eta',\eta'\eta'$ modes. Experimental measurements of $\acp$ in $\pi^-\eta,\pi^-\eta'$ modes are in better agreement with QCDF than pQCD. It is known that power corrections such as penguin annihilation in QCDF are often plagued by the end-point divergence that in turn
breaks the factorization theorem. In the pQCD approach, the endpoint singularity is cured
by including the parton's transverse momentum.  Due to a different treatment of endpoint divergences in penguin annihilation diagrams, some of the \CP puzzles do not occur in the approach of pQCD. For example, pQCD predicts the right sign of \CP asymmetries for $\bar B^0\to \pi^0\pi^0$ and $B^-\to \pi^-\eta$ without invoking soft corrections to $a_2$.

For decays involving $\eta$ and $\eta'$, there are two sets of SCET solutions as there exist two different sets of SCET parameters that minimize $\chi^2$. It is clear from Table \ref{tab:PPCP} that the predicted signs of \CP asymmetries for $K^-\pi^0,\pi^0\pi^0,\pi^-\eta$ disagree with the data and hence the $\Delta A_{K\pi}$ puzzle is not resolved.
Also the predicted \CP violation for $\bar K^0\pi^0$ and $\bar K^{*0}\pi^0$ is of opposite sign to QCDF and pQCD. This is not a surprise because the long-distance charming penguins in SCET mimic the penguin annihilation effects in QCDF. All the $B$-{\it CP} puzzles occurred in QCDF will also manifest in SCET. (The reader can compare the SCET results of $\acp$ in Tables \ref{tab:PPCP} (for $B\to PP$) and \ref{tab:VPtreeCP}-\ref{tab:VPCP} (for $B\to VP$) with the QCDF predictions in the third column of Tables \ref{tab:PP} and \ref{tab:VP}.) This means that one needs other power corrections to resolve the \CP puzzles induced by charming penguins. In the current phenomenological analysis of SCET \cite{SCET1}, the ratio of $C^{(')}/T^{(')}$ is small and real to the leading order. This constraint should be released.

\subsection{Mixing-induced \CP asymmetry}
Possible New Physics beyond the Standard Model is being
intensively searched via the measurements of time-dependent \CP
asymmetries in neutral $B$ meson decays into final \CP eigenstates
defined by
 \be
 {\Gamma(\ov B(t)\to f)-\Gamma(B(t)\to f)\over
 \Gamma(\ov B(t)\to f)+\Gamma(B(t)\to
 f)}=S_f\sin(\Delta mt)-C_f\cos(\Delta mt),
 \en
where $\Delta m$ is the mass difference of the two neutral $B$
eigenstates, $S_f$ monitors mixing-induced \CP asymmetry and
$A_f$ measures direct \CP violation (note that $C_f=-\acp$). The $CP$-violating parameters $C_f$ and
$S_f$ can be expressed as
 \be
 C_f={1-|\lambda_f|^2\over 1+|\lambda_f|^2}, \qquad S_f={2\,{\rm
 Im}\lambda_f\over 1+|\lambda_f|^2},
 \en
where
 \be
 \lambda_f={q_B\over p_B}\,{A(\ov B^0\to f)\over A(B^0\to f)}.
 \en
In the standard model $\lambda_f\approx \eta_f e^{-2i\beta}$ for $b\to s$ penguin-dominated or pure
penguin modes with $\eta_f=1$ ($-1$) for final $CP$-even (odd)
states. Therefore, it is anticipated in the Standard Model that
$-\eta_fS_f\approx \sin 2\beta$ and $A_f\approx 0$.

The predictions of $S_f$ of $B\to PP$ decays in various approaches and the experimental measurements from BaBar and Belle are summarized in Table \ref{tab:SPP}.
It is clear that $\eta'K_S$
appears theoretically very clean in QCDF and SCET and is close to $\sin 2\beta=0.672\pm0.023$ determined from $b\to c\bar cs$ transitions \cite{HFAG}.
Note also that the experimental errors on
$S_{\eta'K_S}$ are the smallest and its branching
fraction is the largest, making it especially suitable for faster
experimental progress in the near future.

\begin{table}[!tb]
\begin{ruledtabular}
\caption{Mixing-induced \CP\ violation $S_f$ in $B\to PP$ decays predicted in various approaches.
 The pQCD results are taken from \cite{LiPP,XiaoKeta,Xiaoetaeta,Xiaopieta}. For final states involving $\eta$ and/or $\eta'$, there are
two solutions with SCET predictions \cite{Zupan}. The parameter $\eta_f=1$ except for $K_S(\pi^0,\eta,\eta')$ modes where $\eta_f=-1$. Experimental results from BaBar (first entry) and Belle (second entry) are listed whenever available. The input values of $\sin 2\beta$ used at the time of theoretical calculations are displayed. }\label{tab:SPP}
\begin{tabular}{l| c c c | c  c}
Mode
       &QCDF (this work) 
       &pQCD 
       &SCET 
       & Expt. \cite{BaBar:etaK0,Belle:etaK0,Belle:piK0,BaBar:pipi,Belle:pipi}
       & Average  
       \\
       \hline
 $\sin 2\beta$ & 0.670 & 0.685 & 0.725 & \\
 \hline
 $\eta' K_S$
       & $0.67^{+0.01+0.01}_{-0.01-0.01}$
       & $0.63^{+0.50}_{-0.91}$
       & $\begin{array}{c}0.706\pm0.008\\ 0.715\pm0.010\end{array}$
       & $\begin{array}{c}0.57\pm0.08\pm0.02 \\ 0.64\pm0.10\pm0.04 \end{array}$
       & $0.59\pm0.07$
       \\
 $\eta K_S$
       & $0.79^{+0.04+0.08}_{-0.06-0.06}$
       & $0.62^{+0.50}_{-0.92}$
       & $\begin{array}{c}0.69\pm0.16\\ 0.79\pm0.15\end{array}$
       &
       \\
 $\pi^0K_S$
       & $0.79^{+0.06+0.04}_{-0.04-0.04}$
       & $0.74^{+0.02}_{-0.03}$
       & $0.80\pm0.03$
       & $\begin{array}{c}0.55\pm0.20\pm0.03\\ 0.67\pm0.31\pm0.08\end{array}$
       & $0.57\pm0.17$
       \\
       \hline
 $\pi^+\pi^-$
       & $-0.69^{+0.08+0.19}_{-0.10-0.09}$ & $-0.42^{+1.00}_{-0.56}$ & $-0.86\pm0.10$ &   $\begin{array}{c} -0.68\pm0.10\pm0.03\\-0.61\pm0.10\pm0.04\end{array}$ & $-0.65\pm0.07$
       \\
 $\pi^0\eta$ & $0.08^{+0.06+0.19}_{-0.12-0.23}$
       & $0.067^{+0.005}_{-0.010}$
       & $\begin{array}{c} -0.90\pm0.24\\-0.67\pm0.82\end{array}$ &
       \\
 $\pi^0\eta'$ & $0.16^{+0.05+0.11}_{-0.07-0.14}$
       &$0.067^{+0.004}_{-0.011}$
       & $\begin{array}{c} -0.96\pm0.12\\-0.60\pm1.31\end{array}$ &
       \\
 $\eta\eta$ & $-0.77^{+0.07+0.12}_{-0.05-0.06}$
       &$0.535^{+0.004}_{-0.004}$
       &
       $\begin{array}{c} -0.98\pm0.11\\-0.78\pm0.31\end{array}$ &
       \\
 $\eta\eta'$ & $-0.76^{+0.07+0.06}_{-0.05-0.03}$
       &$-0.131^{+0.056}_{-0.050}$
       &$\begin{array}{c} -0.82\pm0.77\\-0.71\pm0.37\end{array}$ &
       \\
 $\eta'\eta'$ & $-0.85^{+0.03+0.07}_{-0.02-0.06}$
       &$0.93^{+0.08}_{-0.12}$
       & $\begin{array}{c} -0.59\pm1.10\\-0.78\pm0.31\end{array}$ &
       \\
\end{tabular}
\end{ruledtabular}
\end{table}

\begin{table}[!htbp]
\begin{ruledtabular}
\caption{Same as Table \ref{tab:SPP} except for $\Delta S_f$ for penguin-dominated modes. The QCDF results obtained by Beneke \cite{BenekeS} are displayed  for comparison.}\label{tab:DeltaSPP}
\begin{tabular}{l| c c c c c |c c}
 &  \multicolumn{2}{c}{QCDF (this work)} &
    \\ \cline{2-3}
\raisebox{2.0ex}[0cm][0cm]{Mode} & With $\rho_C,\phi_C$ & W/o $\rho_C$ & \raisebox{2.0ex}[0cm][0cm]{QCDF (Beneke)} & \raisebox{2.0ex}[0cm][0cm]{pQCD} & \raisebox{2.0ex}[0cm][0cm]{SCET} & \raisebox{2.0ex}[0cm][0cm]{Expt.} & \raisebox{2.0ex}[0cm][0cm]{Average} \\ \hline
 $\eta' K_S$
       & $0.00^{+0.01}_{-0.01}$
       & $0.01^{+0.01}_{-0.01}$
       & $0.01^{+0.01}_{-0.01}$
       & $-0.06^{+0.50}_{-0.91}$
       & $\begin{array}{c}-0.02\pm0.01\\ -0.01\pm0.01\end{array}$
       & $\begin{array}{c}-0.10\pm0.08\\ -0.03\pm0.11\end{array}$
       & $-0.08\pm0.07$
       \\
 $\eta K_S$
       & $0.12^{+0.09}_{-0.08}$
       & $0.12^{+0.04}_{-0.03}$
       & $0.10^{+0.11}_{-0.07}$
       & $-0.07^{+0.50}_{-0.92}$
       & $\begin{array}{c}-0.04\pm0.16\\ 0.07\pm0.15\end{array}$
       \\
 $\pi^0K_S$
       & $0.12^{+0.07}_{-0.06}$
       & $0.09^{+0.07}_{-0.06}$
       & $0.07^{+0.05}_{-0.04}$
       & $0.06^{+0.02}_{-0.03}$
       & $0.08\pm0.03$
       & $\begin{array}{c}-0.12\pm0.20\\ 0.00\pm0.32\end{array}$
       & $-0.10\pm0.17$
       \\
\end{tabular}
\end{ruledtabular}
\end{table}

Time-dependent \CP violation in $\bar B^0\to \pi^0 K_S$ has received a great deal of attention. A correlation between $S_{\pi^0 K_S}$ and $\acp(\pi^0K_S)$ has been investigated  in \cite{Fleischer}. Recently, it has been argued that soft corrections to the color-suppressed tree amplitude will reduce
the mixing-induced asymmetry $S_{\pi^0K_S}$ to the level of 0.63 \cite{Li09}. However, we find that it is the other way around in our case.
The asymmetry $S_{\pi^0 K_S}$ is enhanced from 0.76 to $0.79^{+0.06+0.04}_{-0.04-0.04}$ in the presence of power correction effects on $a_2$. Our result of $S_{\pi^0 K_S}$
is consistent with \cite{Chua,Ciuchini,Kagan} where power corrections were studied. \footnote{Since power corrections will affect not only $a_2$, but also other parameters $a_i$ with $i\neq 2$, we have examined such effects by using $\rho_H\approx 4.9$ and $\phi_H\approx -77^\circ$ (see discussions after Eq. (\ref{eq:XH}) ) and obtained the same result as before.}
Although this deviates somewhat from the world average value of $0.57\pm0.17$ \cite{HFAG}, it does agree with the Belle measurement of $0.67\pm0.31\pm0.08$ \cite{Belle:piK0}.

In sharp contrast to QCDF and SCET where the theoretical predictions for $S_{\eta'K_S}$ are very clean,
the theoretical errors in pQCD predictions for both $S_{\eta'K_S}$ and $S_{\eta K_S}$  arising from uncertainties in the CKM angles $\alpha$ and $\gamma$ are very large \cite{XiaoKeta}. This issue should be resolved.

For the mixing-induced asymmetry in $B\to \pi^+\pi^-$, we obtain $S_{\pi^+\pi^-}=-0.69^{+0.08+0.19}_{-0.10-0.09}$, in accordance with the world average of $-0.65\pm0.07$ \cite{HFAG}. For comparison, the SCET prediction  $-0.86\pm0.10$ \cite{Zupan} is too large and the theoretical uncertainty of the pQCD result $-0.42^{+1.00}_{-0.56}$ \cite{LiPP} is too large. For $\pi^0\eta^{(')}$ modes, SCET predictions are opposite to QCDF and pQCD in signs. For $\eta\eta'$, the pQCD result is very small compared to QCDF and SCET.

The reader may wonder why the QCDF result $S_{\eta'K_S}\approx 0.67$ presented in this work is smaller than the previous result  $\approx 0.74$ obtained in \cite{CCSsin2beta,BenekeS,BuchallaS}. This is because the theoretical calculation of $S_f$ depends on the input of the angle $\beta$ or $\sin 2\beta$. For example, $\sin 2\beta\approx 0.725$ was used in the earlier estimate of $S_f$ around 2005, while a smaller value of 0.670 is used in the present work. \footnote{The experimental value of $\sin 2\beta$ determined from all $B$-factory charmonium data is $0.672\pm0.023$ \cite{HFAG}. However, as pointed out by Lunghi and Soni \cite{Lunghi:sin2beta}, one can use some observables to deduce the value of $\sin 2 \beta$: \CP-violating parameter $\epsilon_K$, $\Delta M_s/\Delta M_d$ and $V_{cb}$ from experiment along with the lattice
hadronic matrix elements, namely, the kaon $B$-parameter $B_K$ and the SU(3) breaking ratio $\xi_s$. A prediction $\sin 2 \beta=0.87\pm0.09$ is yielded in the SM. If the ratio $|V_{ub}/V_{cb}|$ is also included as an input, one gets a smaller value $0.75\pm0.04$. The deduced value of $\sin 2 \beta$ thus differs from the directly measured value at the $2\sigma$ level.
If the SM description of \CP violation through the CKM-paradigm with a single {\it CP}-odd phase is correct, then the deduced value of $\sin 2 \beta$ should agree with the directly measured  value of $\sin 2 \beta$ in $B$-factory experiments.}
Therefore, it is more sensible to consider the difference
\be
\Delta S_f\equiv -\eta_f S_f-\sin 2\beta
\en
for penguin-dominated decays. In the SM, $S_f$ for these decays should be nearly the same as the value measured from the $b\to c\bar c s$ decays such
as $\bar B^0\to J/\psi K^0$; there is a
small deviation {\it at most} ${\cal O}(0.1)$ \cite{London}. In Table \ref{tab:SPP} we have listed the values of $\sin 2\beta$ used in the theoretical calculations. Writing the decay amplitude in the form
\be
 M(\ov B^0\to f) = V_{ub}V_{us}^*A_f^u+V_{cb}V_{cs}^*A_f^c
\en
it is known that to the
first order in $r_f\equiv(\lambda_u A_f^u)/(\lambda_c A_f^c)$
\cite{Gronau,Grossman03}
 \be \label{eq:CfSf}
 \Delta S_f=2|r_f|\cos 2\beta\sin\gamma\cos\delta_f,
 \en
with $\delta_f={\rm arg}(A_f^u/A_f^c)$. Hence, the magnitude of
the \CP asymmetry difference $\Delta S_f$ is governed by the size of $A_f^u/A_f^c$. In QCDF the dominant contributions to
$A_f^u/A_f^c$ are given by \cite{BenekeS}
\be
{A^u\over A^c}\bigg|_{\eta' K_S} &\sim& {[-P^u]-[C]\over [-P^c]}\sim {[-(a_4^u+r_\chi a_6^u)]-[a_2^u R_{\eta'K_S}]\over [-(a_4^c+r_\chi a_6^c)]}, \non \\
{A^u\over A^c}\bigg|_{\eta K_S} &\sim& {[P^u]+[C]\over [P^c]}\sim {[-(a_4^u+r_\chi a_6^u)]+[a_2^u R_{\eta K_S}]\over [-(a_4^c+r_\chi a_6^c)]}, \non \\
{A^u\over A^c}\bigg|_{\pi^0 K_S} &\sim& {[-P^u]+[C]\over [-P^c]}\sim {[-(a_4^u+r_\chi a_6^u)]+[a_2^u R_{\pi K_S}]\over [-(a_4^c+r_\chi a_6^c)]},
\en
where $R$'s are real and positive ratios of form factors and decay constants and we have followed \cite{BenekeS} to denote the complex quantities by square brackets if they have real positive parts.
For $\eta'K_S$, $[-P]$ is enhanced because of the constructive interference of various penguin amplitudes. This together with the destructive interference between penguin and color-suppressed tree amplitudes implies the smallness of $\Delta S_{\eta' K_S}$.  As explained before, the penguin amplitude of $\bar B^0\to \eta K_S$ is small because of the destructive interference of two penguin amplitudes [see Eq. (\ref{eq:BtoKeta})]. This together with the fact that the color-suppressed tree amplitude contributes constructively to $A^u/A^c$ explains why
$\Delta S_{\eta K_S}$ is positive and sizable.

Mixing-induced \CP asymmetries in various approaches are listed in Table \ref{tab:DeltaSPP} where the soft effects due to $\rho_C$ and $\phi_C$ are also displayed.
In the QCDF approach, soft corrections to the color-suppressed tree amplitude will enhance $\Delta S_{\pi^0 K_S}$ slightly from ${\cal O}(0.09)$ to ${\cal O}(0.12)$. It is clear that the  QCDF results in the absence of power corrections are consistent with that obtained by Beneke \cite{BenekeS}, by us \cite{CCSsin2beta} and by Buchalla {\it et al.} \cite{BuchallaS}. For example, we obtained $S_{\eta'K_S}\approx 0.737$ in 2005 and $S_{\eta'K_S}\approx 0.674$ this time. But the value of $\Delta S_{\eta'K_S}$ remains the same as the value of $\sin 2\beta$ has been changed since 2005.

 \begin{table}[!tb]
 \caption{Same as Table \ref{tab:PP} except for some  selective $B\to VP$ decays with $\rho_C=0.8$ and $\phi_C=-80^\circ$.
 } \label{tab:VP}
\begin{ruledtabular}
 \begin{tabular}{l c c c c c}
 {Mode}
   &   {\rm W/o}~$\rho_{A,C},\phi_{A,C}$ &  {\rm With}~$\rho_A,\phi_A$ &  {\rm With}~$\rho_{A,C},\phi_{A,C}$ & Expt. \cite{HFAG} \\  \hline
  $\B(\ov B^0\to K^-\rho^+)$ & $6.5^{+5.4+0.4}_{-2.6-0.4}$ & $8.6^{+5.7+7.4}_{-2.8-4.5}$ & $8.6^{+5.7+7.4}_{-2.8-4.5}$ & $8.6^{+0.9}_{-1.1}$ \\
  $\B(\ov B^0\to \bar K^0\rho^0)$ & $4.7^{+3.3+0.3}_{-1.7-0.3}$ & $5.5^{+3.5+4.3}_{-1.8-2.8}$ & $5.4^{+3.3+4.3}_{-1.7-2.8}$ & $4.7\pm0.7$ \\
  $\B(B^-\to \bar K^0\rho^-)$ & $5.5^{+6.1+0.7}_{-2.8-0.5}$ & $7.8^{+6.3+7.3}_{-2.9-4.4}$ & $7.8^{+6.3+7.3}_{-2.9-4.4}$ & $8.0^{+1.5}_{-1.4}$ \\
  $\B(B^-\to K^-\rho^0)$ & $1.9^{+2.5+0.3}_{-1.0-0.2}$ & $3.3^{+2.6+2.9}_{-1.1-1.7}$ & $3.5^{+2.9+2.9}_{-1.2-1.8}$ & $3.81^{+0.48}_{-0.46}$ \\
  $\B(\ov B^0\to K^{*-}\pi^+)$ & $3.7^{+0.5+0.4}_{-0.5-0.4}$ & $9.2^{+1.0+3.7}_{-1.0-3.3}$ & $9.2^{+1.0+3.7}_{-1.0-3.3}$  & $8.6^{+0.9}_{-1.0}$ \\
  $\B(\ov B^0\to \bar K^{*0}\pi^0)$ & $1.1^{+0.2+0.2}_{-0.2-0.2}$ & $3.5^{+0.4+1.7}_{-0.5-1.5}$ & $3.5^{+0.4+1.6}_{-0.4-1.4}$ & $2.4\pm0.7$ \\
  $\B(B^-\to \bar K^{*0}\pi^-)$ & $4.0^{+0.7+0.6}_{-0.9-0.6}$ & $10.4^{+1.3+4.3}_{-1.5-3.9}$ & $10.4^{+1.3+4.3}_{-1.5-3.9}$ & $9.9^{+0.8}_{-0.9}$ \\
  $\B(B^-\to K^{*-}\pi^0)$ & $3.2^{+0.4+0.3}_{-0.4-0.3}$ & $6.8^{+0.7+2.3}_{-0.7-2.2}$ & $6.7^{+0.7+2.4}_{-0.7-2.2}$ & $6.9\pm2.3$ \\
  $\B(\ov B^0\to \bar K^{*0}\eta)$ & $11.0^{+6.9+1.7}_{-3.5-1.0}$ & $15.4^{+7.7+9.4}_{-4.0-7.1}$& $15.6^{+7.9+9.4}_{-4.1-7.1}$ & $15.9\pm1.0$ \\
  $\B(\ov B^0\to \omega \bar K^0)$ & $2.9^{+4.0+0.9}_{-1.6-0.4}$ & $3.9^{+4.0+3.3}_{-1.6-2.2}$& $4.1^{+4.2+3.3}_{-1.7-2.2}$ & $5.0\pm0.6$ \\
  $\B(\bar B^0\to \rho^0\pi^0)$ & $0.76^{+0.96+0.66}_{-0.37-0.31}$ & $0.58^{+0.88+0.60}_{-0.32-0.22}$ & $1.3^{+1.7+1.2}_{-0.6-0.6}$ & $2.0\pm0.5$ \footnotemark[1] \\
  $\B(B^-\to \rho^-\pi^0)$ & $11.6^{+1.2+0.9}_{-0.9-0.5}$ & $11.8^{+1.3+1.0}_{-0.9-0.6}$ & $11.8^{+1.8+1.4}_{-1.1-1.4}$ & $10.9^{+1.4}_{-1.5}$ \\
  $\B(B^-\to \rho^0\pi^-)$ & $8.2^{+1.8+1.2}_{-0.9-0.6}$ & $8.5^{+1.8+1.2}_{-0.9-0.6}$ & $8.7^{+2.7+1.7}_{-1.3-1.4}$ & $8.3^{+1.2}_{-1.3}$ \\
  $\B(\bar B^0\to \rho^-\pi^+)$ & $15.3^{+1.0+0.5}_{-1.5-0.9}$ & $15.9^{+1.1+0.9}_{-1.5-1.1}$ & $15.9^{+1.1+0.9}_{-1.5-1.1}$ & $15.7\pm1.8$ \\
  $\B(\bar B^0\to \rho^+\pi^-)$ & $8.4^{+0.4+0.3}_{-0.7-0.5}$ & $9.2^{+0.4+0.5}_{-0.7-0.7}$ & $9.2^{+0.4+0.5}_{-0.7-0.7}$ & $7.3\pm1.2$ \\
  \hline
  $\acp(\ov B^0\to K^-\rho^+)$ & $-1.3^{+0.7+3.8}_{-0.3-3.8}$ & $31.9^{+11.5+19.6}_{-11.0-12.7}$ & $31.9^{+11.5+19.6}_{-11.0-12.7}$ & $15\pm6$ \\
  $\acp(\ov B^0\to \bar K^0\rho^0)$ & $6.8^{+1.1+4.9}_{-1.2-4.9}$ & $-5.0^{+3.2+6.0}_{-6.4-4.5}$ & $8.7^{+1.2+8.7}_{-1.2-6.8}$ & --\\
  $\acp(B^-\to \bar K^0\rho^-)$ & $0.24^{+0.12+0.08}_{-0.15-0.07}$ & $0.27^{+0.19+0.46}_{-0.27-0.17}$ & $0.27^{+0.19+0.46}_{-0.27-0.17}$ & $-12\pm17$ \\
  $\acp(B^-\to K^-\rho^0)$ & $-8.3^{+3.5+7.0}_{-0.9-7.0}$ & $56.5^{+16.1+30.0}_{-18.2-22.8}$ & $45.4^{+17.8+31.4}_{-19.4-23.2}$  & $37\pm11$  \\
  $\acp(\ov B^0\to K^{*-}\pi^+)$ & $15.6^{+0.9+4.5}_{-0.7-4.7}$ & $-12.1^{+0.5+12.6}_{-0.5-16.0}$ & $-12.1^{+0.5+12.6}_{-0.5-16.0}$ & $-23\pm8$ \\
  $\acp(\ov B^0\to \bar K^{*0}\pi^0)$ & $-12.0^{+2.4+11.3}_{-4.6-~7.6}$ & $-0.87^{+1.71+6.04}_{-0.89-6.79}$ & $-10.7^{+1.8+9.1}_{-2.8-6.3}$  & $-15\pm12$ \\
  $\acp(B^-\to \bar K^{*0}\pi^-)$ & $0.97^{+0.11+0.12}_{-0.07-0.11}$ & $0.39^{+0.04+0.10}_{-0.03-0.12}$ & $0.39^{+0.04+0.10}_{-0.03-0.12}$ & $-3.8\pm4.2$ \\
  $\acp(B^-\to K^{*-}\pi^0)$ & $17.5^{+2.0+6.3}_{-1.3-8.0}$ & $-6.7^{+0.7+11.8}_{-1.1-14.0}$ & $1.6^{+3.1+11.1}_{-1.7-14.3}$ & $4\pm29$ \\
  $\acp(\ov B^0\to \bar K^{*0}\eta)$ & $3.0^{+0.4+1.9}_{-0.4-1.8}$ & $0.20^{+0.51+2.00}_{-1.00-1.21}$ & $3.5^{+0.4+2.7}_{-0.5-2.4}$ & $19\pm5$ \\
  $\acp(\ov B^0\to \omega \bar K^0)$ & $-5.9^{+1.9+3.4}_{-2.3-4.1}$ & $6.6^{+4.7+6.0}_{-3.4-5.3}$& $-4.7^{+1.8+5.5}_{-1.6-5.8}$ & $32\pm17$ \\
  $\acp(\bar B^0\to \rho^0\pi^0)$ & $-2.3^{+2.4+9.9}_{-3.7-9.2}$ & $31.5^{+13.3+21.5}_{-12.5-30.9}$ & $11.0^{+5.0+23.5}_{-5.7-28.8}$ & $-30\pm38$  \\
  $\acp(B^-\to \rho^-\pi^0)$ & $-5.4^{+0.4+2.0}_{-0.3-2.1}$ & $16.3^{+1.1+~7.1}_{-1.2-10.5}$ & $9.7^{+2.1+~8.0}_{-3.1-10.3}$ & $2\pm11$ \\
  $\acp(B^-\to \rho^0\pi^-)$ & $6.7^{+0.5+3.5}_{-0.8-3.1}$ & $-19.8^{+1.7+12.6}_{-1.2-~8.8}$ & $-9.8^{+3.4+11.4}_{-2.6-10.2}$ & $18^{+~9}_{-17}$ \\
  $\acp(\bar B^0\to \rho^-\pi^+)$ & $-3.5^{+0.2+1.0}_{-0.2-0.9}$ & $4.4^{+0.3+5.8}_{-0.3-6.8}$ & $4.4^{+0.3+5.8}_{-0.3-6.8}$ & $11\pm6$ \\
  $\acp(\bar B^0\to \rho^+\pi^-)$ & $0.6^{+0.1+2.2}_{-0.1-2.2}$ & $-22.7^{+0.9+8.2}_{-1.1-4.4}$ & $-22.7^{+0.9+8.2}_{-1.1-4.4}$ & $-18\pm12$ \\
 \end{tabular}
 \footnotetext[1]{If an $S$ factor is included, the average will become $2.0\pm0.8$.}
 \end{ruledtabular}
 \end{table}

\section{$B\to VP$ Decays}

Power corrections to $a_2$ for $B\to VP$ and $B\to VV$
are not the same as that for $B\to PP$
as described by Eq. (\ref{eq:a2PP}). From Table \ref{tab:VP} we see that an enhancement of $a_2$ is needed to improve the rates of $B\to \rho^0\pi^0$ and the direct \CP asymmetry of $\bar B^0\to \bar K^{*0}\eta$. However, it is constrained by the measured rates of $\rho^0\pi^-$ and $\rho^-\pi^0$ modes. The central values of their branching fractions are already saturated even for vanishing $\rho_C(VP)$.
This means that $\rho_C(VP)$ is preferred to be smaller than $\rho_C(PP)=1.3$\,. In Table \ref{tab:VP} we show the dependence of the branching fractions and \CP asymmetries in $B\to VP$ decays with respect to $\rho_{A,C}$ and $\phi_{A,C}$. The corresponding values of $a_2$ for $\rho_C=0.8$ and $\phi_C=-80^\circ$ are
\be \label{eq:a2VP}
&& a_2(\pi\rho)\approx 0.40\,e^{-i51^\circ}, \quad a_2(\rho\pi)\approx 0.38\,e^{-i52^\circ}, \non \\
 && a_2(\rho \bar K)\approx 0.36\,e^{-i52^\circ}, \quad
a_2(\pi \bar K^*)\approx 0.39\,e^{-i51^\circ}.
\en

It is clear from Table \ref{tab:VP} that in the heavy quark limit, the predicted rates for $\bar B\to \bar K^*\pi$ are too small by a factor of $2\sim 3$, while $\B(\bar B\to \bar K\rho)$ are too small by $(15\sim 100)\%$ compared with experiment. The rate deficit for penguin-dominated decays can be accounted by the subleading power corrections from penguin annihilation.
Soft corrections to $a_2$ will enhance
$\B(B\to \rho^0\pi^0)$ to the order of $1.3\times 10^{-6}$, while the BaBar and Belle results, $(1.4\pm0.6\pm0.3)\times 10^{-6}$ \cite{BaBarrho0pi0} and $(3.0\pm0.5\pm0.7)\times 10^{-6}$ \cite{Bellerho0pi0} respectively, differ in their central values by a factor of 2. Improved measurements are certainly needed for this decay mode.

\subsection{Branching fractions}
\vskip 0.1in \noindent{\it \underline{$B\to\rho\pi,\omega\pi$}} \vskip 0.1in
From Table \ref{tab:VPBr} it is evident that the calculated $B\to\rho\pi,~\omega\pi$ rates in QCDF are in good agreement with experiment.
The previous QCDF predictions \cite{BN} for $B\to\rho\pi$ (except $B^0\to\pi^0\rho^0$) are too large because of the large form factor $A_0^{B\rho}(0)=0.37\pm0.06$ adopted in \cite{BN}. In this work we use the updated sum rule result $A_0^{B\rho}(0)=0.303\pm0.029$ \cite{Ball:PP}. It appears that there is no updated pQCD calculation for $B\to\rho\pi$ and $B\to\omega\pi$.

\vskip 0.2in \noindent{\it \underline{$B\to(\rho,\omega,\phi)\eta^{(')}$}} \vskip 0.1in
The relevant decay amplitudes are
\begin{eqnarray} \label{eq:Btorhoeta}
\sqrt{2}A(B^-\to\rho^-\eta) &\approx& A_{\rho\eta_q}\left[\delta_{pu}(\alpha_2+\beta_2)+2 \alpha_3^p+\hat\alpha_4^p\right]+A_{\eta_q\rho}\left[\delta_{pu}(\alpha_1+\beta_2)+\hat\alpha_4^p\right], \non \\
-2A(\bar B^0\to\rho^0\eta) &\approx& A_{\rho\eta_q}\left[\delta_{pu}(\alpha_2-\beta_1)+2 \alpha_3^p+\hat\alpha_4^p\right]+A_{\eta_q\rho}\left[\delta_{pu}(-\alpha_2-\beta_1)+\hat\alpha_4^p\right], \non \\
2A(\bar B^0\to\omega\eta) &\approx& A_{\omega\eta_q}\left[\delta_{pu}(\alpha_2+\beta_1)+2 \alpha_3^p+\hat\alpha_4^p\right]+A_{\eta_q\omega}\left[\delta_{pu}(\alpha_2+\beta_1)+2 \alpha_3^p+\hat\alpha_4^p\right], \non \\
\sqrt{2}A(\bar B^0\to\phi\eta) &\approx& A_{\eta_q\phi} \alpha_3^p+\sqrt{2}B_{\eta_s\phi}b_4^p+\sqrt{2}B_{\phi\eta_s}b_4^p,
\end{eqnarray}
and similar expressions for $\eta'$.
It is clear that
the decays $B^-\to\rho^-\eta^{(')}$ have rates much larger than $\bar B^0\to\rho^0\eta^{(')}$ as the former receive color-allowed tree contributions while the color-suppressed tree amplitudes in the latter cancel each other. Both QCDF and pQCD lead to the pattern $\Gamma(B^-\to \rho^-\eta)>\Gamma(B^-\to \rho^-\eta')$. This should be tested by more accurate measurements. The SCET prediction of
$\B(B^-\to \rho\eta')\sim 0.4\times 10^{-6}$ is far too small and clearly ruled out by experiment. Since the color-suppressed tree amplitudes in the decay $\bar B\to\omega\eta^{(')}$ are added together, one should have $\Gamma(\bar B^0\to \omega\eta^{(')})>\Gamma(\bar B^0\to \rho^0\eta^{(')})$. It appears that SCET predictions for $(\rho^-,\rho^0,\omega)\eta'$ \cite{SCETVP} are at odds with experiment. For example, solution I yields $\Gamma(\bar B^0\to \omega\eta^{'})<\Gamma(\bar B^0\to \rho^0\eta^{'})$ in contradiction to the theoretical expectation and solution II gives $\Gamma(\bar B^0\to \rho^0\eta^{'})>\Gamma(B^-\to \rho^-\eta^{'})$ in disagreement with the data.

The decays $\bar B^0\to\phi\eta^{(')}$ are very suppressed as their amplitudes are governed by  $V_{ub}V_{ud}^*(a_3^u-a_5^u)$. For example, we obtain $\B(\bar B^0\to\phi\eta)\approx 10^{-9}$ in the QCDF approach.  Since the branching fraction of the $\omega\eta$ mode is of order $10^{-6}$, it appears that the $\phi$ meson can be produed from the decay $\bar B^0\to\omega\eta$ followed by $\omega-\phi$ mixing. This will be possible if  $\phi$
is not a pure $s\bar s$ state and contains a tiny $q\bar q$
component. Neglecting isospin violation and the
admixture with the $\rho^0$ meson, one can parametrize the
$\omega$--$\phi$ mixing in terms of an angle $\delta$ such that the
physical $\omega$ and $\phi$ are related to the ideally mixed states
$\omega^I \equiv (u \bar u + d \bar d)/\sqrt{2}$ and $\phi^I \equiv
s \bar s$ by
\begin{eqnarray}\label{mixing}
\left(
\begin{array}{c} \omega \\ \phi \end{array} \right) = \left(
\begin{array}{c c} \cos \delta & \sin \delta \\ - \sin \delta & \cos
\delta
\end{array} \right)
\left( \begin{array}{c} \omega^I \\ \phi^I \end{array} \right),
\end{eqnarray}
and the mixing angle is about $|\delta| \sim 3.3^\circ$
\cite{Benayoun:1999fv} (see \cite{KLOE} for the latest determination of $\delta$). Therefore, the production of $\phi\eta$ through $\omega-\phi$ mixing is expected to be
\begin{eqnarray}
{\cal B}(\bar B^0\to \phi\eta)_{\rm \omega-\phi~ mixing}={\cal B}
(\bar B^0\to \omega\eta)\sin^2\delta \approx 0.85\times
10^{-6}\times (0.08)^2 \sim 5.4\times 10^{-9}.
\end{eqnarray}
It turns out that the $\omega-\phi$ mixing effect dominates over the short-distance contribution.
By the same token, the $\omega-\phi$ mixing effect should also manifest in the decay $B^-\to\phi\pi^-$:
\begin{eqnarray}
{\cal B}(B^-\to \phi\pi^-)_{\rm \omega-\phi~ mixing}={\cal B}
(B^-\to \omega\pi^-)\sin^2\delta \approx 6.7\times
10^{-6}\times (0.08)^2 \sim 4.3\times 10^{-8}.
\end{eqnarray}
For this decay, the short-distance contribution is only of order $2\times 10^{-9}$.

\begin{table}[tb]
\begin{ruledtabular}{\tiny
\caption{Branching fractions (in units of $10^{-6}$) of $B\to VP$
decays induced by the $b\to d$ ($\Delta S =0$) transition.  We also cite the experimental data \cite{HFAG,PDG}
and theoretical results given in
pQCD~\cite{Lupirho,XiaoKKst,Luphipi,XiaoVeta} and in SCET \cite{SCETVP}. }\label{tab:VPtreeBr}}
\begin{tabular} {l c c c c c c}
          Mode & QCDF (this work)  &pQCD& SCET 1 & SCET 2 & Expt.\\ \hline
 $ B^- \to \rho^- \pi^0$                & $11.8^{+1.8+1.4}_{-1.1-1.4}$
                                    & $6\sim9$
                                    & $8.9_{-0.1-1.0}^{+0.3+1.0}$
                                    & $11.4_{-0.6-0.9}^{+0.6+1.1}$
                                    & $10.9^{+1.4}_{-1.5}$\\
 $ B^- \to \rho^0 \pi^-$                & $8.7^{+2.7+1.7}_{-1.3-1.4}$
                                    & $5\sim 6$
                                    & $10.7_{-0.7-0.9}^{+0.7+1.0}$
                                    & $7.9_{-0.1-0.8}^{+0.2+0.8}$
                                    & $8.3^{+1.2}_{-1.3}$ \\
 $\bar B^0 \to \rho^\pm \pi^\mp$     &  $25.1^{+1.5+1.4}_{-2.2-1.8}$
                                    & $18\sim45$
                                    & $13.4_{-0.5-1.2}^{+0.6+1.2}$
                                    & $16.8_{-0.5-1.5}^{+0.5+1.6}$
                                    & $23.0\pm2.3$ \\
 $ \bar B^0 \to \rho^+  \pi^-$
                                    & $9.2^{+0.4+0.5}_{-0.7-0.7}$
                                    &
                                    & $5.9_{-0.5-0.5}^{+0.5+0.5}$
                                    & $6.6_{-0.1-0.7}^{+0.2+0.7}$
                                    & $7.3\pm1.2$ \\
 $ \bar B^0 \to \rho^-  \pi^+ $               & $15.9^{+1.1+0.9}_{-1.5-1.1}$
                                    &
                                    & $7.5_{-0.1-0.8}^{+0.3+0.8}$
                                    & $10.2_{-0.5-0.9}^{+0.4+0.9}$
                                    & $15.7\pm1.8$ \\
 $ \bar  B^0 \to \rho^0 \pi^0$           & $1.3^{+1.7+1.2}_{-0.6-0.6}$
                                    & $0.07\sim 0.11$
                                    & $2.5_{-0.1-0.2}^{+0.2+0.2}$
                                    & $1.5_{-0.1-0.1}^{+0.1+0.1}$
                                    & $2.0\pm0.5$ \\
 $ B^- \to \omega \pi^- $                      & $6.7^{+2.1+1.3}_{-1.0-1.1}$
                                    & $4\sim 8$
                                    & $6.7_{-0.3-0.6}^{+0.4+0.7}$
                                    & $8.5_{-0.3-0.8}^{+0.3+0.8}$
                                    & $6.9\pm0.5$                    \\
 $\bar  B^0 \to \omega \pi^0$                       & $0.01^{+0.02+0.04}_{-0.00-0.01}$
                                    &  $0.10\sim 0.28$
                                    & $0.0003_{-0.0000-0.0000}^{+0.0299+0.0000}$
                                    & $0.015_{-0.000-0.002}^{+0.024+0.002}$ & $<0.5$  \\
 $ B^- \to  K^{*0} K^- $                              & $0.80^{+0.20+0.31}_{-0.17-0.28}$
                                    & $0.32^{+0.12}_{-0.07}$
                                    & $0.49_{-0.20-0.08}^{+0.26+0.09}$
                                    & $0.51_{-0.16-0.06}^{+0.18+0.07}$
                                    & $0.68\pm0.19$ \footnotemark[1] \\
 $ B^- \to  K^{*-} K^0$                                & $0.46^{+0.37+0.42}_{-0.17-0.26}$
                                    & $0.21^{+0.14}_{-0.13}$
                                    & $0.54_{-0.21-0.08}^{+0.26+0.10}$
                                    & $0.51_{-0.17-0.07}^{+0.21+0.08}$
                                    &      \\
 $\bar B^0\to K^{*+}K^-$            &
                                    $0.08^{+0.01+0.02}_{-0.01-0.02}$  & $0.083^{+0.072}_{-0.067}$
                                      \\
 $\bar B^0\to K^{*-}K^+$            &
                                    $0.07^{+0.01+0.04}_{-0.01-0.03}$  & $0.017^{+0.027}_{-0.011}$
                                      \\
 $\bar   B^0 \to K^{*0} \bar K^0$   &
                                    $0.70^{+0.18+0.28}_{-0.15 -0.25}$
                                    & $0.24^{+0.07}_{-0.06}$
                                    & $0.45_{-0.19-0.07}^{+0.24+0.09}$
                                    & $0.47_{-0.14-0.05}^{+0.17+0.06}$
                                     &   \\
 $\bar   B^0 \to \bar K^{*0} K^0$   &
                                    $0.47^{+0.36+0.43}_{-0.17-0.27}$ & $0.49^{+0.15}_{-0.09}$
                                    & $0.51_{-0.20-0.08}^{+0.24+0.09}$
                                    & $0.48_{-0.16-0.06}^{+0.20+0.07}$
                                    &  $<1.9$ \\
 $ B^- \to \phi \pi^- $             & $\approx 0.043$ \footnotemark[2]
                                    &  $0.032^{+0.012}_{-0.014}$
                                    &$\approx0.003$
                                    &$\approx0.003$
                                    & $<0.24$  \\
 $ \bar  B^0 \to \phi \pi^0$                        & $0.01^{+0.03+0.02}_{-0.01-0.01}$
                                    & $0.0068^{+0.0010}_{-0.0008}$  &$\approx0.001$ &$\approx0.001$
                                     & $<0.28$ \\
                                     \hline
 $ B^- \to \rho^- \eta $                         & $8.3^{+1.0+0.9}_{-0.6-0.9}$
                                    & $6.7^{+2.6}_{-1.9}$
                                    & $3.9_{-1.7-0.4}^{+2.0+0.4}$
                                    & $3.3_{-1.6-0.3}^{+1.9+0.3}$
                                    & $6.9\pm1.0$\\
 $ B^- \to \rho^- \eta ^\prime$         & $5.6^{+0.9+0.8}_{-0.5-0.7}$
                                    & $4.6^{+1.6}_{-1.4}$
                                    & $0.37_{-0.22-0.07}^{+2.46+0.08}$
                                    & $0.44_{-0.20-0.05}^{+3.18+0.06}$
                                    & $9.1^{+3.7}_{-2.8}$ \\
 $ \bar  B^0 \to \rho^0 \eta $                        & $0.10^{+0.02+0.04}_{-0.01-0.03}$
                                    & $0.13^{+0.13}_{-0.06}$
                                    & $0.04_{-0.01-0.00}^{+0.20+0.00}$
                                    & $0.14_{-0.13-0.01}^{+0.33+0.01}$
                                    & $<1.5$ \\
 $ \bar  B^0 \to \rho^0 \eta'$      &
                                    $0.09^{+0.10+0.07}_{-0.04-0.03}$
                                    & $0.10^{+0.05}_{-0.05}$
                                    & $0.43_{-0.12-0.05}^{+2.51+0.05}$
                                    & $1.0_{-0.9-0.1}^{+3.5+0.1}$
                                    & $<1.3$ \\
 $ \bar  B^0\to \omega \eta $       &
                                    $0.85^{+0.65+0.40}_{-0.26-0.24}$
                                    & $0.71^{+0.37}_{-0.28}$
                                    & $0.91_{-0.49-0.09}^{+0.66+0.09}$
                                    & $1.4_{-0.6-0.1}^{+0.8+0.1}$
                                    & $0.94^{+0.36}_{-0.31}$ \\
 $ \bar  B^0 \to \omega \eta'$      &
                                    $0.59^{+0.50+0.33}_{-0.20-0.18}$   &
                                    $0.55^{+0.31}_{-0.26}$
                                    & $0.18_{-0.10-0.03}^{+1.31+0.04}$
                                    & $3.1_{-2.6-0.3}^{+4.9+0.3}$
                                    & $1.01^{+0.47}_{-0.39}$  \\
 $ \bar  B^0 \to \phi \eta$                          & $\approx 0.005$ \footnotemark[2]
                                    & $0.011^{+0.062}_{-0.009}$
                                    & $\approx0.0004$
                                    & $\approx0.0008$
                                    & $<0.5$  \\
 $ \bar  B^0 \to  \phi \eta'$                       & $\approx 0.004$
                                    &
                                    $0.017^{+0.161}_{-0.010}$
                                    & $\approx0.0001$
                                    & $\approx0.0007$  & $<0.5$  \\
\end{tabular}
\footnotetext[1]{from the preliminary Belle measurement \cite{Belle:KstK}.}
\footnotetext[2]{due to the $\omega-\phi$ mixing effect.}
\end{ruledtabular}
\end{table}


\begin{table}[tb]
\begin{ruledtabular}
\caption{Branching fractions (in units of $10^{-6}$) of $B\to VP$
decays induced by the $b\to s$ ($\Delta S =1$) transition. We also cite the average of the experimental data \cite{HFAG,PDG}
and theoretical results given in
pQCD~\cite{LiVP,ChenKeta} and in SCET \cite{SCETVP}.  } \label{tab:VPBr}
\begin{tabular}{l c c c c c c}
       Mode & QCDF (this work)  &  pQCD & SCET 1 &  SCET 2 &   Expt. \\ \hline
  $B^-\to  K^{*-}\pi^0$                          & $6.7^{+0.7+2.4}_{-0.7-2.2}$
                                    & $4.3^{+5.0}_{-2.2}$
                                    & $4.2_{-1.7-0.7}^{+2.2+0.8}$
                                    & $6.5_{-1.7-0.7}^{+1.9+0.7}$ &  $6.9\pm2.3$  \\
  $B^-\to \bar K^{*0}\pi^- $                      & $10.4^{+1.3+4.3}_{-1.5-3.9}$
                                    & $6.0^{+2.8}_{-1.5}$
                                    & $8.5_{-3.6-1.4}^{+4.7+1.7}$
                                    & $9.9_{-3.0-1.1}^{+3.5+1.3}$
                                    & $9.9^{+0.8}_{-0.9}$ \\
  $\bar B^0\to  \bar K^{*0}\pi^0$          & $3.5^{+0.4+1.6}_{-0.4-1.4}$
                                    & $2.0^{+1.2}_{-0.6}$
                                    & $4.6_{-1.8-0.7}^{+2.3+0.9}$
                                    & $3.7_{-1.2-0.5}^{+1.4+0.5}$
                                    & $2.4\pm0.7$ \\
  $\bar B^0\to  K^{*-}\pi^+$                  & $9.2^{+1.0+3.7}_{-1.0-3.3}$
                                    & $6.0^{+6.8}_{-2.6}$
                                    & $8.4_{-3.4-1.3}^{+4.4+1.6}$
                                    & $9.5_{-2.8-1.1}^{+3.2+1.2}$ & $8.6^{+0.9}_{-1.0}$   \\
  $B^-\to \rho^0  K^{-}$                 & $3.5^{+2.9+2.9}_{-1.2-1.8}$
                                    & $5.1^{+4.1}_{-2.8}$
                                    & $6.7_{-2.2-0.9}^{+2.7+1.0}$
                                    & $4.6_{-1.5-0.6}^{+1.8+0.7}$   & $3.81^{+0.48}_{-0.46}$ \\
  $B^-\to \rho^-  \bar K^{0}$             & $7.8^{+6.3+7.3}_{-2.9-4.4}$
                                    & $8.7^{+6.8}_{-4.4}$
                                    & $9.3_{-3.7-1.4}^{+4.7+1.7}$
                                    & $10.1_{-3.3-1.3}^{+4.0+1.5}$
                                    & $8.0^{+1.5}_{-1.4}$   \\
  $\bar B^0\to  \rho^0\bar K^{0}$          & $5.4^{+3.4+4.3}_{-1.7-2.8}$
                                    & $4.8^{+4.3}_{-2.3}$
                                    & $3.5_{-1.5-0.6}^{+2.0+0.7}$
                                    & $5.8_{-1.8-0.7}^{+2.1+0.8}$
                                    & $4.7\pm0.7$\\
  $\bar B^0\to  \rho^+K^{-}$               & $8.6^{+5.7+7.4}_{-2.8-4.5}$
                                    & $8.8^{+6.8}_{-4.5}$
                                    & $9.8_{-3.7-1.4}^{+4.6+1.7}$
                                    & $10.2_{-3.2-1.2}^{+3.8+1.5}$   & $8.6^{+0.9}_{-1.1}$ \\
  $B^-\to \omega K^{-}$                            & $4.8^{+4.4+3.5}_{-1.9-2.3}$
                                    & $10.6^{+10.4}_{-~5.8}$
                                    & $5.1_{-1.9-0.8}^{+2.4+0.9}$
                                    & $5.9_{-1.7-0.7}^{+2.1+0.8}$
                                    & $6.7\pm0.5$  \\
  $\bar B^0\to \omega\bar K^{0}$                   & $4.1^{+4.2+3.3}_{-1.7-2.2}$
                                    & $9.8^{+8.6}_{-4.9}$
                                    & $4.1_{-1.7-0.7}^{+2.1+0.8}$
                                    & $4.9_{-1.6-0.6}^{+1.9+0.7}$
                                    & $5.0\pm0.6$  \\
  $B^-\to \phi K^{-}$                            & $8.8^{+2.8+4.7}_{-2.7-3.6}$
                                    & $7.8^{+5.9}_{-1.8}$
                                    & $9.7_{-3.9-1.5}^{+4.9+1.8}$
                                    & $8.6_{-2.7-1.0}^{+3.2+1.2}$    & $8.30\pm0.65$  \\

  $\bar B^0\to  \phi \bar K^{0}$          & $8.1^{+2.6+4.4}_{-2.5-3.3}$
                                    & $7.3^{+5.4}_{-1.6}$
                                    & $9.1_{-3.6-1.4}^{+4.6+1.7}$
                                    & $8.0_{-2.5-1.0}^{+3.0+1.1}$
                                    & $8.3^{+1.2}_{-1.0}$  \\ \hline
  $B^-\to K^{*-}\eta $                           & $15.7^{+8.5+9.4}_{-4.3-7.1}$
                                    & $22.13^{+0.26}_{-0.27}$
                                    & $17.9_{-5.4-2.9}^{+5.5+3.5}$
                                    & $18.6_{-4.8-2.2}^{+4.5+2.5}$      & $19.3\pm1.6$  \\
  $B^-\to K^{*-}\eta' $                    & $1.7^{+2.7+4.1}_{-0.4-1.6}$
                                    & $6.38\pm0.26$
                                    & $4.5_{-3.9-0.8}^{+6.6+0.9}$
                                    & $4.8_{-3.7-0.6}^{+5.3+0.8}$
                                    & $4.9^{+2.1}_{-1.9}$ \footnotemark[1]\\
  $\bar B^0\to \bar K^{*0}\eta$                   & $15.6^{+7.9+9.4}_{-4.1-7.1}$
                                    & $22.31^{+0.28}_{-0.29}$
                                    & $16.6_{-5.0-2.7}^{+5.1+3.2}$
                                    & $16.5_{-4.3-2.0}^{+4.1+2.3}$
                                    & $15.9\pm1.0$ \\
  $\bar B^0\to \bar K^{*0}\eta'$                 & $1.5^{+2.4+3.9}_{-0.4-1.7}$
                                    & $3.35^{+0.29}_{-0.27}$
                                    & $4.1_{-3.6-0.7}^{+6.2+0.9}$
                                    & $4.0_{-3.4-0.6}^{+4.7+0.7}$
                                    & $3.8\pm1.2$ \footnotemark[2]  \\
\end{tabular}
\footnotetext[1]{This is from the BaBar data \cite{BaBar:Ksteta'}. Belle obtained an upper limit  $2.9\times 10^{-6}$ \cite{Belle:Ksteta'}.}
\footnotetext[2]{This is from the BaBar data \cite{BaBar:Ksteta'}. Belle obtained an upper limit  $2.6\times 10^{-6}$ \cite{Belle:Ksteta'}.}
\end{ruledtabular}
\end{table}

\vskip 0.2in \noindent{\it \underline{$B\to K^*\bar K, K\bar K^*$}} \vskip 0.1in
The decays $B^-\to K^{*-}K^0, K^{*0}K^-$ and $\bar B^0\to \bar K^{*0} K^0, K^{*0}\bar K^0$ are governed by $b\to d$ penguin contributions and $\bar B^0\to K^{*+}K^-, K^{*-}K^+$  proceed only through weak annihilation. Hence, the last two modes are suppressed relative to the first four decays by one order of magnitude. The recent preliminary measurement by Belle \cite{Belle:KstK}, $\B(B^-\to K^{*0}K^-)=(0.68\pm0.16\pm0.10)\times 10^{-6}$, is in agreement with the QCDF prediction (see Table \ref{tab:VPtreeBr}).

\vskip 0.2in \noindent{\it \underline{$B\to K^*\pi,\rho K$}} \vskip 0.1in
The relevant decay amplitudes are
\be
A(\bar B\to \rho \bar K)&=& A_{\rho K}(a_4^c-r_\chi^K a_6^c+\beta_3^c+\cdots), \non \\
A(\bar B\to \pi \bar K^*)&=& A_{\pi K^*}(a_4^c+r_\chi^{K^*} a_6^c+\beta_3^c+\cdots).
\en
Since the chiral factor $r_\chi^K$ is of order unity and $r_\chi^{K^*}$ is small, it turns out numerically $\alpha_4^c(\rho K)\sim -\alpha_4^c(\pi K^*)$. Fortunately, $\beta_3^c(\rho K)$ and $\beta_3^c(\pi K^*)$ are also of opposite sign so that penguin annihilation will contribute constructively. As noted before, in order to accommodate the data,
penguin annihilation should enhance the rates  by $(15\sim 100)\%$ for $\rho K$ modes and by a factor of $2\sim 3$ for $K^*\pi$ ones. 
A fit to the $K^*\pi$ and $K\rho$ data including \CP asymmetries yields $\rho_A(VP)\approx 1.07$, $\phi_A(VP)\approx -70^\circ$, $\rho_A(PV)\approx 0.87$ and $\phi_A(PV)\approx -30^\circ$ as shown in Table \ref{tab:rhoA}.

The pQCD predictions  are too small for the branching fractions of $\bar K^{*0}\pi^-$ and $K^{*-}\pi^+$, and too large for $\omega K^-$ and $\omega\bar K^0$.

\vskip 0.2in \noindent{\it \underline{$B\to \phi K$}} \vskip 0.1in
A direct use of the parameter set $\rho_A(PV)\approx 0.87$ and $\phi_A(PV)\approx -30^\circ$ gives $\B(B^-\to K^-\phi)\approx 13\times 10^{-6}$ which is too large compared to the measured value $(8.30\pm0.65)\times 10^{-6}$ \cite{HFAG}. This means that penguin-annihilation effects should be smaller for the $\phi K$ case. The values of $\rho_A(K\phi)$ and $\phi_A(K\phi)$ are shown in Table \ref{tab:rhoA}. It is interesting to notice that a smaller $\rho_A$ for the $\phi$ meson production also occurs again in $VV$ decays.

\vskip 0.2in \noindent{\it \underline{$B\to K^*\eta^{(')}$}} \vskip 0.1in
In the $PP$ sector we learn that  $\Gamma(B\to K\eta')\gg\Gamma(B\to K\eta)$. It is the other way around in the $VP$ sector, namely, $\Gamma(B\to K^*\eta)\gg\Gamma(B\to K^*\eta')$. This is due to an additional sign difference between $\alpha_4(\eta_q K^*)$ and $\alpha_4(K^*\eta_s)$ as discussed before.

The QCDF prediction for the branching fraction of $B\to K^*\eta'$, of order $1.5\times 10^{-6}$, is smaller compared to pQCD and SCET. The experimental averages quoted in Table \ref{tab:VPBr} are dominated by the BaBar data \cite{BaBar:Ksteta'}. Belle obtained only the upper bounds \cite{Belle:Ksteta'}: $\B(B^-\to K^{*-}\eta')<2.9\times 10^{-6}$ and $\B(B^-\to \bar K^{*0}\eta')<2.6\times 10^{-6}$. Therefore, although our predictions are smaller compared to BaBar, they are consistent with Belle. It will be of importance to measure them to discriminate between various model predictions.

\begin{table}[tb]
\begin{ruledtabular}
\caption{Same as Table \ref{tab:VPtreeBr} except for direct \CP asymmetries involving $b\to d$ ($\Delta S =0$)
transitions. }\label{tab:VPtreeCP}
\begin{tabular} { l|c c c c c c}
     Mode & QCDF (this work)   &pQCD& SCET 1 & SCET 2 &   Expt.  \\\hline
 $ B^- \to \rho^- \pi^0$                          & $9.7^{+2.1+~8.0}_{-3.1-10.3}$
                                    & $0\sim 20$
                                    & $15.5_{-18.9-1.4}^{+16.9+1.6}$
                                    & $12.3_{-10.0-1.1}^{+~9.4+0.9}$
                                    & $2\pm11$  \\
 $ B^- \to \rho^0 \pi^-$             & $-9.8^{+3.4+11.4}_{-2.6-10.2}$
                                    & $-20\sim 0$
                                    & $-10.8_{-12.7-0.7}^{+13.1+0.9}$
                                    & $-19.2_{-13.4-1.9}^{+15.5+1.7}$
                                    & $18^{+~9}_{-17}$ \\
 $ \bar B^0 \to \rho^+  \pi^-$                & $-22.7^{+0.9+8.2}_{-1.1-4.4}$
                                    &
                                    & $-9.9_{-16.7-0.7}^{+17.2+0.9}$
                                    & $-12.4_{-15.3-1.2}^{+17.6+1.1}$
                                    & $-18\pm12$  \\
 $ \bar B^0 \to \rho^-  \pi^+ $               & $4.4^{+0.3+5.8}_{-0.3-6.8}$
                                    &
                                    & $11.8_{-20.0-1.1}^{+17.5+1.2}$
                                    & $10.8_{-10.2-1.0}^{+~9.4+0.9}$
                                    & $11\pm6$  \\
 $ \bar  B^0 \to \rho^0 \pi^0$              & $11.0^{+5.0+23.5}_{-5.7-28.8}$
                                    & $-75\sim 0$
                                    & $-0.6_{-21.9-0.1}^{+21.4+0.1}$
                                    & $-3.5_{-20.3-0.3}^{+21.4+0.3}$
                                    & $-30\pm38$   \\
 $ B^- \to \omega \pi^- $                   & $-13.2^{+3.2+12.0}_{-2.1-10.7}$
                                    & $\sim0$
                                    & $0.5_{-19.6-0.0}^{+19.1+0.1}$
                                    & $2.3_{-13.2-0.2}^{+13.4+0.2}$
                                    & $-4\pm6$   \\
 $\bar  B^0 \to \omega \pi^0$       & $-17.0^{+55.4+98.6}_{-22.8-82.3}$
                                    & $-20\sim 75$
                                    & $-9.4_{-0.0-0.9}^{+24.0+1.1}$
                                    & $39.5_{-185.5-3.1}^{+~79.1+3.4}$
                                    &  \\
 $ B^- \to  K^{*0} K^- $            & $-8.9^{+1.1+2.8}_{-1.1-2.4}$
                                    & $-6.9^{+5.6+1.0+9.2+4.0}_{-5.3-0.3-6.5-6.0}$
                                    & $-3.6_{-5.3-0.4}^{+6.1+0.4}$
                                    & $-4.4_{-4.1-0.2}^{+4.1+0.2}$
                                    &   \\
 $ B^- \to  K^{*-} K^0$                                 & $-7.8^{+5.9+~4.1}_{-4.1-10.0}$
                                    & $6.5^{+7.9+1.1+9.1+2.1}_{-7.3-1.4-7.7-3.9}$
                                    & $-1.5_{-2.3-0.1}^{+2.6+0.1}$
                                    & $-1.2_{-1.7-0.1}^{+1.7+0.1}$
                                    &  \\
 $\bar B^0\to K^{*+}K^-$             &
                                    $-4.7^{+0.1+4.7}_{-0.2-2.7}$    \\
 $\bar B^0\to K^{*-}K^+$             &
                                    $5.5^{+0.2+7.0}_{-0.2-5.5}$
                                    \\
 $\bar   B^0 \to K^{*0} \bar K^0$                       & $-13.5^{+1.6+1.4}_{-1.7-2.3}$&
                                    & $-3.6_{-5.3-0.4}^{+6.1+0.4}$
                                    & $-4.4_{-4.1-0.2}^{+4.1+0.2}$
                                    &   \\
 $\bar   B^0 \to \bar K^{*0} K^0$                        & $-3.5^{+1.3+0.7}_{-1.7+2.0}$ &
                                    & $-1.5_{-2.3-0.1}^{+2.6+0.1}$
                                    & $-1.2_{-1.7-0.1}^{+1.7+0.1}$
                                    &  \\
 $ B^- \to \phi \pi^- $             &
                                    0 & $-8.0^{+0.9+1.5}_{-1.0-0.1}$                     \\
 $ \bar B^0 \to \phi \pi^0 $        &
                                    0
                                    & $-6.3^{+0.7+2.5}_{-0.5-2.5}$ \\
                                    \hline
 $ B^- \to \rho^- \eta $                       & $-8.5^{+0.4+6.5}_{-0.4-5.3}$
                                    & $1.9^{+0.1+0.2+0.1+0.6}_{-0.0-0.3-0.0-0.5}$
                                    & $-6.6_{-21.3-0.7}^{+21.5+0.6}$
                                    & $-9.1_{-15.8-0.8}^{+16.7+0.9}$
                                    & $11\pm11$ \\
 $ B^- \to \rho^- \eta ^\prime$              & $1.4^{+0.8+14.0}_{-2.2-11.7}$
                                    & $-25.0^{+0.4+4.1+0.8+2.1}_{-0.3-1.6-0.7-1.8}$
                                    & $-19.8_{-37.5-3.1}^{+66.5+2.8}$
                                    & $-21.7_{-~24.3-1.7}^{+135.9+2.1}$
                                    & $4\pm28$ \\
 $ \bar  B^0 \to \rho^0 \eta $      & $86.2^{+3.7+10.4}_{-5.8-21.4}$
                                    & $-89.6^{+1.9+13.7+0.7+4.6}_{-0.9-~3.9-0.1-9.0}$
                                    & $-46.7_{-~74.3-3.7}^{+170.4+2.9}$
                                    & $33.3_{-62.4-2.8}^{+66.9+3.1}$
                                    &  \\
 $ \bar  B^0 \to \rho^0 \eta'$      & $53.5^{+4.5+39.5}_{-7.9-57.6}$
                                    & $-75.7^{+5.6+13.1+6.3+12.9}_{-4.8-~7.0-4.0-~9.9}$
                                    & $-51.7_{-~42.9-3.9}^{+103.3+3.4}$
                                    & $52.2_{-80.6-4.1}^{+19.9+4.4}$
                                    &  \\
 $ \bar  B^0\to \omega \eta $                            & $-44.7^{+13.1+17.7}_{-~9.9-11.6}$
                                    & $33.5^{+1.0+0.8+5.9+3.9}_{-1.4-4.6-6.8-4.4}$
                                    & $-9.4_{-30.2-1.0}^{+30.7+0.9}$
                                    & $-9.6_{-16.8-0.9}^{+17.8+0.9}$
                                    &   \\
 $ \bar  B^0 \to \omega \eta'$                           & $-41.4^{+2.5+19.5}_{-2.4-14.4}$
                                    & $16.0^{+0.1+3.3+2.2+1.7}_{-0.9-3.9-3.2-2.0}$
                                    & $-43.0_{-38.8-5.1}^{+87.5+4.8}$
                                    & $-27.2_{-29.7-2.2}^{+18.1+2.4}$
                                    &  \\
 $ \bar  B^0 \to \phi \eta$         & 0     &0                 & \\
 $ \bar  B^0 \to  \phi \eta'$       & 0      & 0                & \\
\end{tabular}\end{ruledtabular}
\end{table}

\begin{table}[tb]
\begin{ruledtabular}
\caption{Same as Table \ref{tab:VPBr} except for direct \CP asymmetries (in $\%$) involving $\Delta S=1$ processes.
 } \label{tab:VPCP}
\begin{tabular}{ l c c c c c c}
          Mode& QCDF (this work)   &  pQCD & SCET 1 &  SCET 2 &   Expt.  \\ \hline
  $B^-\to  K^{*-}\pi^0$                               & $1.6^{+3.1+11.1}_{-1.7-14.4}$
                                    & $-32^{+21}_{-28}$
                                    & $-17.8_{-24.6-2.0}^{+30.3+2.2}$
                                    & $-12.9_{-12.2-0.8}^{+12.0+0.8}$ & $4\pm29$ \\
  $B^-\to \bar K^{*0}\pi^- $                      & $0.4^{+1.3+4.3}_{-1.6-3.9}$
                                    & $-1^{+1}_{-0}$
                                    &    $0$           & $0$
                                    & $-3.8\pm4.2$ \\
  $\bar B^0\to  \bar K^{*0}\pi^0$                     & $-10.8^{+1.8+9.1}_{-2.8-6.3}$
                                    & $-11^{+7}_{-5}$
                                    & $5.0_{-8.4-0.5}^{+7.5+0.5}$
                                    & $5.4_{-5.1-0.5}^{+4.8+0.4}$  & $-15\pm12$  \\
  $\bar B^0\to  K^{*-}\pi^+$                    & $-12.1^{+0.5+12.6}_{-0.5-16.0}$
                                    & $-60^{+32}_{-19}$
                                    & $-11.2_{-16.2-1.3}^{+19.0+1.3}$
                                    & $-12.2_{-11.3-0.8}^{+11.4+0.8}$     & $-23\pm8$   \\
  $B^-\to \rho^0  K^{-}$                      & $45.4^{+17.8+31.4}_{-19.4-23.2}$
                                    & $71^{+25}_{-35}$
                                    & $9.2_{-16.1-0.7}^{+15.2+0.7}$
                                    & $16.0_{-22.4-1.6}^{+20.5+1.3}$     & $37\pm11$  \\
  $B^-\to \rho^-  \bar K^{0}$                       & $0.3^{+0.2+0.5}_{-0.3-0.2}$
                                    & $1\pm1$
                                    &    $0$             &  $0$
                                    & $-12\pm17$ \\
  $\bar B^0\to  \rho^0\bar K^{0}$            & $8.7^{+1.2+8.7}_{-1.2-6.8}$
                                    & $7^{+8}_{-5}$
                                    &   $-6.6_{-9.7-0.9}^{+11.6+0.8}$
                                    & $-3.5_{-4.8-0.2}^{+4.8+0.3}$
                                    & $6\pm20$ \\
  $\bar B^0\to  \rho^+K^{-}$                      & $31.9^{+11.5+19.6}_{-11.0-12.7}$
                                    & $64^{+24}_{-30}$
                                    & $7.1_{-12.4-0.7}^{+11.2+0.7}$
                                    &  $9.6_{-13.5-0.9}^{+13.0+0.7}$  & $15\pm6$    \\
  $B^-\to \omega K^{-}$                                 & $22.1^{+13.7+14.0}_{-12.8-13.0}$
                                    & $32^{+15}_{-17}$
                                    & $11.6_{-20.4-1.1}^{+18.2+1.1}$
                                    & $12.3_{-17.3-1.1}^{+16.6+0.8}$ & $2\pm5$ \\
  $\bar B^0\to \omega\bar K^{0}$                         & $-4.7^{+1.8+5.5}_{-1.6-5.8}$
                                    & $-3^{+2}_{-4}$
                                    & $5.2_{-9.2-0.6}^{+8.0+0.6}$
                                    & $3.8_{-5.4-0.3}^{+5.2+0.3}$
                                    & $32\pm17$  \footnotemark[1] \\
  $B^-\to \phi K^{-}$                             & $0.6^{+0.1+0.1}_{-0.1-0.1}$
                                    & $1^{+0}_{-1}$              &   $0$            &  $0$      & $-1\pm6$  \\
  $\bar B^0\to  \phi \bar K^{0}$                     & $0.9^{+0.2+0.2}_{-0.1-0.1}$
                                    & $3^{+1}_{-2}$
                                    &  $0$              &$0$
                                    & $23\pm15$  \\\hline
  $B^-\to K^{*-}\eta $                                 & $-9.7^{+3.9+6.2}_{-3.7-7.1}$
                                    & $-24.57^{+0.72}_{-0.27}$
                                    & $-2.6_{-5.5-0.3}^{+5.4+0.3}$
                                    & $-1.9_{-3.6-0.1}^{+3.4+0.1}$    & $2\pm6$  \\
  $B^-\to K^{*-}\eta' $                       & $65.5^{+10.1+34.2}_{-39.5-50.2}$
                                    & $4.60^{+1.16}_{-1.32}$
                                    & $2.7_{-19.5-0.3}^{+27.4+0.4}$
                                    & $2.6_{-32.9-0.2}^{+26.7+0.2}$
                                    & $-30^{+37}_{-33}$ \\
  $\bar B^0\to \bar K^{*0}\eta$                       & $3.5^{+0.4+2.7}_{-0.5-2.4}$
                                    & $0.57\pm0.011$
                                    & $-1.1_{-2.4-0.1}^{+2.3+0.1}$
                                    & $-0.7_{-1.3-0.0}^{+1.2+0.1}$
                                    & $19\pm5$ \\
  $\bar B^0\to \bar K^{*0}\eta'$                 & $6.8^{+10.7+33.2}_{-~9.2-50.2}$
                                    & $-1.30\pm0.08$
                                    & $9.6_{-11.0-1.2}^{+~8.9+1.3}$
                                    & $9.9_{-4.3-0.9}^{+6.2+0.9}$ &  $8\pm25$    \\
\end{tabular}
\footnotetext[1]{Note that the measurements of $52^{+22}_{-20}\pm3$ by BaBar \cite{BaBar:omegaK} and $-9\pm29\pm6$ by Belle \cite{Belle:omegaK} are of opposite sign.}
\end{ruledtabular}
\end{table}

\subsection{Direct \CP asymmetries}
\vskip 0.1in \noindent{\it \underline{$\acp(K^*\pi)$ and $\acp(K\rho)$}} \vskip 0.1in

First of all, \CP violation for $\bar K^{*0}\pi^-$ and $\rho^-\bar K^0$ is expected to be very small as they are pure penguin processes (apart from a $W$-annihilation contribution). From Table \ref{tab:VP} we see that \CP asymmetries for $\rho^0 K^-$, $\rho^+K^-$ and $K^{*-}\pi^+$ predicted in the heavy quark limit are all wrong in signs when confronted with experiment. For the last two modes, \CP asymmetries are governed by the quantity $r_{\rm FM}$ defined in Eq. (\ref{eq:rFM}) except that $PP$ is replaced by $VP$ or $PV$. Since $\hat\alpha_4^c(\rho K)$ and $\hat\alpha_4^c(\pi K^*)$ are of opposite sign, this means that $\acp(\rho^+ K^-)$ and $\acp(K^{*-}\pi^+)$ should have different signs. This is indeed borne out by experiment (see Table \ref{tab:VPCP}).
Numerically, we have  $\alpha_4^c(\rho K)=0.041+0.001i$, $\hat\alpha_4^c(\rho K)=\alpha_4^c(\rho K)+\beta_3^c(\rho K)=0.045-0.046i$, $\alpha_4^c(\pi K^*)=-0.034+0.009i$ and $\hat\alpha_4^c(\pi K^*)=-0.066+0.013i$.
Therefore, one needs the $\beta_3^c$ terms (i.e. penguin annihilation) to get correct signs for \CP violation of above-mentioned three modes.
One can check from Eqs. (\ref{eq:acpKpi}) and (\ref{eq:rFM}) that $\acp(\rho^+ K^-)$ is positive, while $\acp(K^{*-}\pi^+)$ is negative.

In order to see the effects of soft corrections to $a_2$, we consider the following quantities
\be \label{eq:acpKstpi}
\Delta A_{K^*\pi} &\equiv &\acp(K^{*-}\pi^0)-\acp(K^{*-}\pi^+)=0.036^{+0.002+0.035}_{-0.003-0.045}-2\sin\gamma\, {\rm Im}\,r_C(K^*\pi)+\cdots, \non \\
\Delta A'_{K^*\pi}&\equiv& \acp(\bar K^{*0}\pi^0)-\acp(\bar K^{*0}\pi^-)=(-0.23^{+0.01+0.01}_{-0.01-0.04})\%+
2\sin\gamma\, {\rm Im}\,r_C(K^*\pi)+\cdots,
\en
defined in analog to $\Delta A_{K\pi}$ and $\Delta A'_{K\pi}$ with
\be
r_C(K^*\pi) &=& \left|{\lambda_u^{(s)}\over \lambda_c^{(s)}}\right|\,{f_\pi A_0^{BK^*}(0)\over f_{K^*} F_1^{B\pi}(0)}{\alpha_2(K^*\pi)\over -\alpha_4^c(\pi\bar K^*)-\beta_3^c(\pi\bar K^*)}.
\en
The first terms on the r.h.s. of  Eq. (\ref{eq:acpKstpi}) come from the interference between QCD and electroweak penguins. We will not consider similar quantities for $K\rho$ modes as the first term there will become large. In other words, as far as \CP violation is concerned, $K^*\pi$ mimics $K\pi$ more than $K\rho$.
We obtain Im\,$r_C(K^*\pi)=-0.057$ and Im\,$r_C(K\rho)=0.023$ and
predict that $\Delta A_{K^*\pi}=(13.7^{+2.9+3.6}_{-1.4-6.9})\%$
and $\Delta A'_{K^*\pi}=(-11.1^{+1.7+9.1}_{-2.8-6.3})\%$, while it is naively expected that $K^{*-}\pi^0$ and $K^{*-}\pi^+$ have similar {\it CP}-violating effects. It will be very important to measure \CP asymmetries of these two modes to test our prediction. It is clear from Eqs. (\ref{eq:acpKstpi}) and (\ref{eq:acpKpi}) (see also Tables \ref{tab:PP} and \ref{tab:VP}) that \CP asymmetries of $\bar K^{*0}\pi^0$ and $\bar K^0\pi^0$ are of order $-0.10$ and arise dominantly from soft corrections to $a_2$. As for $\acp(\bar K^0\rho^0)$, it is predicted to be $\approx 0.09$ ($\approx -0.05$) with (without) soft corrections to $a_2$ (cf. Table \ref{tab:VP}).

Power corrections to the color-suppressed tree amplitude is needed to improve the prediction for $\acp(\bar K^{*0}\eta)$. The current experimental measurement $\acp(\bar K^{*0}\eta)=0.19\pm0.05$ is in better agreement with QCDF than pQCD and SCET.

In the pQCD approach, the predictions for some of the $VP$ modes, e.g. $\acp(K^{*-}\pi^+)$, $\acp(\rho^0K^-)$ and $\acp(\rho^+ K^-)$ are very large, above 50\%. This is because QCD penguin contributions in these modes are small, and direct \CP violation arises from the interference between tree and annihilation diagrams. The strong phase comes mainly from the annihilation diagram in this approach. On the other hand, the predicted $\acp(\bar K^{*0}\eta)$ is too small. So far the pQCD results for $\acp(K^*\eta^{(')})$ are quoted from \cite{ChenKeta} where $m_{qq}=0.22$ GeV is used. Since the pQCD study of $B\to K\eta^{(')}$ has been carried to the (partial) NLO and a drastic different prediction for $\acp(K^-\eta)$ has been found, it will be crucial to generalize the NLO calculation to the $K^*\eta^{(')}$ sector.

We would like to point out the \CP violation of $\bar B^0\to \omega \bar K^0$. It is clear from Table \ref{tab:VP} that power correction on $a_2$ will flip the sign of  $\acp(\omega\bar K^0)$ to a negative one. The pQCD estimate is similar to the QCDF one . At first sight, it seems that QCDF and pQCD predictions are ruled out by the data $\acp(\omega\bar K^0)=0.032\pm0.017$. However, the BaBar and Belle measurements $0.52^{+0.22}_{-0.20}\pm0.03$ \cite{BaBar:omegaK} and $-0.09\pm0.29\pm0.06$ \cite{Belle:omegaK}, respectively, are opposite in sign. Hence, we need to await more accurate experimental studies to test theory predictions.

As for the approach of SCET, the predicted \CP asymmetries for the neutral modes $\bar K^{*0}\pi^0,\rho^0\bar K^0,\omega\bar K^0$ and $\bar K^{*0}\eta$ have signs opposite to QCDF and pQCD. Especially, the predicted $\acp(\bar K^{*0}\eta)$ is already ruled out by experiment.

\vskip 0.1in \noindent{\it \underline{$\acp(\rho\pi)$}} \vskip 0.1in

The decay amplitudes of $\bar B^0\to \rho^\pm\pi^\mp$ are given by
\be
A(\bar B^0\to \rho^-\pi^+) &=& A_{\pi\rho} \left[ \delta_{pu}\alpha_1+\alpha_4^p+\beta_3^p+\cdots\right], \non \\
A(\bar B^0\to \rho^+\pi^-) &=& A_{\rho\pi} \left[ \delta_{pu}\alpha_1+\alpha_4^p+\beta_3^p+\cdots\right].
\en
Since the penguin contribution is small compared to the tree one, its \CP asymmetry is approximately given by
\be
\acp(\rho^-\pi^+)\approx 2\sin\gamma\,{\rm Im}\,r_{\pi\rho}, \qquad \acp(\rho^+\pi^-)\approx 2\sin\gamma\,{\rm Im}\,r_{\rho\pi},
\en
with
\be
 r_{\pi\rho}=\left|{\lambda_c^{(d)}\over \lambda_u^{(d)}}\right|\,{\alpha_4^c(\pi \rho)+\beta_3^c(\pi\rho)\over \alpha_1(\pi\rho)}, \quad
 r_{\rho\pi}=\left|{\lambda_c^{(d)}\over \lambda_u^{(d)}}\right|\,{\alpha_4^c(\rho \pi)+\beta_3^c(\rho\pi)\over \alpha_1(\rho\pi)}.
\en
We obtain the values Im$r_{\pi\rho}=0.037$ and Im$r_{\rho\pi}=-0.134$. Therefore, \CP asymmetries for $\rho^+\pi^-$ and $\rho^-\pi^+$ are opposite in signs and the former is much bigger than the latter. We see from Table \ref{tab:VPtreeCP} that the predicted signs for \CP violation of $\rho^+\pi^-$ and $\rho^-\pi^+$ agree with experiment. The $B^-\to\rho^0\pi^-$ decay amplitude reads
\be
A(B^-\to \rho^0\pi^-) = A_{\rho\pi} \left[ \delta_{pu}\alpha_1+\alpha_4^p+\beta_3^p\right]+A_{\pi\rho} \left[ \delta_{pu}\alpha_2-\alpha_4^p-\beta_3^p\right].
\en
As far as the sign is concerned, it suffices to keep terms in the first square bracket on the r.h.s. and obtain a negative $\acp(\rho^0\pi^-)$.
By the same token, $\acp(\rho^-\pi^0)$ is predicted to be positive.

\CP violation of $\bar B^0\to\rho^0\pi^0$ is predicted to be of order $0.11$ by QCDF and  negative by pQCD and SCET. The current data are $0.10\pm0.40\pm0.53$ by BaBar \cite{BaBar:rho0pi0} and $-0.49\pm0.36\pm0.28$ by Belle \cite{Belle:rho0pi0}.
It is of interest to notice that QCDF and pQCD predictions for \CP asymmetries of $B\to (\rho,\omega)\eta^{(')}$ are opposite in signs.

\begin{table}[!tbp]
\begin{ruledtabular}
\caption{Mixing-induced \CP\ violation $S_f$ in $\bar B\to VP$ decays predicted in various approaches.
 The pQCD results are taken from \cite{LiVP,XiaoVeta}.  There are
two solutions with SCET predictions \cite{SCETVP}. The parameter $\eta_f=1$ except for $(\phi,\rho,\omega)K_S$ modes where $\eta_f=-1$. Experimental results from BaBar (first entry) and Belle (second entry) are listed whenever available. The input values of $\sin 2\beta$ used at the time of theoretical calculations which are needed for the calculation of $\Delta S_f$ are displayed. }\label{tab:SVP}
\begin{tabular}{l| c c c | c  c}
Decay
       &QCDF (this work) 
       &pQCD 
       &SCET 
       &Expt. \cite{BaBar:phiK0,Belle:etaK0,BaBar:rhoK0,Belle:rhoK0,BaBar:etaK0,Belle:omegaK0,BaBar:rhopi,Belle:rhopi}
       & Average  
       \\
       \hline
 $\sin2\beta$ & 0.670 & 0.687 & 0.687 & \\     \hline
 $\phi K_S$
       & $0.692^{+0.003+0.002}_{-0.000-0.002}$
       & $0.71\pm0.01$
       & $\begin{array}{c} 0.69\\0.69\end{array}$
       & $\begin{array}{c} 0.26\pm0.26\pm0.03\\0.67^{+0.22}_{-0.32}\end{array}$
       & $0.44^{+0.17}_{-0.18}$
       \\
 $\omega K_S$
       & $0.84^{+0.05+0.04}_{-0.05-0.06}$
       & $0.84^{+0.03}_{-0.07}$
       & $\begin{array}{c} 0.51^{+0.05+0.02}_{-0.06-0.02}\\0.80^{+0.02+0.01}_{-0.02-0.01}\end{array}$
       & $\begin{array}{c} 0.55^{+0.26}_{-0.29}\pm0.02\\0.11\pm0.46\pm0.07\end{array}$
       & $0.45\pm{0.24}$
       \\
 $\rho^0K_S$
       & $0.50^{+0.07+0.06}_{-0.14-0.12}$
       & $0.50^{+0.10}_{-0.06}$
       & $\begin{array}{c}0.85^{+0.04+0.01}_{-0.05-0.01}\\ 0.56^{+0.02+0.01}_{-0.03-0.01}\end{array}$
       & $\begin{array}{c}0.35^{+0.26}_{-0.31}\pm0.06\pm0.03\\ 0.64^{+0.19}_{-0.25}\pm0.09\pm0.10\end{array}$
       & $0.54_{-0.21}^{+0.18}$
       \\
    \hline
  $\rho^0\pi^0$ & $-0.24^{+0.15+0.20}_{-0.14-0.22}$
                &
                & $\begin{array}{c} -0.11^{+0.14+0.10}_{-0.14-0.15}\\-0.19^{+0.14+0.10}_{-0.14-0.15}\end{array}$
                & $\begin{array}{c} 0.04\pm0.44\pm0.18\\0.17\pm0.57\pm0.35\end{array}$
                &$0.12\pm0.38$ \\
  $\omega\pi^0$ & $0.78^{+0.14+0.20}_{-0.20-1.39}$
                &
                & $\begin{array}{c} -0.87^{+0.44+0.02}_{-0.00-0.01}\\0.72^{+0.36+0.07}_{-1.54-0.11}\end{array}$                & \\
  $\rho^0\eta$  & $0.51^{+0.08+0.19}_{-0.07-0.32}$
                & $0.23^{+0.30}_{-0.37}$
                & $\begin{array}{c} 0.86^{+0.15+0.03}_{-2.03-0.07}\\0.29^{+0.36+0.09}_{-0.44-0.15}\end{array}$
                & \\
  $\rho^0\eta'$ & $0.80^{+0.04+0.24}_{-0.09-0.43}$
                & $-0.49^{+0.25}_{-0.20}$
                & $\begin{array}{c} 0.79^{+0.20+0.05}_{-1.73-0.09}\\0.38^{+0.22+0.09}_{-1.24-0.14}\end{array}$
                & \\
  $\omega\eta$  & $-0.16^{+0.13+0.17}_{-0.13-0.16}$
                & $0.39^{+0.51}_{-0.66}$
                & $\begin{array}{c} 0.12^{+0.19+0.10}_{-0.20-0.17}\\-0.16^{+0.14+0.10}_{-0.15-0.15}\end{array}$
                & \\
  $\omega\eta'$ & $-0.28^{+0.14+0.16}_{-0.13-0.13}$
                & $0.77^{+0.22}_{-0.53}$
                & $\begin{array}{c} 0.23^{+0.59+0.10}_{-1.10-0.10}\\-0.27^{+0.17+0.09}_{-0.33-0.14}\end{array}$
                & \\
\end{tabular}
\end{ruledtabular}
\end{table}

\begin{table}[!htbp]
\begin{ruledtabular}
\caption{Same as Table \ref{tab:SVP} except for $\Delta S_f$ for penguin-dominated modes. The QCDF results obtained by Beneke \cite{BenekeS} are included for comparison.}\label{tab:DeltaSVP}
\begin{tabular}{l| c c c c c |c c}
 &  \multicolumn{2}{c}{QCDF (this work)} &
    \\ \cline{2-3}
\raisebox{2.0ex}[0cm][0cm]{Decay} & With $\rho_C,\phi_C$ & W/o $\rho_C$ & \raisebox{2.0ex}[0cm][0cm]{QCDF (Beneke)} & \raisebox{2.0ex}[0cm][0cm]{pQCD} & \raisebox{2.0ex}[0cm][0cm]{SCET} & \raisebox{2.0ex}[0cm][0cm]{Expt.} & \raisebox{2.0ex}[0cm][0cm]{Average} \\ \hline
 $\phi K_S$
       & $0.022^{+0.004}_{-0.002}$
       & $0.022^{+0.004}_{-0.002}$
       & $0.02^{+0.01}_{-0.01}$
       & $0.02\pm0.01$
       & $\begin{array}{c}\sim 0\\ \sim 0\end{array}$
       & $\begin{array}{c}-0.43\pm0.26\\ 0.02^{+0.22}_{-0.32}\end{array}$
       & $-0.25^{+0.17}_{-0.18}$
       \\
 $\omega K_S$
       & $0.17^{+0.06}_{-0.08}$
       & $0.13^{+0.06}_{-0.04}$
       & $0.13^{+0.08}_{-0.08}$
       & $0.15^{+0.03}_{-0.07}$
       & $\begin{array}{c}-0.18^{+0.05}_{-0.06}\\ 0.11^{+0.02}_{-0.02}\end{array}$
       & $\begin{array}{c}-0.43\pm0.26\\ 0.02^{+0.22}_{-0.32}\end{array}$
       & $-0.14^{+0.26}_{-0.29}$
       \\
 $\rho^0K_S$
       & $-0.17^{+0.09}_{-0.18}$
       & $-0.11^{+0.07}_{-0.11}$
       & $-0.08^{+0.08}_{-0.12}$
       & $-0.19^{+0.10}_{-0.06}$
       & $\begin{array}{c}0.16^{+0.04}_{-0.05}\\ -0.13^{+0.02}_{-0.03}\end{array}$
       & $\begin{array}{c}-0.34^{+0.27}_{-0.31}\\ -0.05^{+0.23}_{-0.28}\end{array}$
       & $-0.10\pm0.17$
       \\
\end{tabular}
\end{ruledtabular}
\end{table}

\subsection{Mixing-induced \CP asymmetries}
Mixing-induced \CP asymmetries $S_f$ and $\Delta S_f$ of $B\to VP$ in various approaches are listed in Tables \ref{tab:SVP} and \ref{tab:DeltaSVP}, respectively. Just as the $\eta'K_S$ mode, $\phi K_S$ is also theoretically very clean as it is a pure penguin process.
Although the prediction of $S_{\phi K_S}\sim 0.69$ has some deviation from the world average of $0.44^{+0.17}_{-0.18}$, it does agree with one of the $B$ factory measurements, namely, $0.67^{+0.22}_{-0.32}$ by Belle \cite{Belle:etaK0}. In short, it appears that the theoretical predictions of $S_f$ for several penguin-dominated $B\to PP,VP$ decays deviate from the world averages and hence may indicate some New Physics effects. However, if we look at the individual measurement made by BaBar or Belle, the theory prediction actually agrees with one of the measurements. Hence, in order to uncover New Physics effects through the time evolution of \CP violation, we certainly need more accurate measurements of time-dependent \CP violation and better theoretical estimates of $S_f$.
This poses a great challenge to both theorists and experimentalists.

The ratio of $A^u/A^c$ for the penguin-dominated decays $(\phi,\omega,\rho^0)K_S$ has the expressions \cite{BenekeS}
\be
{A^u\over A^c}\bigg|_{\phi K_S} &\sim& {[-P^u]\over [-P^c]}\sim {[-(a_4^u+r_\chi^\phi a_6^u)]\over [-(a_4^c+r_\chi^\phi a_6^c)]}, \non \\
{A^u\over A^c}\bigg|_{\omega K_S} &\sim& {[P^u]+[C]\over [P^c]}\sim {[(a_4^u-r_\chi^K a_6^u)]+[a_2^u R_{\omega K_S}]\over [(a_4^c-r_\chi^K a_6^c)]}, \non \\
{A^u\over A^c}\bigg|_{\rho^0 K_S} &\sim& {[P^u]-[C]\over [P^c]}\sim {[(a_4^u-r_\chi^K a_6^u)]-[a_2^u R_{\rho K_S}]\over [(a_4^c-r_\chi^K a_6^c)]},
\en
As discussed before, the quantity $(a_4^c-r_\chi^K a_6^c)$ in above equations is positive and has a magnitude similar to $|a_4^c|$. Since $a_2$ is larger than $-a_4^c$, $\Delta S_f$ is positive for $\omega K_S$ but negative for $\rho^0 K_S$ and both have large magnitude due to the small denominator of $\Delta S_f$. From Table \ref{tab:DeltaSVP} we see that $\Delta S_{\omega K_S}={\cal O}(0.17)$, while $\Delta S_{\rho^0 K_S}={\cal O}(-0.17)$. Effects of soft corrections on them are sizable. For example, $\Delta S_{\rho^0 K_S}$ is shifted from $\approx -0.11$ to $\approx -0.17$ in the presence of power corrections. This explains why our prediction of $\Delta S_{\rho^0 K_S}$  is substantially different from the Beneke's estimate \cite{BenekeS} and our previous calculation \cite{CCSsin2beta}.

For tree-dominated decays, so far there is only one measurement, namely, $S_{\rho^0\pi^0}$ with a sign opposite to the theoretical predictions of QCDF and SCET.

\vskip 0.1in \noindent{\it \underline{Time-dependent \CP violation of the $\rho^\pm\pi^\mp$ systems}} \vskip 0.1in

The study of \CP violation for
$\bar B^0\to \rho^+\pi^-$ and $\rho^-\pi^+$ becomes more complicated as $\rho^\pm \pi^\mp$ are not  \CP eigenstates. The time-dependent \CP asymmetries are given by
 \be \label{eq:CPKstK}
 \A(t) &\equiv & {\Gamma(\ov
B^0(t)\to \rho^{\pm}\pi^\mp)-\Gamma(B^0(t)\to \rho^{\pm}\pi^\mp)\over
\Gamma(\ov
B^0(t)\to \rho^{\pm}\pi^\mp)+\Gamma(B^0(t)\to \rho^{\pm}\pi^\mp)} \non \\
 &=& (S\pm \Delta S)\sin(\Delta
 m t)-(C\pm \Delta C)\cos(\Delta m t),
 \en
where $\Delta m$ is the mass difference of the two neutral $B^0$
eigenstates, $S$ is referred to as mixing-induced \CP asymmetry
and $C$ is the direct \CP asymmetry, while $\Delta S$ and $\Delta
C$ are {\it CP}-conserving quantities.  Defining
\begin{eqnarray}
 A_{+-} & \equiv & A(B^0\to \rho^{+}\pi^-)~,~~~
A_{-+}  \equiv A(B^0\to \rho^{-}\pi^+)~,\nonumber\\
\bar{A}_{-+} & \equiv & A(\overline B^0\to \rho^{-}\pi^+)~,~~~
\bar{A}_{+-} \equiv A(\overline B^0 \to \rho^{+}\pi^-),
\end{eqnarray}
and
 \be
 \lambda_{+-}={q_{_{B}}\over p_{_{B}}}\,{\bar A_{+-}\over A_{+-}}, \qquad
 \lambda_{-+}={q_{_{B}}\over p_{_{B}}}\,{\bar A_{-+}\over A_{-+}},
 \en
with $q_{_{B}}/ p_{_{B}}\approx e^{-2i\beta}$,
we have
 \be \label{eq:C}
 C+\Delta C={1-|\lambda_{+-}|^2\over 1+|\lambda_{+-}|^2}=
 {|A_{+-}|^2-|\bar A_{+-}|^2\over |A_{+-}|^2+|\bar A_{+-}|^2}, \quad
 C-\Delta C={1-|\lambda_{-+}|^2\over 1+|\lambda_{-+}|^2}={|A_{-+}|^2-|\bar A_{-+}|^2\over |A_{-+}|^2+|\bar A_{-+}|^2},
 \en
and
 \be
 S+\Delta S\equiv {2\,{\rm Im}\lambda_{+-}\over 1+|\lambda_{+-}|^2}={2\,{\rm Im}(e^{2i\beta}\bar A_{+-}A_{+-}^*)\over
 |A_{+-}|^2+|\bar A_{+-}|^2}, \non \\
  S-\Delta S\equiv {2\,{\rm Im}\lambda_{-+}\over 1+|\lambda_{-+}|^2}={2\,{\rm Im}(e^{2i\beta}\bar A_{-+}A_{-+}^*)\over
 |A_{-+}|^2+|\bar A_{-+}|^2}.
 \en
Hence we see that $\Delta S$ describes the strong phase difference
between the amplitudes contributing to $B^0\to \rho^{\pm}\pi^\mp$ and
$\Delta C$ measures the asymmetry between $\Gamma(B^0\to
\rho^{+}\pi^-)+\Gamma(\ov B^0\to \rho^{-}\pi^+)$ and $\Gamma(B^0\to
\rho^{-}\pi^+)+\Gamma(\ov B^0\to \rho^{+}\pi^-)$.

\begin{table}[tb]
\caption{Various {\it CP}-violating parameters in the decays $\bar B^0\to \rho^{\pm} \pi^\mp$. SCET results are quoted from \cite{SCETVP}. Experimental results are taken from \cite{BaBar:rho0pi0,Belle:rho0pi0} and the world average from \cite{HFAG}.
} \label{tab:mixingBtorhopi}
\begin{ruledtabular}
{\footnotesize
\begin{tabular}{c r r r r}
  Parameter  &   QCDF (this work) &   SCET 1                               & SCET 2 & Expt. \\ \hline
    $\A_{\rho\pi}$   &  $-0.11^{+0.00+0.07}_{-0.00-0.05}$
                                          & $-0.12_{-0.05-0.03}^{+0.04+0.04}$
                                          & $-0.21_{-0.02-0.03}^{+0.03+0.02}$ & $-0.13\pm0.04$
                                          \\
  $C$                                     & $0.09^{+0.00+0.05}_{-0.00-0.07}$
                                          & $-0.01_{-0.12-0.00}^{+0.13+0.00}$
                                          & $0.01_{-0.10-0.00}^{+0.09+0.00}$ &$0.01\pm0.07$
                                          \\
  $S$                                     &
                                          $-0.04^{+0.01+0.10}_{-0.01-0.09}$
                                          & $-0.11_{-0.08-0.13}^{+0.07+0.08}$
                                          & $-0.01_{-0.07-0.14}^{+0.06+0.08}$ & $0.01\pm0.09$
                                          \\
  $\Delta C$                              &
                                          $0.26^{+0.02+0.02}_{-0.02-0.02}$
                                          & $0.11_{-0.13-0.01}^{+0.12+0.01}$
                                          & $0.12_{-0.10-0.01}^{+0.09+0.01}$ &$0.37\pm0.08$
                                          \\
  $\Delta S$                              &
                                          $-0.02^{+0.00+0.03}_{-0.00-0.02}$
                                          & $-0.47_{-0.06-0.04}^{+0.08+0.05}$
                                          & $0.43_{-0.07-0.03}^{+0.05+0.03}$ & $-0.04\pm0.10$
                                          \\
\end{tabular}}
\end{ruledtabular}
\end{table}

Next consider the time- and flavor-integrated charge asymmetry
 \be \label{eq:chargeA}
 \A_{\rho\pi}\equiv {|A_{+-}|^2+|\bar A_{+-}|^2-|A_{-+}|^2-|\bar
 A_{-+}|^2\over |A_{+-}|^2+|\bar A_{+-}|^2+|A_{-+}|^2+|\bar
 A_{-+}|^2}.
 \en
Then, following \cite{CKMfitter} one can transform the
experimentally motivated \CP parameters $\A_{\rho\pi}$ and
$C_{\rho\pi}$ into the physically motivated choices
 \be
 \acp(\rho^{+}\pi^-) &\equiv& {|\kappa^{-+}|^2-1\over |\kappa^{-+}|^2+1},
 \qquad  \acp(\rho^{-}\pi^+) \equiv {|\kappa^{+-}|^2-1\over |\kappa^{+-}|^2+1},
 \en
with
 \be
 \kappa^{+-}={q_{_{B}}\over p_{_{B}}}\,{\bar A_{-+}\over A_{+-}}, \qquad
  \kappa^{-+}={q_{_{B}}\over p_{_{B}}}\,{\bar A_{+-}\over A_{-+}}.
  \en
Hence,
 \be
 \acp(\rho^{+}\pi^-) &=& {\Gamma(\ov B^0\to \rho^{+}\pi^-)-\Gamma(B^0\to
 \rho^{-}\pi^+)\over \Gamma(\ov B^0\to \rho^{+}\pi^-)+\Gamma(B^0\to
 \rho^{-}\pi^+)}={\A_{\rho\pi}-C_{\rho\pi}-\A_{\rho\pi}\Delta C_{\rho\pi}\over 1-\Delta
 C_{\rho\pi}-\A_{\rho\pi} C_{\rho\pi}},   \non \\
 \acp(\rho^{-}\pi^+) &=& {\Gamma(\ov B^0\to \rho^{-}\pi^+)-\Gamma(B^0\to
 \rho^{+}\pi^-)\over \Gamma(\ov B^0\to \rho^{-}\pi^+)+\Gamma(B^0\to
 \rho^{+}\pi^-)}=-{\A_{\rho\pi}+C_{\rho\pi}+\A_{\rho\pi}\Delta C_{\rho\pi}\over 1+\Delta
 C_{\rho\pi}+\A_{\rho\pi} C_{\rho\pi}}. \non\\
 \en
Therefore,
direct \CP asymmetries $\acp(\rho^{+}\pi^-)$ and $\acp(\rho^{-}\pi^+)$ are determined from the above two equations and shown in Tables \ref{tab:VP} and \ref{tab:VPtreeCP}. Results for various {\it CP}-violating parameters in the decays $\bar B^0\to \rho^{\pm} \pi^\mp$ are displayed in Table \ref{tab:mixingBtorhopi}. The {\it CP}-violating quantity $\A_{\rho\pi}$ with the experimental value $-0.13\pm0.04$ is different from zero by 3.3$\sigma$ deviations. The QCDF prediction is in good agreement with experiment.

\section{$B\to VV$ decays}

\subsection{Branching fractions}
In two-body decays $B_{u,d}\to PP,VP,VV$, we have the pattern $VV>PV>VP> PP$ for the branching fractions of tree-dominated modes and $PP>PV\sim VV>VP$ for penguin-dominated ones, where  $B\to VP(PV)$ here means that the factorizable amplitude is given by $\la V(P)|J_\mu|B\ra\la P(V)|J^\mu|0\ra$. For example,
\be
&& \B(B^-\to \rho^-\rho^0)>\B(B^-\to \rho^-\pi^0)>\B(B^-\to \rho^0\pi^-)>\B(B^-\to\pi^-\pi^0), \non \\
&& \B(B^-\to  \bar K^0\pi^-)>\B(B^-\to \bar K^{*0}\pi^-)\sim\B( B^-\to \bar K^{*0}\rho^-)>\B(B^-\to \bar K^0\rho^-),
\en
for tree- and penguin-dominated $B^-$ decays, respectively.
The first hierarchy is due to the pattern of decay constants $f_V>f_P$ and the second hierarchy stems from the fact that the penguin amplitudes  are proportional to $a_4+r_\chi^P a_6$, $a_4+r_\chi^{V}a_6$,  $a_4-r_\chi^P a_6$ $a_4+r_\chi^{V}a_6$, respectively, for $B\to PP,PV,VP,VV$. Recall that $r_\chi^P\sim {\cal O}(1)\gg r_\chi^V$. There are a few exceptions to the above hierarchy patterns. For example, $\B(B^0\to\rho^0\rho^0)\lsim \B(B^0\to\pi^0\pi^0)$ is observed. This is ascribed to the fact that the latter receives a large soft correction to $a_2$, while the former does not.

There exist three QCDF calculations of $B\to VV$ \cite{BRY,ChengVV,BuchallaVV}. However, only the longitudinal polarization states of $B\to VV$ were considered in \cite{BuchallaVV}. The analyses in \cite{BRY,ChengVV} differ mainly in  (i) the values of the parameters $\rho_A$ and $\phi_A$ and (ii) the treatment of penguin annihilation contributions characterized by the parameters $\beta_i$ [see Eq. (\ref{eq:beta})] for penguin-dominated $VV$ modes. Beneke, Rohrer and Yang (BRY) applied the values $\rho_A(K^*\phi)=0.6$ and $\phi_A(K^*\phi)=-40^\circ$ obtained from a fit to the data of
$B\to K^*\phi$ to study other $\bar B\to VV$ decays. However, as pointed out in \cite{ChengVV}, the parameters  $\rho_A(K^*\rho)\approx 0.78$ and
$\phi_A(K^*\rho)\approx -43^\circ$ fit to the data of $B\to K^*\rho$ decays are slightly different from the ones $\rho_A(K^*\phi)$ and $\phi_A(K^*\phi)$. Indeed, we have noticed before that phenomenologically penguin annihilation should contribute less to $\phi K$ than $\rho K$ and $\pi K^*$.
This explains why the $K^*\rho$ branching fractions obtained by BRY are systematically below the measurements. Second, as noticed in \cite{ChengVV}, there
are sign errors in the expressions of the annihilation terms $A_3^{f,0}$ and $A_3^{i,0}$ obtained by BRY. As a consequence, BRY claimed (wrongly) that the longitudinal penguin annihilation amplitude $\beta_3^0$ is strongly
suppressed, while the $\beta_3^-$ term receives a sizable penguin annihilation contribution. This will affect the decay rates and longitudinal polarization fractions in some of $B\to K^*\rho$ modes, as discussed in details in \cite{ChengVV}.

In Table \ref{tab:VVBr}, QCDF results are taken from \cite{ChengVV} except that (i) a new channel $\bar B^0\to \omega\omega$ is added, and (ii) branching fractions and $f_L$ for $B\to (\rho,K^*)\omega$ decays are updated. \footnote{The $B\to\omega$ transition form factors were mistakenly treated to be the same as that of $B\to\rho$ ones in the computer code of \cite{ChengVV}. Here we use the light-cone sum rule results from \cite{Ball:BV} for $B\to\omega$ form factors.}
We see that the overall agreement between QCDF and experiment is excellent. In QCDF, the decay $\bar B^0\to\omega\rho^0$ has a very small rate
\begin{eqnarray} \label{eq:Btoomegarho}
-2A(\bar B^0\to\omega\rho^0) &\approx& A_{\rho\omega}\left[\delta_{pu}(\alpha_2-\beta_1)+2\hat \alpha_3^p+\hat\alpha_4^p\right]+A_{\omega\rho}\left[\delta_{pu}(-\alpha_2-\beta_1)+\hat\alpha_4^p\right],
\end{eqnarray}
due to a near cancelation of the the color-suppressed tree amplitudes. In view of this, it seems rather peculiar that
the rate of $\bar B^0\to\rho^0\omega$ predicted by pQCD \cite{Lu:rhoomega} is larger than QCDF by a factor of 20 and exceeds the current experimental upper bound. Likewise, $\B(B^-\to \bar K^{*0}\rho^-)$ obtained by pQCD is slightly too large.

\begin{table}[t]
\caption{$CP$-averaged branching fractions (in units of $10^{-6}$) and polarization fractions for $\bar B\to VV$ decays. For QCDF, the
annihilation parameters are specified to be $\rho_A=0.78$ and
$\phi_A=-43^\circ$ for $K^*\rho, K^*\bar K^*$ and $\rho_A=0.65$ and
$\phi_A=-53^\circ$ for $K^*\phi$ and $K^*\omega$ by default.
The world averages of experimental results are taken from
 \cite{HFAG}. The pQCD results are taken from \cite{Li:rhorho,Lu:rhoomega,Lu:KstV,Lu:KstKst}.
There are two distinct pQCD predictions for the branching fractions and longitudinal polarization fractions of $B\to K^*(\rho,\phi,\omega)$ decays, depending on the type of wave functions. Numbers in parentheses are for asymptotic wave functions. Estimates of uncertainties are not available in many of pQCD predictions.
} \label{tab:VVBr}
\begin{ruledtabular}
\begin{tabular}{l c c c c c c}
 &  \multicolumn{3}{c}{$\B$}
 &   \multicolumn{3}{c}{$f_L$} \\ \cline{2-4} \cline{5-7}
\raisebox{2.0ex}[0cm][0cm]{Decay}  & QCDF  & pQCD & Expt. &
QCDF & pQCD & Expt. \\ \hline
$B^-\to\rho^-\rho^0$                                       &
                                                           $20.0^{+4.0+2.0}_{-1.9-0.9}$ & $16.0^{+15.0}_{-~8.1}$  \footnotemark[1] & $24.0^{+1.9}_{-2.0}$ & $0.96^{+0.01+0.02}_{-0.01-0.02}$ & & $0.950\pm0.016$ \\
$\ov B^0\to\rho^+\rho^-$                                   &
                                                           $25.5^{+1.5+2.4}_{-2.6-1.5}$ & $25.3^{+25.3}_{-13.8}$  \footnotemark[1] & $24.2^{+3.1}_{-3.2}$ & $0.92^{+0.01+0.01}_{-0.02-0.02}$ & & $0.978^{+0.025}_{-0.022}$ \\
$\ov B^0\to\rho^0\rho^0$                                   &
                                                           $0.9^{+1.5+1.1}_{-0.4-0.2}$ &  $0.92^{+1.10}_{-0.56}$  \footnotemark[1] & $0.73^{+0.27}_{-0.28}$ & $0.92^{+0.03+0.06}_{-0.04-0.37}$ & 0.78 & $0.75^{+0.12}_{-0.15}$ \\
$B^-\to\rho^-\omega$                                       &
                                                           $16.9^{+3.2+1.7}_{-1.6-0.9}$ & $19\pm2\pm1$ & $15.9\pm2.1$ & $0.96^{+0.01+0.02}_{-0.01-0.03}$ & 0.97 & $0.90\pm0.06$ \\
$\ov B^0\to\rho^0\omega$                                   &
                                                           $0.08^{+0.02+0.36}_{-0.02-0.00}$ & $1.9\pm0.2\pm0.2$ & $<1.5$ & $0.52^{+0.11+0.50}_{-0.25-0.36}$ & 0.87 & $$ \\
$\ov B^0\to\omega\omega$                                   &
                                                           $0.7^{+0.9+0.7}_{-0.3-0.2}$ & $1.2\pm0.2\pm0.2$ & $<4.0$ & $0.94^{+0.01+0.04}_{-0.01-0.20}$ & 0.82 $$ \\
$B^-\to K^{*0}K^{*-}$                                      &
                                                           $0.6^{+0.1+0.3}_{-0.1-0.3}$ & $0.48^{+0.12}_{-0.08}$ & $1.2\pm0.5$ & $0.45^{+0.02+0.55}_{-0.04-0.38}$ & 0.82 & $0.75^{+0.16}_{-0.26}$ \\
$\ov B^0\to K^{*-}K^{*+}$                                  &
                                                           $0.1^{+0.0+0.1}_{-0.0-0.1}$ & $0.064^{+0.005}_{-0.010}$ & $<2.0$ & $\approx1$ & 0.99
                                                           \\
$\ov B^0\to K^{*0}\bar K^{*0}$                             &
                                                           $0.6^{+0.1+0.2}_{-0.1-0.3}$ & $0.35^{+0.13}_{-0.07}$ & $1.28^{+0.37}_{-0.32}$ \footnotemark[2] &  $0.52^{+0.04+0.48}_{-0.07-0.48}$ & 0.78 & $0.80^{+0.12}_{-0.13}$ \\
\hline
$B^-\to \bar K^{*0}\rho^-$ \footnotemark[3]                                &
                                                           $9.2^{+1.2+3.6}_{-1.1-5.4}$ & $17~(13)$ & $9.2\pm1.5$ & $0.48^{+0.03+0.52}_{-0.04-0.40}$ & $0.82~(0.76)$ & $0.48\pm0.08$ \\
$B^-\to K^{*-}\rho^0$                                      &
                                                           $5.5^{+0.6+1.3}_{-0.5-2.5}$ & 9.0 (6.4)& $<6.1$ & $0.67^{+0.02+0.31}_{-0.03-0.48}$ & 0.85 (0.78) & $0.96^{+0.06}_{-0.16}$ \footnotemark[4]  \\
$\ov B^0\to K^{*-}\rho^+$                                  &
                                                           $8.9^{+1.1+4.8}_{-1.0-5.5}$ & 13 (9.8)& $<12$ & $0.53^{+0.02+0.45}_{-0.03-0.32}$ & 0.78 (0.71)\\
$\ov B^0\to \bar K^{*0}\rho^0$                             &
                                                           $4.6^{+0.6+3.5}_{-0.5-3.5}$ & 5.9 (4.7) & $3.4\pm1.0$ & $0.39^{+0.00+0.60}_{-0.00-0.31}$ & 0.74 (0.68)& $0.57\pm0.12$ \\
$B^-\to K^{*-}\phi$ \footnotemark[5]                                        &
                                                           $10.0^{+1.4+12.3}_{-1.3-~6.1}$ & \footnotemark[6] & $10.0\pm1.1$ & $0.49^{+0.04+0.51}_{-0.07-0.42}$ & \footnotemark[6] & $0.50\pm0.05$
                                                           \\
$\ov B^0\to \bar K^{*0}\phi$                               &
                                                           $9.5^{+1.3+11.9}_{-1.2-~5.9}$ & \footnotemark[6]  & $9.8\pm0.7$ & $0.50^{+0.04+0.51}_{-0.06-0.43}$ & \footnotemark[6] & $0.480\pm0.030$ \\
$B^-\to K^{*-}\omega$                                      &
                                                           $3.0^{+0.4+2.5}_{-0.3-1.5}$ & 7.9 (5.5)& $<7.4$ & $0.67^{+0.03+0.32}_{-0.04-0.39}$ & $0.81~(0.73)$ & $0.41\pm0.19$
                                                           \\
$\ov B^0\to \bar K^{*0}\omega$                             &
                                                           $2.5^{+0.4+2.5}_{-0.4-1.5}$ & 9.6 (6.6)& $2.0\pm0.5$ & $0.58^{+0.07+0.43}_{-0.10-0.14}$ & $0.82~(0.74)$ & $0.70\pm0.13$ \\
\end{tabular}
\footnotetext[1]{There exist several pQCD calculations for $\rho\rho$  modes \cite{Lu:rhoomega,ChenVV,Li:rhorho}. Here we cite the NLO results from \cite{Li:rhorho}. }
\footnotetext[2]{This is from the BaBar data \cite{BaBar:KstKst}. The Belle's new measurement yields $(0.3\pm0.3\pm0.1)\times 10^{-6}$ \cite{Belle:KstK}.}
\footnotetext[3]{This mode is employed as an input for extracting the parameters $\rho_A$ and $\phi_A$ for $B\to K^*\rho$ decays.}
\footnotetext[4]{A recent BaBar measurement gives $f_L(K^{*-}\rho^0)=0.9\pm0.2$ \cite{BaBar:KVrhonew}, but it has only $2.5\sigma$ significance. }
\footnotetext[5]{This mode is employed as an input for extracting the parameters $\rho_A$ and $\phi_A$ for $B\to K^*\phi$ decays.}
\footnotetext[6]{See footnote 6 in Sec.VI.B.}
\end{ruledtabular}
\end{table}

We notice that the calculated  $B^0\to \rho^0\rho^0$ rate in QCDF  is $\B(B^0\to \rho^0\rho^0)=(0.88^{+1.46+1.06}_{-0.41-0.20})\times 10^{-6}$ for $\rho_C=0$  \cite{ChengVV},  while BaBar and Belle obtained $(0.92\pm0.32\pm0.14)\times 10^{-6}$ \cite{BaBarrho0rho0} and $(0.4\pm0.4^{+0.2}_{-0.3})\times 10^{-6}$ \cite{Bellerho0rho0}, respectively. Therefore, soft corrections to $a_2$ i.e. $\rho_C(VV)$ should be very small for $B^0\to\rho^0\rho^0$. Consequently, a pattern follows: Power corrections to $a_2$ are large for $PP$ modes, moderate for $VP$ ones and very small for $VV$ cases.
This is consistent with the observation made in \cite{Kagan} that soft power correction dominance is much larger for $PP$ than $VP$ and $VV$ final states.
It has been argued that this has to do with the special nature of the pion which is a $q\bar q$ bound state on the one hand and a nearly massless Nambu-Goldstone boson on the other hand \cite{Kagan}. The two seemingly distinct pictures of the pion can be reconciled by considering a soft cloud of higher Fock states surrounding the bound valence quarks.  From the FSI point of view, since $B\to\rho^+\rho^-$ has a rate much larger than $B\to \pi^+\pi^-$, it is natural to expect that $B\to\pi^0\pi^0$ receives a large enhancement from the weak decay $B\to\rho^+\rho^-$ followed by the rescattering of $\rho^+\rho^-$ to $\pi^0\pi^0$ through the exchange of the $\rho$ particle. Likewise, it is anticipated that
$B\to \rho^0\rho^0$ will receive a large enhancement via isospin final-state
interactions from $B\to \rho^+\rho^-$. The fact that the branching fraction of this mode is rather small
and is consistent with the theory prediction implies that the isospin phase
difference of $\delta_0^\rho$ and $\delta_2^\rho$ and the final-state interaction must be negligible
 \cite{Vysotsky}.

Both $\ov B^0\to \bar K^{*0}K^{*0}$ and $B^-\to K^{*0}K^{*-}$ are $b\to d$ penguin-dominated decays, while $\ov B^0\to K^{*-}K^{*+}$ proceeds only through weak annihilation. Hence, their branching ratios are expected to be small, of order $\lsim 10^{-6}$. However, the predicted rates for $\bar K^{*0}K^{*0}$ and $K^{*0}K^{*-}$ modes are slightly smaller than the data. Note that a new Belle measurement of $\B(\bar B^0\to K^{*0}\bar K^{*0})=(0.3\pm0.3\pm0.1)\times 10^{-6}<0.8\times 10^{-6}$ \cite{Belle:KstK} is smaller than the BaBar result $\B(\bar B^0\to K^{*0}\bar K^{*0})=(1.28^{+0.37}_{-0.32})\times 10^{-6}$ \cite{BaBar:KstKst}. Hence, the experimental issue with $B\to K^{*}\bar K^*$ decays needs to be resolved.

\subsection{Polarization fractions}
For charmless $\ov B\to VV$
decays, it is naively expected that the helicity
amplitudes $\bar \A_h$ (helicities $h=0,-,+$ ) for both tree- and penguin-dominated $\ov B \to VV$ decays respect the
hierarchy pattern
\be \label{eq:hierarchy}
\bar \A_0:\bar \A_-:\bar\A_+=1:\left({\Lambda_{\rm QCD}\over m_b}\right):\left({\Lambda_{\rm QCD}\over
m_b}\right)^2.
\en
Hence,  they are dominated by the longitudinal polarization
states and satisfy the scaling law, namely \cite{KaganVV},
 \be \label{eq:scaling}
f_T\equiv 1-f_L={\cal O}\left({m^2_V\over m^2_B}\right), \qquad {f_\bot\over f_\parallel}=1+{\cal
O}\left({m_V\over m_B}\right),
 \en
with $f_L,f_\bot$, $f_\parallel$ and $f_T$ being the longitudinal, perpendicular, parallel and transverse polarization fractions, respectively, defined as
 \be
 f_\alpha\equiv \frac{\Gamma_\alpha}{\Gamma}
                     =\frac{|\bar \A_\alpha|^2}{|\bar\A_0|^2+|\bar\A_\parallel|^2+|\bar\A_\bot|^2},
 \label{eq:f}
 \en
with $\alpha=L,\parallel,\bot$.
In sharp contrast to the
$\rho\rho$ case, the large fraction of transverse polarization of order 0.5 observed in $\bar B\to \bar K^*\phi$ and
$\bar B\to \bar K^*\rho$  decays at $B$ factories  is thus a surprise and poses an interesting challenge for
any theoretical interpretation.  Therefore, in order to obtain a large transverse polarization in $\bar B\to \bar K^*\phi,\bar K^*\rho$, this scaling law must be
circumvented in one way or another. Various mechanisms  such as sizable penguin-induced annihilation contributions \cite{KaganVV}, final-state
interactions \cite{Colangelo,CCSfsi}, form-factor tuning \cite{HNLi} and new
physics \cite{Yang&Das,YDYangnew,NP-tensor,newphysics}  have been proposed for solving
the $\bar B\to VV$ polarization puzzle.

As pointed out by Yang and one of us (HYC) \cite{ChengVV}, in the presence of NLO nonfactorizable corrections e.g. vertex,
penguin and hard spectator scattering contributions, effective Wilson coefficients $a_i^h$ are helicity dependent.
Although the factorizable helicity amplitudes $X^0$, $X^-$ and $X^+$ defined by Eq. (\ref{eq:Xh}) respect the scaling law (\ref{eq:hierarchy}) with $\Lambda_{\rm QCD}/m_b$ replaced by $2m_V/m_B$ for the light vector meson production, one needs to consider the effects of helicity-dependent Wilson coefficients: $\A^-/\A^0= f(a_i^-)X^-/[f(a_i^0)X^0]$.
For some penguin-dominated modes, the constructive (destructive) interference in the negative-helicity (longitudinal-helicity) amplitude of the $\ov B\to VV$ decay will render $f(a_i^-)\gg f(a_i^0)$ so that $\A^-$ is comparable to $\A^0$ and the transverse polarization is enhanced. For example, $f_L(\bar K^{*0}\rho^0)\sim 0.91$ is predicted in the absence of NLO corrections. When NLO effects are turned on,  their corrections on $a_i^-$ will render the negative helicity amplitude $\A^-(\bar B^0\to\bar K^{*0}\rho^0)$ comparable to the longitudinal one $\A^0(\bar B^0\to\bar K^{*0}\rho^0)$ so that even at the short-distance level, $f_L$ for $\ov B^0\to \bar K^{*0}\rho^0$ can be as low as 50\%. However, this does not mean that the polarization anomaly is resolved. This is because the calculations based on naive factorization
often predict too small rates for penguin-dominated $\bar B\to VV$ decays, e.g. $\bar B\to \bar K^*\phi$ and $\bar B\to \bar K^*\rho$, by a factor of
$2\sim 3$. Obviously, it does not make sense to compare theory with experiment for $f_{L,T}$ as the definition of polarization fractions depends on the partial rate and hence the prediction can be easily off by a factor of $2\sim 3$. Thus, the first important task is to have some mechanism to bring
up the rates. While the QCD factorization and pQCD \cite{Mishima} approaches
rely on penguin annihilation, soft-collinear effective theory invokes charming
penguin \cite{SCET} and the final-state interaction model considers final-state
rescattering of intermediate charm states \cite{Colangelo,Ladisa,CCSfsi}.
A nice feature of the $(S-P)(S+P)$ penguin annihilation is that it contributes to $\A^0$ and $\A^-$ with similar amount. This together with the NLO corrections will lead to $f_L\sim 0.5$ for penguin-dominated $VV$ modes. Hence, within the framework of QCDF
 we shall assume weak
annihilation to account for the discrepancy between theory and experiment, and fit the existing data of branching fractions and $f_L$ simultaneously by adjusting the
parameters $\rho_A$ and $\phi_A$.

For the longitudinal fractions in $\bar B\to \bar K^*\rho$ decays, we have the pattern (see also \cite{BRY})
\be \label{eq:KVrhofL}
f_L(K^{*-}\rho^0)> f_L(K^{*-}\rho^+)> f_L(\bar K^{*0}\rho^-)> f_L(\bar
K^{*0}\rho^0).
\en
Note that the quoted experimental value $f_L(K^{*-}\rho^0)=0.96^{+0.06}_{-0.16}$ in
Table \ref{tab:VVBr}  was obtained by BaBar in a previous
measurement where $K^{*-}\rho^0$ and $K^{*-}f_0(980)$ were not separated
\cite{BaBar:KVV}. This has been overcome in a recent BaBar measurement, but the
resultant value $f_L(K^{*-}\rho^0)=0.9\pm0.2$ has only 2.5$\sigma$ significance \cite{BaBar:KVrhonew}. At any rate,
 it would be important to have a refined
measurement of the longitudinal polarization fraction for $K^{*-}\rho^0$ and $\bar
K^{*0}\rho^0$ and a new measurement of $f_L(K^{*-}\rho^+)$ to test the hierarchy pattern (\ref{eq:KVrhofL}).

In the QCDF approach, we expect that the $b\to d$ penguin-dominated modes $K^{*0}K^{*-}$ and $K^{*0}\bar K^{*0}$ have $f_L\sim 1/2$ similar to the $\Delta S=1$ penguin-dominated channels. However, the data seem to prefer to $f_L\sim {\cal O}(0.75-0.80)$.
Due to the near cancelation of the color-suppressed tree amplitudes, the decay $\bar B^0\to \omega\rho^0$ is actually dominated by $b\to d$ penguin transitions. Hence, it is expected that $f_L(\rho^0\omega)\sim 0.52$. It will be interesting to measure $f_L$ for this mode.

For $\Delta S=1$ penguin-dominated modes, the pQCD approach predicts $f_L\sim 0.70-0.80$\,. \footnote{Early pQCD calculations of $B\to K^*\phi$ tend to give a large  branching fraction of order $15\times 10^{-6}$ and the polarization fraction $f_L\sim 0.75$ \cite{Chen:phiKst,Mishima}. Two possible remedies have been considered: a small form factor $A_0^{BK^*}(0)=0.32$ \cite{HNLi} and a proper choice of the hard scale $\bar\Lambda$ in $B$ decays \cite{ChenVV}. As shown in \cite{ChenVV}, the branching fraction of $\bar B^0\to \bar K^{*0}\phi$ becomes $8.9\times 10^{-6}$ and $f_L\sim 0.63$ for $\bar\Lambda=1.3$ GeV.}

\subsection{Direct \CP asymmetries}
Direct \CP asymmetries of $B\to VV$ decays are displayed in Table \ref{tab:VVCPdir}. They are small for color-allowed tree-dominated processes and large for penguin-dominated decays. Direct \CP violation is very small for the pure penguin processes $\bar K^{*0}\rho^-$ and $K^*\phi$.

\begin{table}[t]
\caption{Direct \CP asymmetries (in \%) of $\bar B\to VV$ decays. The pQCD results are taken from \cite{Lu:rhoomega} for $\rho\rho,\rho\omega, K^*\bar K^*$.   \CP asymmetries of $K^*\rho$ and $K^*\omega$ in the pQCD approach are shown in Fig. 4 of \cite{Lu:KstV} as a function of $\gamma$ and  only the signs of $\acp(K^*\rho)$ and $\acp(K^*\omega)$ are displayed here. Note that the definition of $\acp$ in \cite{Lu:KstV} has a sign opposite to the usual convention. } \label{tab:VVCPdir}
\begin{ruledtabular}
\begin{tabular}{l c c c }
{Decay} & QCDF (this work) & pQCD & Expt. \cite{HFAG} \\ \hline
$B^-\to\rho^-\rho^0$                                       &
                                                           $0.06$ & 0 & $-5.1\pm5.4$ \\
$\ov B^0\to\rho^+\rho^-$                                   &
                                                           $-4^{+0+3}_{-0-3}$ & $-7$ & $6\pm13$ \\
$\ov B^0\to\rho^0\rho^0$                                   &
                                                           $30^{+17+14}_{-16-26}$ & 80 & \\
$B^-\to\rho^-\omega$                                       &
                                                           $-8^{+1+3}_{-1-4}$ & $-23\pm7$ & $-20\pm9$ \\
$\ov B^0\to\rho^0\omega$                                   & $3^{+2+51}_{-6-76}$ &
                                                             & $$ \\
$\ov B^0\to\omega\omega$                                   &
                                                           $-30^{+15+16}_{-14-18}$ &  $$ \\
$B^-\to K^{*0}K^{*-}$                                      &
                                                           $16^{+1+17}_{-3-34}$ & $-15$ & \\
$\ov B^0\to K^{*-}K^{*+}$                                  &
                                                           $0$ & $-65$ &
                                                           \\
$\ov B^0\to K^{*0}\bar K^{*0}$                             &
                                                           $-14^{+1+6}_{-1-2}$
                                                           & 0 & \\
\hline
$B^-\to \bar K^{*0}\rho^-$                                 &
                                                           $-0.3^{+0+2}_{-0-0}$ & $+$ & $-1\pm16$ \\
$B^-\to K^{*-}\rho^0$                                      &
                                                           $43^{+6+12}_{-3-28}$ & $+$ & $20^{+32}_{-29}$  \\
$\ov B^0\to K^{*-}\rho^+$                                  &
                                                           $32^{+1+~2}_{-3-14}$ & $+$ & $$  \\
$\ov B^0\to \bar K^{*0}\rho^0$                             &
                                                           $-15^{+4+16}_{-8-14}$ & $-$ & $9\pm19$ \\
$B^-\to K^{*-}\phi$                                        &
                                                           $0.05$ & & $-1\pm8$
                                                           \\
$\ov B^0\to \bar K^{*0}\phi$                               &
                                                           $0.8^{+0+0.4}_{-0-0.5}$ & & $1\pm5$ \\
$B^-\to K^{*-}\omega$                                      &
                                                           $56^{+3+~4}_{-4-43}$ & $+$ & $29\pm35$
                                                           \\
$\ov B^0\to \bar K^{*0}\omega$                             &
                                                           $23^{+9+~5}_{-5-18}$ & $+$ & $45\pm25$ \\
\end{tabular}
\end{ruledtabular}
\end{table}

\subsection{Time-dependent \CP violation}
In principle, one can study time-dependent \CP asymmetries
for each helicity component,
 \be
 \A_h(t) &\equiv & {\Gamma(\ov
B^0(t)\to V_hV_h)-\Gamma(B^0(t)\to V_hV_h)\over
\Gamma(\ov
B^0(t)\to V_hV_h)+\Gamma(B^0(t)\to V_hV_h)} \non \\
 &=& S_h\sin(\Delta
 m t)-C_h\cos(\Delta m t).
 \en
Time-dependent \CP violation has been measured for the longitudinally polarized components of $\bar B^0\to \rho^+\rho^-$ and $\rho^0\rho^0$ with the results \cite{BaBar:Srhorho,Belle:Srhorho}:
\be
 S^{\rho^+\rho^-}_L=-0.05\pm0.17,  &\quad& C^{\rho^+\rho^-}_L=-0.06\pm0.13, \non \\
 S^{\rho^0\rho^0}_L=-0.3\pm0.7\pm0.2, &\quad& C^{\rho^0\rho^0}_L=0.2\pm0.8\pm0.3\,.
\en
In the QCDF approach we obtain
\be \label{eq:Srhorho}
&& \B(\rho^+\rho^-)_L=(24.7^{+1.6+1.3}_{-2.8-2.8})\times 10^{-6}, \quad S^{\rho^+\rho^-}_L=-0.19^{+0.01+0.09}_{-0.00-0.10},  \quad C^{\rho^+\rho^-}_L=0.11^{+0.01+0.11}_{-0.01-0.04}, \non \\
&& \B(\rho^0\rho^0)_L=(0.6^{+1.3+0.8}_{-0.3-0.3})\times 10^{-6}, \quad S^{\rho^0\rho^0}_L=0.16^{+0.05+0.50}_{-0.11-0.48}, \quad C^{\rho^0\rho^0}_L=-0.53^{+0.23+0.12}_{-0.25-0.48}.
\en
As pointed out in \cite{BN}, since (see Eq. (33) of \cite{BRY} and Eq. (106) of \cite{BN})
\be
S^{\rho^+\rho^-}_L=\sin 2 \alpha+2r_P\,\cos\delta_P\,\sin\gamma\,\cos 2\alpha+{\cal O}(r_P^2),
\en
with $P=|T|\,r_P\cos\delta_P$ and $\alpha=\pi-\beta-\gamma$, the measurement of $S^{\rho^+\rho^-}_L$ can be used to fix the angle $\gamma$ with good accuracy. For the QCDF predictions in Eq. (\ref{eq:Srhorho}) we have used $\beta=(21.6^{+0.9}_{-0.8})^\circ$ and $\gamma=(67.8^{+4.2}_{-3.9})^\circ$ \cite{CKMfitter}.

\section{Conclusion and Discussion}
We have re-examined the branching fractions and {\it CP}-violating asymmetries of charmless $\bar B\to PP,~VP,~VV$ decays  in the framework of QCD factorization. We have included subleading $1/m_b$ power corrections to the penguin annihilation topology and to color-suppressed tree amplitudes that are crucial for explaining the decay rates of penguin-dominated decays, color-suppressed tree-dominated $\pi^0\pi^0$, $\rho^0\pi^0$ modes and the measured \CP asymmetries in the $B_{u,d}$ sectors. A solution to the $\Delta A_{K\pi}$ puzzle requires  a large complex color-suppressed tree amplitude and/or a large complex electroweak penguin. These two possibilities can be discriminated in tree-dominated $B$ decays. The \CP puzzles with $\pi^-\eta$, $\pi^0\pi^0$ and the rate deficit problems with $\pi^0\pi^0,~\rho^0\pi^0$ can only be resolved by having a large complex color-suppressed tree topology $C$. While the New Physics solution to the $B\to K\pi$ \CP puzzle is interesting, it is irrelevant for tree-dominated decays.

The main results of the present paper are:

\vskip 0.1in \noindent{\it \underline{Branching fractions}} \vskip 0.1in

\renewcommand{\theenumi}{\roman{enumi})}
\begin{enumerate}


\item The observed abnormally large rates of $B\to K\eta'$ decays are naturally explained in QCDF without invoking additional contributions, such as flavor-singlet terms. It is important to have more accurate measurements of $B\to \pi\eta^{(')}$ to confirm the pattern $\B(B^-\to \pi^-\eta')\gg \B(\bar B^0\to \pi^0\eta')$.

\item The observed large rates of the color-suppressed tree-dominated decays
$\bar B^0\to\pi^0\pi^0,\rho^0\pi^0$ can be accommodated due to
the enhancement of $|a_2(\pi\pi)|\sim {\cal O}(0.6)$ and $|a_2(\pi\rho)|\sim {\cal O}(0.4)$.

\item The decays $\bar B^0\to\phi\eta$ and $B^-\to \phi\pi^-$ are dominated by the $\omega-\phi$ mixing effect. They proceed through the weak decays $\bar B^0\to\omega\eta$ and $B^-\to \omega\pi^-$, respectively, followed by $\omega-\phi$ mixing.


\item QCDF predictions for charmless $B\to VV$ rates are in excellent agreement with experiment.

\end{enumerate}

\vskip 0.1in \noindent{\it \underline{Direct \CP asymmetries}} \vskip 0.1in

\renewcommand{\theenumi}{\arabic{enumi})}
\begin{enumerate}
\item In the heavy quark limit, the predicted \CP asymmetries for the penguin-dominated modes $K^-\pi^+$, $K^{*-}\pi^+$, $K^-\rho^+$, $K^-\rho^0$, and tree-dominated modes $\pi^+\pi^-$, $\rho^\pm\pi^\mp$ (with $\acp$ defined in Eq. (\ref{eq:chargeA})) and $\rho^-\pi^+$ are wrong in signs when confronted with experiment. Their signs can be flipped into the right direction by the power corrections from penguin annihilation.

\item  On the contrary, the decays $K^-\pi^0$, $K^-\eta$, $\bar K^{*0}\eta$, $\pi^0\pi^0$ and $\pi^-\eta$ get wrong signs for their direct \CP violation when  penguin annihilation is turned on. These \CP puzzles can be resolved by having soft corrections to the color-suppressed tree coefficient $a_2$ so that $a_2$ is large and complex.

\item The smallness of the \CP asymmetry in $B^-\to \pi^-\pi^0$ is not affected by the soft corrections under consideration. This is different from the topological quark diagram approach where the color-suppressed tree topology is also large and complex, but $\acp(\pi^-\pi^0)$ is predicted to be of order a few percent.

\item
If the color-suppressed tree and electroweak penguin amplitudes are negligible compared to QCD penguins, \CP asymmetry differences of $K^-\pi^0$ and $K^-\pi^+$, $\bar K^0\pi^0$ and $\bar K^0\pi^-$, $K^{*-}\pi^0$ and $K^{*-}\pi^+$, $\bar K^{*0}\pi^0$ and $\bar K^{*0}\pi^-$ will be expected to be small. Defining $\Delta A_{K^{(*)}\pi}\equiv \acp(K^{(*)-}\pi^0)-\acp(K^{(*)-}\pi^+)$ and  $\Delta A'_{K^{(*)}\pi}\equiv \acp(\bar K^{(*)0}\pi^0)-\acp(\bar K^{(*)0}\pi^-)$, we found $\Delta A_{K\pi}=(12.3^{+3.0}_{-4.8})\%$, $\Delta A'_{K\pi}=(-11.0^{+6.4}_{-5.7})\%$, $\Delta A_{K^*\pi}=(13.7^{+4.6}_{-7.0})\%$ and $\Delta A'_{K^*\pi}=(-11.1^{+9.3}_{-6.9})\%$, while they are very small (less than 2\%) in the absence of power corrections to  the topological amplitude $c'$. Experimentally, it will be important to measure the last three \CP asymmetry differences.

\item For both $\bar B^0\to \bar K^0\pi^0$ and $\bar B^0\to \bar K^{*0}\pi^0$ decays, their \CP asymmetries are predicted to be of order $-0.10$ (less  than 1\%) in the presence (absence) of power corrections to $a_2$.
    The relation $\Delta A'_{K\pi}\approx -\Delta A_{K\pi}$ and the smallness of $\acp(\bar K^0\pi^-)$ give a model-independent statement that $\acp(\bar K^0\pi^0)$ is roughly of order $-0.15$.
    Hence, an observation of $\acp(\bar K^0\pi^0)$ at the level of $-(0.10\sim 0.15) $ will give a strong support for the presence of soft corrections to $c'$. It is also in agreement with the value inferred from the {\it CP}-asymmetry sum rule, or SU(3) relation or the diagrammatical approach. For $\bar B^0\to\bar K^0\rho^0$, we obtained $\acp(\bar K^0\rho^0)=0.087^{+0.088}_{-0.069}$.

\item
Power corrections to the color-suppressed tree amplitude is needed to improve the prediction for $\acp(\bar K^{*0}\eta)$. The current measurement $\acp(\bar K^{*0}\eta)=0.19\pm0.05$ is in better agreement with QCDF than pQCD and SCET.

\item There are 6 modes in which direct \CP asymmetries have been measured with significance above $3\sigma$: $K^-\pi^+,\pi^+\pi^-,K^-\eta,\bar K^{*0}\eta,K^-\rho^0$ and $\rho^\pm\pi^\mp$. There are also 7 channels with significance between $3.0\sigma$ and $1.8\sigma$ for \CP violation: $\rho^+K^-,K^{*-}\pi^+,K^-\pi^0, \pi^-\eta,\omega\bar K^0,\pi^0\pi^0$ and $\rho^-\pi^+$. We have shown in this work that the QCDF predictions of $\acp$ for aforementioned 13 decays are in agreement with experiment except the decay $\bar B^0\to \omega \bar K^0$. The QCDF prediction $\acp(\omega\bar K^0)=-0.047^{+0.058}_{-0.060}$ is not consistent with the experimental average, $0.32\pm0.17$. However, we notice that BaBar and Belle measurements of  $\acp(\omega\bar K^0)$ are of opposite sign.

\end{enumerate}

\vskip 0.1in \noindent{\it \underline{Mixing-induced \CP asymmetries}} \vskip 0.1in

\renewcommand{\theenumi}{\alph{enumi})}
\begin{enumerate}
\item
The decay modes $\eta'K_S$ and $\phi K_S$ appear theoretically very clean in QCDF;  for these modes
the central value of $\Delta S_f$ as well as the uncertainties are rather small.
\item
The QCDF approach predicts $\Delta S_{\pi^0 K_S}\approx 0.12$, $\Delta S_{\omega K_S}\approx 0.17$, and $\Delta S_{\rho^0 K_S}\approx -0.17$. Soft corrections to $a_2$ have significant effects on these three observables, especially the last one.
\item
For tree-dominated modes, the predicted $S_{\pi^+\pi^-}\approx -0.69$ agrees well with experiment, while $S_{\rho^0\pi^0}\approx -0.24$ disagrees with the data in sign.

\end{enumerate}

\vskip 0.1in \noindent{\it \underline{Puzzles to be resolved}} \vskip 0.1in

\renewcommand{\theenumi}{\roman{enumi})}
\begin{enumerate}
\item
Both QCDF and pQCD can manage to lead to a correct sign for $\acp(K^-\eta)$, but the predicted magnitude still falls short of the measurement $-0.37\pm0.09$. The same is also true for $\acp(\pi^+\pi^-)$.
\item
The QCDF prediction for the branching fraction of $B\to K^*\eta'$, of order $1.5\times 10^{-6}$, is smaller compared to pQCD and SCET.  Moreover, although the QCDF results are smaller than the BaBar measurements, they are consistent with Belle's upper limits. It will be crucial to measure them to discriminate between various predictions.

\item
\CP asymmetry of $\bar B^0\to\omega \bar K^0$ is estimated to be of order $-0.047$.  The current data $0.52^{+0.22}_{-0.20}\pm0.03$ by BaBar and $-0.09\pm0.29\pm0.06$ by Belle seem to favor a positive $\acp(\omega\bar K^0)$. This should be clarified by more accurate measurements.

\item

\CP violation of $\bar B^0\to\rho^0\pi^0$ is predicted to be of order $0.11$ by QCDF and  negative by pQCD and SCET. The current data are $0.10\pm0.40\pm0.53$ by BaBar and $-0.49\pm0.36\pm0.28$ by Belle. This issue needs to be resolved.

\end{enumerate}

In this work we have collected all the pQCD and SCET predictions whenever available and made a detailed comparison with the QCDF results. In general, QCDF predictions
for the branching fractions and direct \CP asymmetries of $\bar B\to PP,VP,VV$ decays are in good agreement with experiment except for a few remaining puzzles mentioned above. For the pQCD approach, predictions on the penguin-dominated $VV$ modes and tree-dominated $VP$ channels should be updated. Since the sign of $\acp(K^-\eta)$ gets modified by the NLO effects, it appears that all pQCD calculations should be carried out systematically to the complete NLO (not just partial NLO) in order to have  reliable estimates of \CP violation.

As for the approach of SCET, its phenomenological analysis so far is not quite successful in several places. For example, the predicted branching fraction $\B(B^-\to\rho^-\eta')\sim 0.4\times 10^{-6}$ is far below the experimental value of $\sim 9\times 10^{-6}$. The most serious ones are the \CP asymmetries for $K^-\pi^0,\pi^0\pi^0,\pi^-\eta$ and $\bar K^{*0}\eta$. The predicted signs of them disagree with the data (so the $\Delta A_{K\pi}$ puzzle is not resolved).
Also the predicted \CP violation for $\bar K^0\pi^0$ and $\bar K^{*0}\pi^0$ is of opposite sign to QCDF and pQCD. As stressed before, all the $B$-{\it CP} puzzles occurred in QCDF will also manifest in SCET  because the long-distance charming penguins in the latter mimic the penguin annihilation effects in the former.
This means that one needs other large and complex power corrections to resolve the \CP puzzles induced by charming penguins. For example, in the current phenomenological analysis of SCET, the ratio of $C^{(')}/T^{(')}$ is small and real to the leading order. This constraint should be released somehow.

\vskip 1.71cm {\bf Acknowledgments}

We are grateful to Chuan-Hung Chen, Cheng-Wei Chiang, Hsiang-nan
Li, Tri-Nang Pham and Amarjit Soni for valuable discussions. One of us (H.Y.C.) wishes to thank the
hospitality of the Physics Department, Brookhaven National
Laboratory. This research was supported in part by the National
Science Council of R.O.C. under Grant Nos. NSC97-2112-M-001-004-MY3 and NSC97-2112-M-033-002-MY3.


\end{document}